 \documentclass{aa}  

\usepackage{graphicx}
\usepackage{txfonts}
\usepackage{natbib}
\usepackage{mathtools} 
\usepackage{ amssymb }
\usepackage[usenames,dvipsnames]{color} 
\usepackage{multirow}

\definecolor{linkcolor}{rgb}{0.0,0.3,0.5}
\usepackage[unicode, colorlinks=true, linkcolor=linkcolor, citecolor=linkcolor, filecolor=linkcolor,urlcolor=linkcolor, pdfusetitle]{hyperref}

\def\Rsolar{$R_{\odot}$}
\def\Msolar{$M_{\odot}$}
\newcommand{\Mo}{\rm{M}_{\odot}}
\newcommand{\Ro}{R_{\odot}}

\begin{document} 

\title{Stellar triples on the edge}
   
    \subtitle{Comprehensive overview of the evolution of destabilised triples leading to stellar and binary exotica}

   \author{  S. Toonen \inst{1,2} \fnmsep\thanks{First and second author contributed equally } \fnmsep \thanks{\email{toonen@uva.nl}},
           T. C. N. Boekholt
 \inst{3,4,5}
\fnmsep $^\star$
\fnmsep \thanks{\email{tjardaboekholt@gmail.com}},
           S. Portegies Zwart
          \inst{6}
          }

   \institute{ 
   Anton Pannekoek Institute for Astronomy, University of Amsterdam, 1090 GE Amsterdam, The Netherlands
   \and
   Institute for Gravitational Wave Astronomy, School of Physics and Astronomy, University of Birmingham, Birmingham, B15 2TT, UK 
             \and 
             Rudolf Peierls Centre for Theoretical Physics, Clarendon Laboratory, Parks Road, Oxford, OX1 3PU, UK
            \and 
             CFisUC, Department of Physics, University of Coimbra, 3004-516 Coimbra, Portugal
             \and
             Instituto de Telecomunicações, Campus Universitário de Santiago, 3810-193 Aveiro, Portugal
             \and 
             Leiden Observatory, Leiden University, PO Box 9513, 2300 RA, Leiden, The Netherlands
             }

   \date{Received September 15, 1996; accepted March 16, 1997}

 
  \abstract
   { Hierarchical triple stars are ideal laboratories for studying the interplay between orbital dynamics and stellar evolution. Both mass loss from stellar winds and strong gravitational perturbations between the inner and outer orbit cooperate to destabilise triple systems.  }
   { Our current understanding of the evolution of unstable triple systems is mainly built upon results from extensive binary-single scattering experiments.
However, destabilised hierarchical triples cover a different region of phase space. Therefore, we aim to construct a comprehensive overview of the evolutionary pathways of destabilised triple-star systems. }
   { Starting from generic initial conditions, we evolved an extensive set of hierarchical triples
using the code TRES, combining secular dynamics and stellar evolution. We detected those triples that
destabilise due to stellar winds and/or gravitational perturbations. Their evolution was
continued with a direct N-body integrator coupled to stellar evolution.     }
   {The majority of triples (54-69\%) preserve their hierarchy throughout their evolution, 
which is in contradiction with the commonly adopted picture that unstable triples always experience a chaotic, democratic resonant interaction.
The duration of the unstable phase was found to be longer than expected ($10^{3-4}$ crossing times, reaching up to millions), so that long-term stellar evolution effects cannot be neglected. 
The most probable outcome is dissolution of the triple into a single star and binary (42-45\%). This occurs through the commonly known democratic channel, during which the initial hierarchy is lost and the lightest body usually escapes, but also through a hierarchical channel, during which the tertiary is ejected in a slingshot, independent of its mass.
Collisions are common (13-24\% of destabilised triples), and they mostly involve the two original inner binary components still on the main sequence (77-94\%). This contradicts the idea that collisions with a giant during democratic encounters dominate (only 5-12\%). 
Together with collisions in stable triples, we find that triple evolution is the dominant mechanism for stellar collisions in the Milky Way. Lastly, our simulations produce runaway and walk-away  stars  with  speeds  up  to  several  tens of km/s, with a maximum of a few 100km/s. We suggest that destabilised triples can explain --- or at least alleviate the tension behind --- the origin of the observed (massive) runaway stars.
}    
   {A promising indicator for distinguishing triples that will follow the democratic or hierarchical route, is the relative inclination between the inner and outer orbits. Its influence can be summed up in two rules of thumb: 1) prograde triples tend to evolve towards hierarchical collisions and ejections, and 2) retrograde triples tend to evolve towards democratic encounters and a loss of initial hierarchy, unless the system is compact, which experience collision preferentially. The trends found in this work complement those found previously from binary-single scattering experiments, and together they will help to generalise and improve our understanding on the evolution of unstable triple systems of various origins.  }

   \keywords{Methods: numerical --
                Stars: Binaries (including multiples) --
                Stars: kinematics and dynamics --
                Stars: evolution -- Stars: blue stragglers --
                Stars: white dwarfs --
               }

   \maketitle

\section{Introduction}
\label{sec:intro}

While the Sun is a single star, other stars are often found as part of small hierarchical stellar systems, consisting of two, three, or even more stars. 
While stellar evolution has been studied for centuries, and binary evolution has been studied for decades, our understanding of the evolution of triple stars and higher-order multiples is still limited. It was \cite{Harri68, Harri69} who studied the Von Zeipel-Lidov-Kozai mechanism for three stars, describing the Hamiltonian up to third order, determining orbital stability and demonstrating eccentricity-inclination variations in the inner orbit. Later, the octuple term in the three-body Hamiltonian was derived and its influence was investigated by \cite{Ford00} and \cite{Blaes02}, who also added relativistic precession in the inner orbit. It was demonstrated by \cite{Naoz11} that octuple order oscillations in the eccentricity and inclination can lead to orbital flips, that is a transition from prograde to retrograde inner orbits and vice versa. They also included tidal effects in the equations of motion. Further progress was made through the many  efforts of the community \citep[e.g.][]{Kis98,Kat11,Tho11,Mic14, Nao16b,Too20,Ham20}. It is an important step as hierarchical triples are abundant in the Galactic field. For solar-type stars, one in four binaries has a third companion \citep{Rag10,Tok14b}. And as these triples evolve into interacting (mass-transferring) systems  three times more often  than binaries \citep{Too20}, triples should play an important role in the formation of many close binaries. Moreover, for O- and B-type stars, the fraction of triples is much higher \citep{Eva11,San12} than for solar-type stars. The triple fraction of massive stars even surpasses that of massive-star  binaries \citep{Moe17}.

The reason why triples evolve so different from binaries is that the presence of the tertiary star can act as a catalyst for stellar interactions; three-body dynamics \citep[e.g.][]{Von1910,Koz62, Lid62, Harri68, Szebehely77, Egg95, Nao13, Nao16b, Luo16} lead to mass transfer in eccentric orbits  \citep[e.g.][]{Tho11, Shappee13, Mic14, Salas19, Stephan19, Too20}, enhanced tidal interactions  \citep[e.g.][]{Maz79, Kis98,Fab07, Liu15, Bat18}, and accelerates stellar mergers that give rise to bright transients such as supernova type Ia, gamma-ray bursts and gravitation wave emission  \citep[e.g.][]{Wen03, Per09, Tho11, Kat12, Ham13, Kimpson16, Stephan16, Por16, Antonini16, Ant17,  Too18b}. 
The dynamical impact of the tertiary is strongest when the system is less hierarchical and close to dynamically unstable. The hierarchy of an
 isolated stable triple can decrease and even breakdown because of its own evolution (see Sect.\,\ref{sec:ev} for an overview). The breakdown and destabilisation of the triples is also known as the triple evolution induced instability \citep[TEDI,][]{Per12}, and is the topic of this paper. 

Until recently, unstable triples were mostly considered in the context of stellar clusters. A binary captures an interloper star temporarily, that is a democratic resonance, after which the system dissolves again \citep{Heg75}. 
In-depth studies on the energy exchange of the binary with passing by single stars, and outcomes of democratic triple interactions, have been performed extensively both numerically \citep[e.g.][]{Hills75, Valtonen79, Hut83, Leigh12, TB15, Samsing18} and theoretically \citep[e.g.][]{Heg75, Monaghan76, Stone19, Ginat2020, Kol21}. 
They have provided us with some statistical intuition for the outcome. For example, we know that the least massive body is most likely to escape (but not always), thus leaving the two more massive bodies in a binary system \citep[e.g.][]{Ano86,Mikkola88,Ste98}. The eccentricity of the resulting binary is thermal if the triple was initially close to being virial, or superthermal if it was rather cold \citep{Heg75}. Its binding energy is independent of the mass ratio and follows a power law with an index determined by the total angular momentum \citep[e.g.][]{TB15}. The trajectories of the stellar components evolve on dynamical timescales, such that stellar evolution (SE) is typically of secondary importance. 

In this paper we show that these expectations can not simply be extrapolated to destabilised triples in the field. 
Due to the presence of chaos, most individual binary-single scattering experiments tend to be unpredictable in the sense that a small perturbation, such as a numerical error, is magnified exponentially, resulting in diverged trajectories in phase space \citep{TB20}, for example the triple loses memory of its initial condition. 
Statistical distributions however seem to be predictable nevertheless and only depend on the global conserved quantities, such as energy, angular momentum and mass ratios \citep{Valtonen06}. 
However, the destabilised triples in the field have a very different type of initial condition compared to unstable triples formed in clusters.
 First of all, the single third body is on an elliptical orbit instead of a hyperbolic orbit with respect to the binary. 
Secondly, the inner binary and the outer binary are usually separated from each other such that they live on the stable side of the instability border for hierarchical triples \citep{Mar01b}. Generally, the distributions of energy and angular momentum are thus expected to be different for hierarchical triples compared to binary-single scatterings.

We follow-up on the work of \cite{Too20} who simulated the evolution of isolated hierarchical triples with realistic initial conditions of young populations. Here we focus on those isolated triples that are born dynamically stable, but become unstable as a result of the internal evolution of the stars and by the dynamical interplay among the three stars. We perform self-consistent simulations of the stable phase as well as the unstable phase (Sect.\,\ref{sec:method}). This includes full triple evolution calculations with \texttt{TRES} \citep{Too16} for the former \citep[similar to][]{Too20}, and N-body calculations linked with stellar evolution for the latter. 
We explore and quantify the various outcomes in Sect.\,\ref{sec:results}, such as stellar collisions and runaway stars, place them in the observational context in Sect.\,\ref{sec:obs}, and summarise our results in Sect.\,\ref{sec:concl}.

\section{Evolutionary pathway}
\label{sec:ev}

Except for periodic braids \citep{0951-7715-11-2-011}, dynamically stable triples are hierarchical, in which two of the stars form a binary pair, and the tertiary orbits the centre of mass of the binary with an orbital period much exceeding that of the inner binary. The inner and the outer orbit can be approximated by Keplerian orbits, and three-body dynamics manifests itself as an perturbation on and interaction between these two orbits. At the lowest order of the secular approximation (i.e. the quadrupole order) these are Lidov-Kozai cycles \citep{Von1910,Lid62,Koz62} in which the mutual inclination and inner eccentricity vary periodically. At higher orders such as the octupole level, more extreme inner eccentricities can be achieved, as well as flips in the orbital orientation \citep[see][ for a comprehensive review]{Nao16}. The long-term evolution of hierarchical systems is governed by the interplay between three-body dynamics and stellar evolution \citep{Too16,Too20}.

The boundary between stable and unstable systems is not easily defined (or definable), and therefore various stability criteria exist, which are based a.o. on the concept of chaos, on the escape of one of the stellar components, on zero velocity surfaces of the restricted three body problem, or on numerical integrations \citep[][for a review]{Geo08}. Here we have adopt the stability criterion of \citet{Mar99} including a factor that reflects the dependence on inclination $i$ \citep{Aar01}: 

\begin{eqnarray}
\begin{array}{l c l}
\dfrac{a_{\rm out}}{a_{\rm in}}|_{\rm crit} &= &\dfrac{2.8}{1-e_{\rm out}} (1-\dfrac{0.3i}{\pi})  \\
&&\\
&&\left( \dfrac{(1.0+q_{\rm out}) (1+e_{\rm out})}{\sqrt{1-e_{\rm out}}} \right)^{2/5}, \\
 \end{array} 
\label{eq:stab_crit}
\end{eqnarray}
with $a_{\rm in}$ the orbital separation of the inner orbit,
$a_{\rm out}$ the orbital separation of the outer orbit, 
$e_{\rm out}$ the eccentricity of the outer orbit, and 
the mass ratio $q_{\rm out}\equiv \dfrac{m_3}{m_1+m_2}$ with 
$m_3$ the mass of the star in the outer orbit (hereafter tertiary), and $m_1$ and $m_2$ the masses of the two stars in the inner orbit (hereafter primary and secondary respectively). By definition, the primary refers to the initially most massive star, such that $m_1 > m_2$ initially. Also 
$q_{\rm in}\equiv \dfrac{m_2}{m_1}$.
Triple systems are unstable if $\dfrac{a_{\rm out}}{a_{\rm in}} < \dfrac{a_{\rm out}}{a_{\rm in}}|_{\rm crit}$. 
This stability criterion is based on the concept of chaos and the consequence of overlapping resonances, and later confirmed with direct n-body integrations to work well for a wide range of parameters \citep{Aar01, Aar04,He18}. 

Eq.\,\ref{eq:stab_crit} shows that in general a hierarchical system can move over the stability limit (or better cross into the instability region) when 1) $a_{\rm out}/a_{\rm in}$ decreases, 2) the outer mass ratio $q_{\rm out}$ increases, 3) the outer eccentricity $e_{\rm out}$ increases, or 4) the mutual inclination decreases.  We mention five processes that can cause these orbital parameters to change in a given triple:
\begin{enumerate}
    \item \textbf{Three-body interactions}: If we consider pure three-body dynamics, and focus on the lowest order of the secular approximation, that is the quadrupole order, the outer orbit does not change. This is generally not the case anymore for higher order approximations. For example, when including the octupole term \citep[see review of][]{Nao16} the outer eccentricity varies on a timescale\footnote{in the limit of circular outer orbits} of:

\begin{equation}
\tau_{\rm oct} \sim \frac{256\sqrt{10}}{15\pi\sqrt{\epsilon_{\rm oct}}}\,  \tau_{LK},
\label{eq:t_oct}
\end{equation}
\citep{Ant15} with the timescale of the Lidov-Kozai cycles \citep{Kin99}
\begin{equation}
t_{\rm LK} \sim \alpha  
\frac{P_{\rm out}^2}{P_{\rm in}} 
\frac{m_1+m_2+m_3}{m_3} \left(1-e_{\rm out}^2\right)^{3/2},
\label{eq:t_kozai}
\end{equation}
 and the octupole parameter \citep{Lit11, Kat11, Tey13,Li14}. 
\begin{equation}
\epsilon_{\rm oct} \equiv \frac{m_1-m_2}{m_1+m_2} \frac{a_{\rm in}}{a_{\rm out}} \frac{e_{\rm out}}{1-e_{\rm out}^2}.
\label{eq:e_oct}
\end{equation}
Generally the octupole term is of importance if  $|\epsilon_{\rm oct}| \gtrsim 0.001-0.01$.

    \item \textbf{Stellar winds:} 
    Fast spherically symmetric winds that do not interact with the stellar companions of the donor star have an adiabatic effect of the orbit. From an angular momentum balance, one can show that the orbit widens as 
    
\begin{equation}
\frac{a'}{a} = \frac{m_1+m_2}{m_1'+m_2'},
\label{eq:wind}
\end{equation}
    where ' denotes quantities after the wind mass loss. If mass is lost from the inner binary of the triple, then
    
 \begin{equation}
\frac{a_{\rm out}'}{a_{\rm in}'} = \frac{a_{\rm out}}{a_{\rm in}} \, \frac{m_1+m_2+m_3}{m_1'+m_2'+m_3'} \, \frac{m_1'+m_2'}{m_1+m_2} < \frac{a_{\rm out}}{a_{\rm in}}.
\label{eq:wind_triple}
\end{equation}
In other words, as the fractional mass loss in the inner binary is larger than that of the triple as a whole, the relative widening of the inner orbit is larger than that of the outer orbit, such that the ratio $ a_{\rm out}/ a_{\rm in}$ reduces in response to wind mass loss. When the two orbits approach one another, the triple system can become unstable as described by Eq.\,\ref{eq:stab_crit}. In this way dynamically stable triple systems can become unstable due to their internal stellar winds \citep[e.g.][]{Kis94,Ibe99,Per12}. \\
We also note that the wind mass loss from the inner binary directly affects the critical orbital separation ratio $\dfrac{a_{\rm out}}{a_{\rm in}}|_{\rm crit}$  through its mass (i.e. $1+q_{\rm out}$) dependence (Eq.\ref{eq:stab_crit}). As the stability criterion is only mildly dependent on mass, the direct effect of mass loss on the stability of the triple is of secondary importance compared to the orbital effect of stellar wind mass losses.     

Another scenario induced by stellar wind mass loss is mass loss induced eccentric Kozai cycles (MIEK) \citep{Shappee13}. In this case, the triple exhibits no Von Zeipel-Lidov-Kozai cycles initially, but the mass loss pushes the triple into a new region of parameter space where the cycles do occur. During phases of extreme eccentricity, tidal forces and large radii during the giant phase facilitate stellar collisions \citep[e.g.][]{Mic14, Nao16b, Stephan16, Stephan21}. Therefore, depending on the initial configuration of the triple, and the amount of mass loss, the triple might follow the MIEK-track, or become dynamically unstable and even disrupt. 

    \item \textbf{Mass transfer:}
Another internal process that can make a triple cross the stability limit, is mass transfer in the inner binary \citep[e.g.][]{Kis94,Ibe99,Fre11,Por11}. If we simplistically assume that the mass transfer is conservative, and that stellar rotational angular momentum can be neglected compared to orbital angular momentum, then angular momentum conservation dictates

 \begin{equation}
\frac{a'}{a} = \left( \frac{m_{\rm d}\, m_{\rm a}} {m_{\rm d}'\,m_{\rm a}'}  \right) ^2,
\label{eq:mt}
\end{equation}
where $m_{\rm d}$ is the mass of the donor star and 
 $m_{\rm a}$ that of the accretor star. For a full description (including less stringent assumptions), see \cite{Sob97} and \cite{Too16}. If the inner binary widens in response to the conservative mass transfer, and approaches the non-changing outer orbit, the triple system can cross the stability limit.

    \item \textbf{Supernova explosions} 
    During the formation of a neutron star or black hole, a kick may be imparted to the compact object, as suggested by Galactic NS studies \citep{Gun70,Cor93, Lyn94,Cha05,Hob05,  Bec12, Ver17, Igo20}. Depending on the magnitude and orientation of the kick it may unbind an orbit all together or simply change it. Besides this direct affect on the inner and outer orbit of a triple, there is the additional affect that a supernova explosion in the inner binary changes its centre-of-mass (position, velocity, mass) which affects the outer orbit \citep{Pij12}. As a result the post-supernova trajectories of the stars may result in unbound triples, unstable triples, or dynamically more interesting triples (i.e. in which the importance of three-body dynamics has increased) \citep{Pij12, Tau14, Lu19}. 
   An interesting prospect of a supernova explosion of the tertiary star is the birth of a runaway binary \citep{Gao19}.

    \item \textbf{External effects}
    External effects can be separated in two classes, 
    the perturbations induced by the background galactic tidal field with the occasional relatively wide encounter with another star, and the strong influence of nearby stars in a dense stellar cluster.
    \begin{itemize}
        \item \textbf{in the birth cluster}
    The latter is most important in the earliest phase when the triple system is still embedded in its birth clustered environment.
    This phase was studied by \cite{2007MNRAS.379..111V}, who performed direct N-body simulations of star clusters with primoridal triples, and binary evolution (mass transfer in triples was not taken into account in these calculations).
    Subsequent studies \citep{Lei13, Fra20} studied the effect of encounters in a clustered environment on wide outer triple orbits.
    \item \textbf{in the galactic field}
Due to their large cross sections,  wide triples  may be disturbed by external effects even in the field. These include torques from the Milky Way's tide, rare, but strong encounters with passing field stars, or a cumulative effect from many weak encounters which change the periastron of the widest orbits   through eccentricity pumping \citep{Kai14,Mic16}. Recently, especially the latter effect has been provoked to lead to the formation of compact binaries and mergers \citep{Mic16, Mic19, Ham19,  Saf20, Mic20}. 
\end{itemize}

\end{enumerate}

In this paper we consider triples that become unstable due to the first and second process.

\begin{table*}[ht!]
\centering
\caption{Distributions of the initial stellar masses, rotation, and orbital parameters. See also Fig.\,1-3 in \cite{Too20} for a graphical representation of these parameters.}
\begin{tabular}{|l|c|ccc|}
\hline
Parameter & Range & Model OBin & Model T14$^{(1)} $  & Model E09$^{(2)} $ \\
\hline \hline
Mass of primary $m_1$ & 1-7.5\Msolar & Kroupa IMF$^{(3)}$& Kroupa IMF$^{(3)}$& Kroupa IMF$^{(3)}$  \\
\multirow{2}{*}{Mass ratio}&$0-1^{*}$ & \multirow{2}{*}{Uniform$^{(4,5,6)}$} & \multirow{2}{*}{Uniform$^{(4,5,6)}$} & \multirow{2}{*}{Model E09$^{(2)}$}\\
 &$m_2$, $m_3 > 0.008$\Msolar & & &\\
Stellar rotation & & Hurley rotation$^{(7)} $ &Hurley rotation$^{(7)}$  & Hurley rotation$^{(7)}$ \\
Orbital separation & $5-5\cdot 10^6$\Rsolar$^{**}$ & Uniform in log($a$)$^{(8,9)}$ & Log-normal$^{***(1,4,11)}$ & Model E09$^{(2)}$\\
Eccentricity & $0-1$ & Thermal distribution$^{(10)} $ & Thermal distribution$^{(10)} $& Thermal distribution$^{(10)} $\\
Inclination & $0-\pi$ &Uniform in cos($i$)$^{(12)}$  & Uniform in cos($i$)$^{(12)}$  &Uniform in cos($i$)$^{(12)}$   \\
Argument of pericentre & --$\pi-\pi$ & Uniform& Uniform& Uniform\\
\hline
\end{tabular}
\label{tbl:init_param}
\begin{flushleft}
\tablefoot{
$^{*}$ Except for the outer mass ratio $q_{\rm out}$ in model E09; 
$^{**}$ Formally the maximum orbital separation for Model T14 is $10^{12}$\Rsolar. Due to the shape of the log-normal distribution, the contribution from systems with $a_{\rm out} \gtrsim 5 \cdot 10^{7}$\Rsolar\, is negligible; 
$^{***}$ A log-normal distribution with periods in days, mean $\mu = 5$ and dispersion $\sigma=2.3$;
References:
$^{(1)}$\citet{Tok14b};        
$^{(2)}$\citet{Egg09};         
$^{(3)}$ \citet{Kro93};        
$^{(4)}$\citet{Rag10};         
$^{(5)}$ \citet{Duc13};        
$^{(6)}$ \citet{San12};        
$^{(7)}$ \citet{Hur00};        
$^{(8)}$\citet{Abt83};         
$^{(9)}$ \citet{Kou07};        
$^{(10)}$ \citet{Heg75};       
$^{(11)}$\citet{Duq91};        
$^{(12)}$\citet{Tok17};        
}
\end{flushleft}
\end{table*}

\section{Method}
\label{sec:method}
In this study we simulate the evolution of a population of stellar triples with a two-step approach. In the first part of the triple's evolution when the system is dynamically stable, we use the triple population synthesis code \texttt{TRES} \citep{Too16} which is based on the secular approach \citep{Von1910,Lid62,Koz62,Nao16}. In this phase-averaged approach, the dynamics follow from a perturbative expansion of the Hamiltonian in terms of $a_{rm in}/a_{\rm out}$.   
When the triple crosses the stability boundary given by Eq.\,\ref{eq:stab_crit}, and the secular approach is no longer valid, we switch to a N-body based approach. Below we describe the set-up of each approach in detail.

\subsection{Hierarchical phase}
\label{sec:hier}
In this work we employ the triple evolution code \texttt{TRES} to generate large populations of triple systems on the zero-age main-sequence (MS), simulate their subsequent evolution and extract those systems that become dynamically unstable within 13.5 Gyr.

At every time-step, processes such as stellar evolution (including stellar winds),  tidal interaction, mass transfer, and three-body dynamics are considered with the appropriate prescriptions. The (single) stellar evolution is calculated with the \texttt{SeBa}-code \citep{Por96, Too12} which is based on the stellar evolutionary tracks of \cite{Hur00}. The tidal method used by \texttt{TRES} is based on the weak-friction equilibrium tide model \citep{Hut81, Hur02} and is appropriate for moderate eccentricities \citep{Moe18} (see sect.\,\ref{sec:discussion} for a discussion). 
Besides precession due to the secular dynamics, precession due to general relativistic effects \citep{Bla02}, by tides \citep{Sme01} and by intrinsic stellar rotation \citep{Fab07} are included in \texttt{TRES} as well. Three-body dynamics is included based on the double-averaged approach up to and including the octupole term \citep{Har68,For00,Nao13}. 
\texttt{TRES} is incorporated into the Astrophysics MUltipurpose Software Environment, or AMUSE \citep[][see also amusecode.org]{Por18}. For a full description of \texttt{TRES}, see \cite{Too16}.

To generate the initial population of triples, we apply three models (i.e. OBin, T14 and E09) that are similar to those used in \cite{Too20}. Model T14 and E09 are based on observed samples of triples constructed in \cite{Tok14b} and \cite{Egg09} respectively, whereas model OBin is based on our understanding of primordial binaries. We focus on primary stars in the mass range $1\Mo\leq m_1<7.5\Mo$, such that the stars experience significant stellar evolution in a Hubble time, while avoiding a supernova explosion. An overview of the most important ingredients of the models is given in Tbl.\,\ref{tbl:init_param}. For a detailed discussion on the models, see \cite{Too20}. After drawing the initial stellar and orbital parameters from the distributions given in Tbl.\,\ref{tbl:init_param},  we require that the triple is dynamically stable initially, according to Eq.\,\ref{eq:stab_crit}. 
In doing so, we neglect the dynamical and stellar evolution on the pre-MS and its effect on the stability of the triple system \citep[but see e.g. ][]{Moe18}. 
The adopted requirement regarding dynamical stability biases the population towards larger ratios of $a_{\rm out}/a_{\rm in}$ and away from high eccentricities \citep[see also Fig.\,1-3 in][]{Too20}. The latter is in line with observations of wide binaries \citep{Tok16, Rag10, Moe17}.

To calculate Galactic event rates, we assume a constant star formation history of 3\Msolar per year, a maximum stellar mass of 100\Msolar, and a minimum primary mass of 0.08\Msolar. Furthermore, we assume that 15\% of stellar systems are triples, and 50\% are binaries, leaving 35\% of stellar systems to be truly single stars \citep{Tok14b}.

\subsection{ Dynamically unstable phase}
\label{sec:dyn}

Once a triple has become unstable according to the stability criterion, we halt the integration with \texttt{TRES}, and continue with a direct N-body integration. The last snapshot of \texttt{TRES} gives the age, masses, radii, wind mass loss rates, and orbital elements, which serve as the initial condition for the direct N-body code. Due to the secular nature of \texttt{TRES}, no mean anomaly is available. In order to extend our sample and to investigate any dependencies on the mean anomaly, we create 10 instances of each triple where the mean anomaly of the outer orbit is uniformly sampled between 0 and 360 degrees, while the inner orbit starts at apocentre. 

We distinguish three types of direct simulations:

\begin{enumerate}
    \item pure N-body,
    \item N-body with uniform wind,
    \item N-body with stellar evolution (i.e. radius evolution and time-dependent mass loss).
\end{enumerate}
    
\noindent Type 1 is appropriate if the unstable phase is relatively short, such that winds play a negligible role. It is often assumed that once a hierarchical triple becomes unstable, the final disintegration follows after only a few dynamical time scales. We show, however, that this assumption is somewhat naive, as the duration of the unstable phase is on average thousands of crossing times, even reaching up to tens of millions. The crossing time is the time for three stars to encounter each other once during a democratic resonance Here we define it as $t_{\rm cross} \equiv R_v/V$, with the characteristic size $R_v = \frac{GM_{\rm tot}^2}{-2E_{\rm tot}}$ and velocity $V = \sqrt{-2E_{\rm tot}/M_{\rm tot}}$, where the total mass $M_{\rm tot}=m_1+m_2+m_3$ and total energy $E_{\rm tot}$ is the sum of the kinetic and potential energy of the system.

In our type 2 ensemble, we therefore include a uniform wind mass loss rate. We adopt the last value given by \texttt{TRES}, and assume it to be constant until the mass has reached a value of $0.5\,\Mo$. Its radius, which was constant up to that point, is then set to $0.01\,\Ro$ and the object is deemed a white dwarf. The main difference with the type I simulations is thus the mass loss experienced by stars with winds, and the finite time frame within which the giant stars retain their large radii and corresponding collisional cross section.

Our type 3 ensemble can be considered as the most realistic one, as this implements detailed mass-radius tracks obtained from the stellar evolution code \texttt{SeBa} \citep{Por96, Too12} as used in \texttt{TRES} as well. This also allows stars which initially do not have any wind, to become giants later on during the simulation. This was not possible for the type 1 and 2 simulations. By comparing results from the three types of simulations we can determine the role of stellar evolution during the unstable phase of hierarchical triple stars. 

The N-body code is the commonly used fourth-order Hermite integrator \citep{1991ApJ...369..200M}. Each time step, we update the masses and radii of the stars according to the uniform wind mass loss rate, or the tracks from \texttt{SeBa}. For the type 3 ensemble, we first run the \texttt{SeBa} code separately, and store the mass-radius tracks of the three stars in a table. This table is then fed to the N-body code, which adds the wind at the time step level. 
Here the time step criterion is adaptive and resolves both changes in the orbit (by taking the minimum time scale proportional to relative distance divided by relative speed over all pairs), and in the mass (by taking the minimum time scale proportional to mass divided by wind mass loss rate).
The direct integrations do not include relativity, tides or other phenomena such as mass transfer or tidal disruption. Although each of these ingredients affect the outcome of triple evolution, it is our main aim here to study the interplay between dynamics and stellar evolution. The results we present here also serve as a benchmark for follow up studies on destabilised triples in which more physical ingredients are included.

The stopping conditions for a simulation are the following: 
\begin{itemize}
    \item Age has reached a Hubble time (Bound),
    \item Unbound binary-single pair (Escape),
    \item Bound binary-single pair separated by a parsec (Drift),
    \item Overlap of stellar radii (Collision),
    \item Three unbound stars (Ionisation),
    \item CPU time exceeded maximum of 24 hours (CPU) (see app.~\ref{app:tcpu} for motivation).
\end{itemize}

The criterion for escape is based on three constraints \citep{Standish1971}: the single body is a certain distance away from the binary's centre of mass ($\Delta r \ge 100a$, with $a$ the semimajor axis), the single body is moving away from the binary's centre of mass, and finally the single body has a positive energy.
If the third constraint was not met, but the separation between the single body and the binary reached one parsec, then we also stop the simulation and deem the triple to be a drifter. At these large separations we can no longer treat the triple as being isolated. Drifters are expected to occur when the single body is ejected to large separations, but would turnaround at only very large distances \citep{Orlov10}.
Each time step we check if any two stars overlap their radii, in which case we categorise the triple as a collision. To analyse the collisional triples in further detail, we store the collision components (primary, secondary, tertiary), their stellar types at the moment of collision, and whether the non-collision component would still be bound or not to the collision product, if replaced by a conservative, centre of mass body. It is also theoretically possible that the triple ionises into three single, unbound bodies. This outcome is very rare and can only occur if the triple is already loosely bound, such that stellar winds in the inner binary can become impulsive \citep{Hills83, Ver11, Too17}. The final stopping condition is not a physically motivated one, but a practical one. From previous studies on the lifetime of unstable triple systems \citep[e.g.][]{Orlov10}, it is known that the lifetime distribution has an algebraic tail towards exceedingly long lifetimes. Using direct integration, these systems will not fulfil any physical stopping condition within a reasonable simulation time (see App.~\ref{app:tcpu}). In an iterative manner, we increase the maximum CPU time, and we show that if it is set to 24 hours, we achieve a percentage of unfinished systems below 10\% in the worst case. We discuss this and other limitations of our direct model in Sec.~\ref{sec:limit} and App.~\ref{app:tcpu}. 

During the simulation we keep track of whether the initial hierarchy of the triple is preserved or not.
If at any instance this is not the case, we tag the triple as having lost its initial hierarchy, and deem it to be democratic \citep{Bahcall83}. 
If the outcome of the simulation produced an escaper or a drifter, we determine which initial components is now the single star. We tag the triple as `Democratic-X', with X denoting the single star (1= primary, 2=secondary, 3=tertiary).  
It is important to note however, that we found our criterion to be rather strict as some triples are able to preserve their initial hierarchy, even though the tertiary might temporarily have a close encounter with either of the inner binary components during pericentre passage. Therefore, some triples tagged as `Democratic-3' could also have been tagged as `Hierarchical'. The flag proved particularly useful as it shows that the majority of triple systems tend to preserve their hierarchy, both in the cases of a collision and a dynamical dissolution.

\begin{figure*}[]
    \centering
\includegraphics[width=\textwidth,clip=true, trim =0mm 80mm 0mm 0mm,]{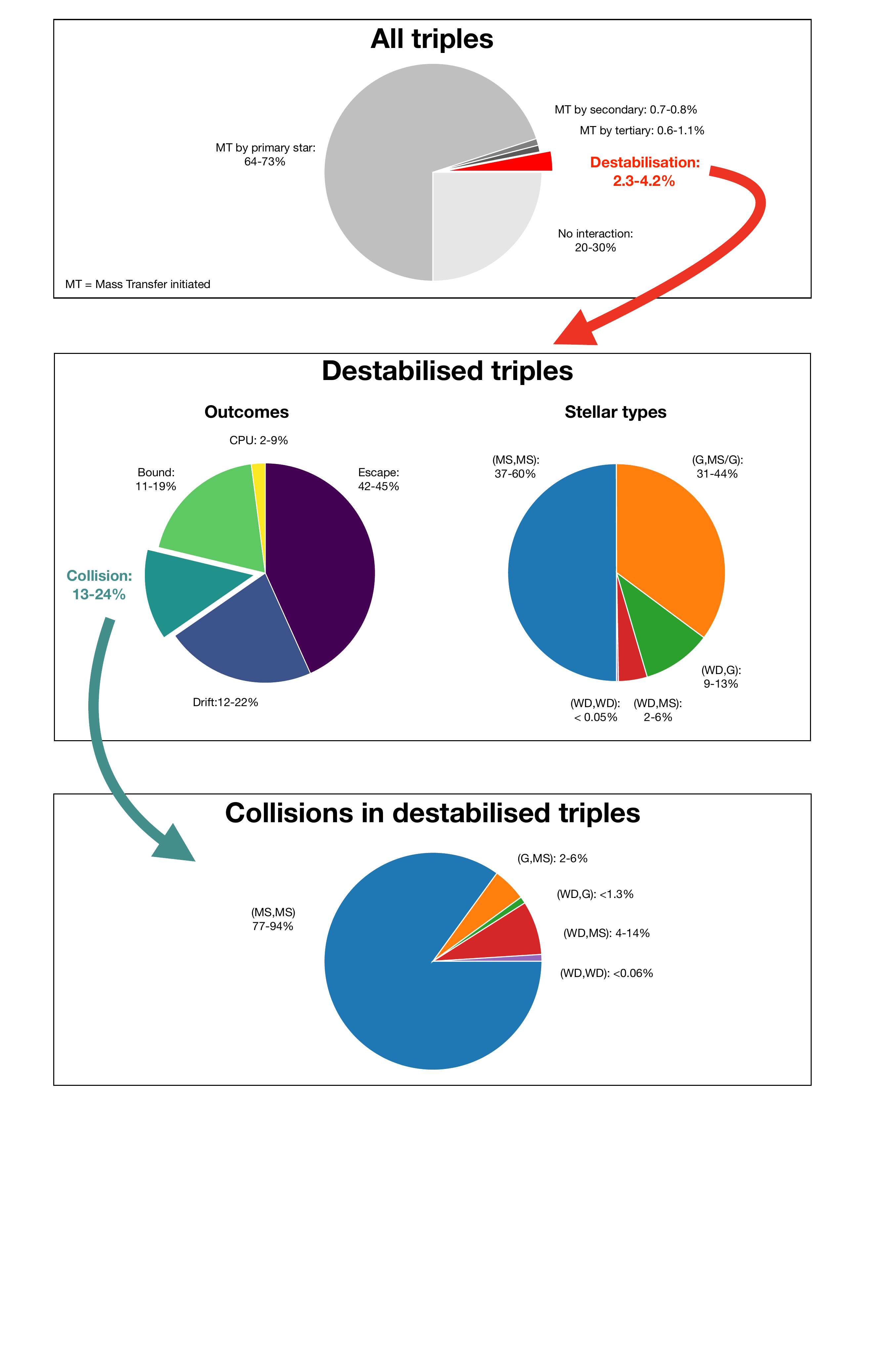}  
    \caption{Overview of the outcome of triple evolution. About 2-4\% of triples destabilise by themselves during their evolution (top panel). These triples commonly consist of three MS stars or  contain one or two giant stars (middle right panel). The outcome of the destabilised phase is often an ejection of one of the stars, but collisions are common as well (middle left panel). Collisions preferably occur between MS stars (bottom panel). Fractions are based on our most advanced N-body simulations that includes stellar evolution.}
    \label{fig:overview}
    \end{figure*}

  \begin{figure*}[t]
    \centering
    \begin{tabular}{cccc}
    model OBin & model T14 & model E09 & \\
\includegraphics[clip=true, trim =45mm 5mm 50mm 5mm, width=0.265\textwidth]{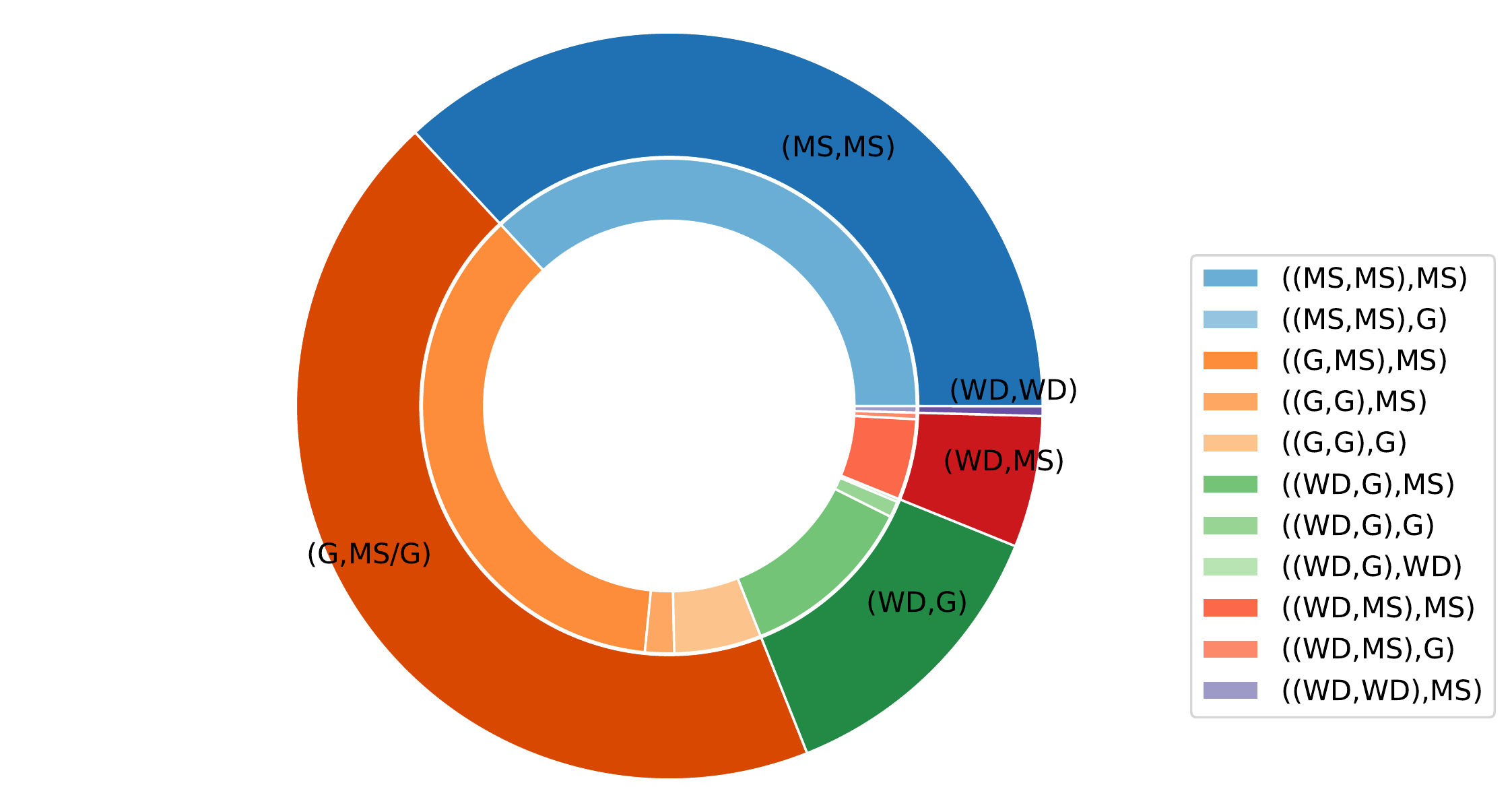}  &
 \includegraphics[clip=true, trim =45mm 5mm 50mm 5mm, width=0.265\textwidth]{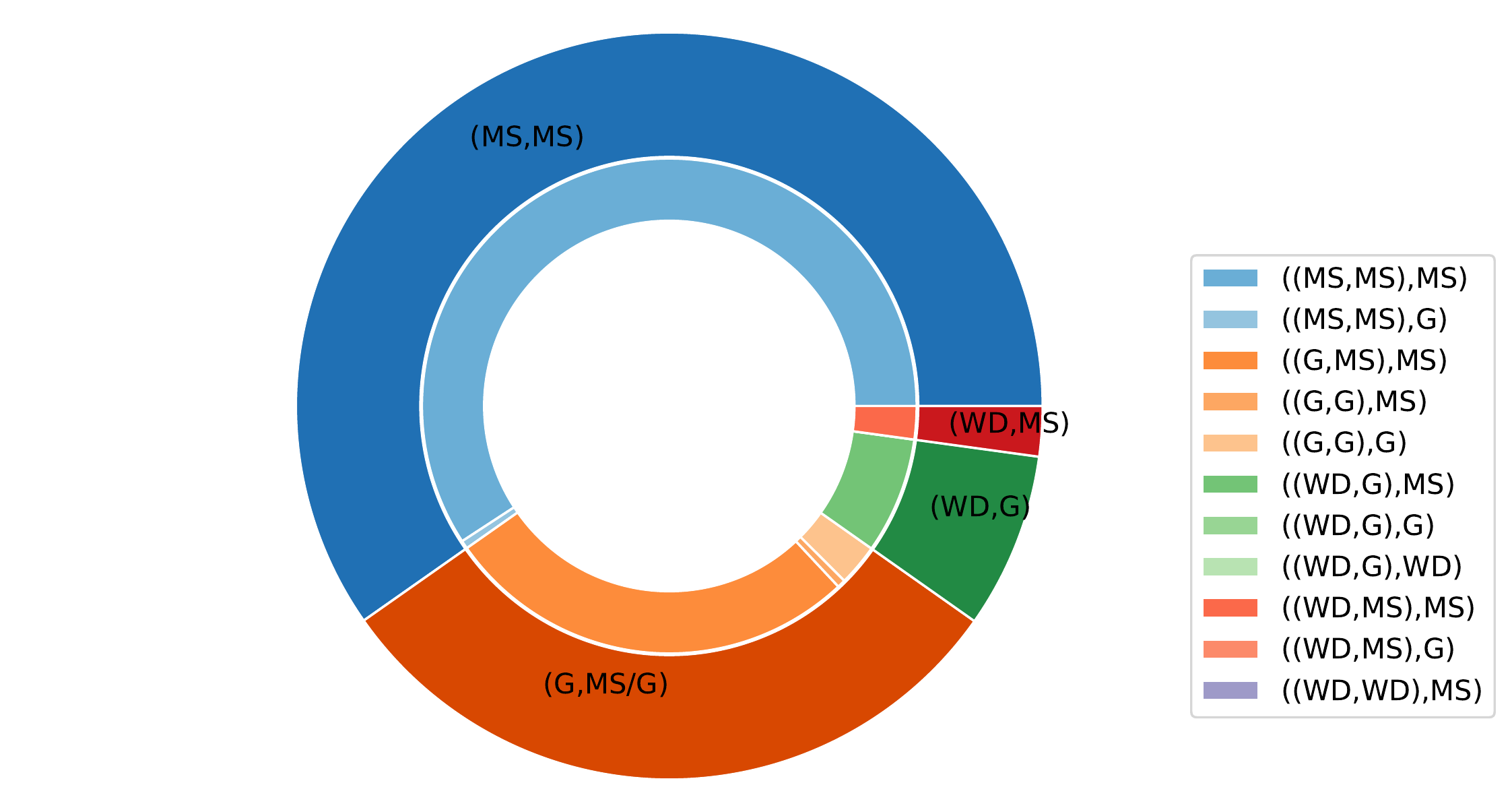} & 
 \includegraphics[clip=true, trim =45mm 5mm 50mm 5mm, width=0.265\textwidth]{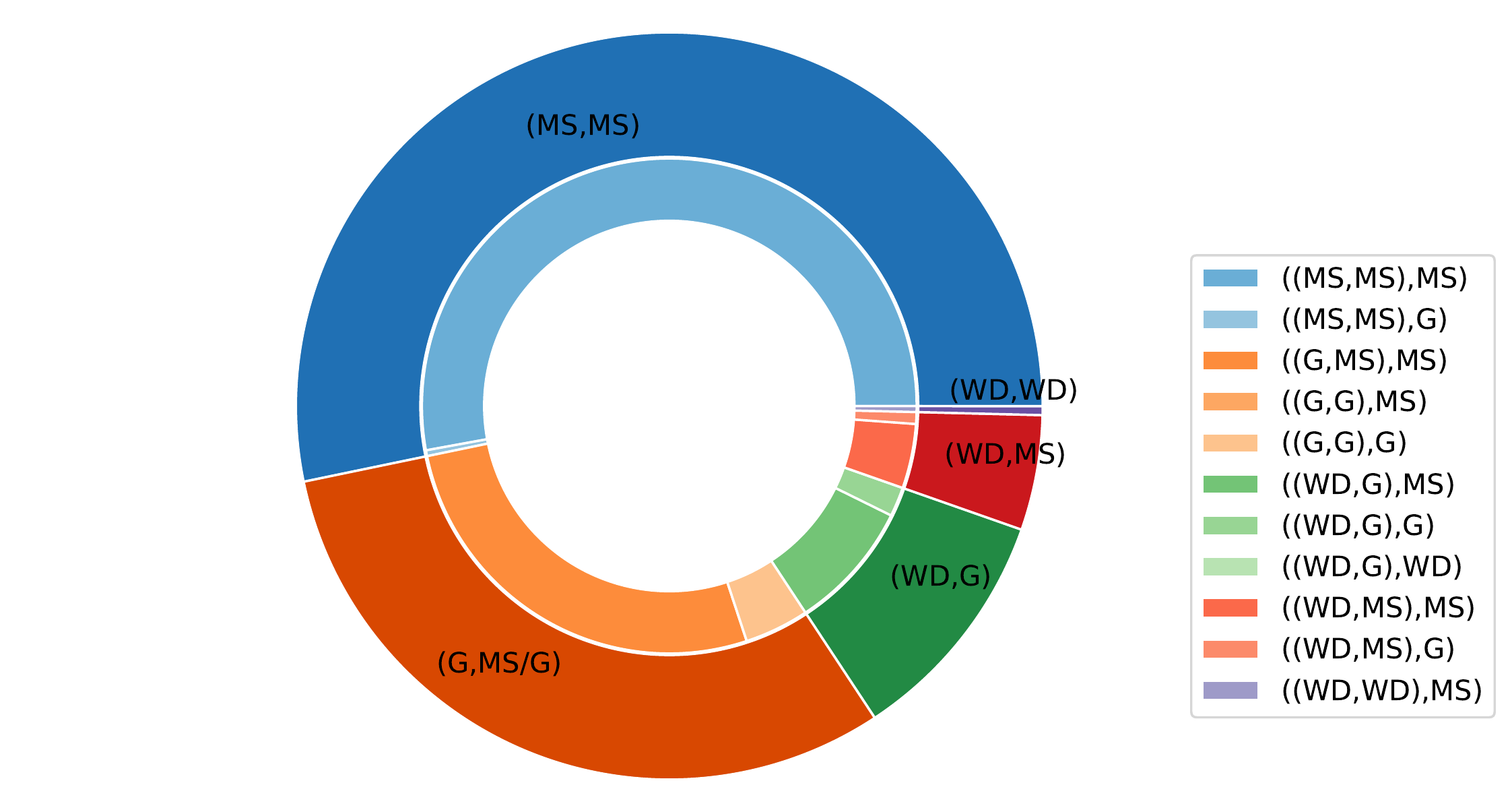} &
\includegraphics[clip=true, trim =175mm 10mm 0mm 25mm, width=0.15\textwidth]{pie_OBin.pdf} 
\\
	\end{tabular}
    \caption{
    Frequencies of the most abundant types of realistic triples that become dynamically unstable during their evolution. The five main types are shown in the outer ring, and subtypes in the inner ring as indicated by the legend. The letters represent the stellar phases of the stars at the moment of the dynamical instability; MS for a main-sequence star, G for a giant, WD for a white dwarf. The first part of the name of the types represents the primary star, the second part the secondary. The name of the subtypes have a third part that represent the evolutionary state of the tertiary. 
    The three pie charts reflect the three models for the initial population of triples on the zero-age main-sequence. 
    }
    \label{fig:pie_hier}
    \end{figure*}

\section{Results}
\label{sec:results}

\subsection{Hierarchical evolution}
\label{sec:res_hier}

\begin{figure}
    \centering
    \begin{tabular}{c}
    model OBin  \\
\includegraphics[width=\columnwidth, clip=true, trim =0mm 0mm 0mm 11mm]{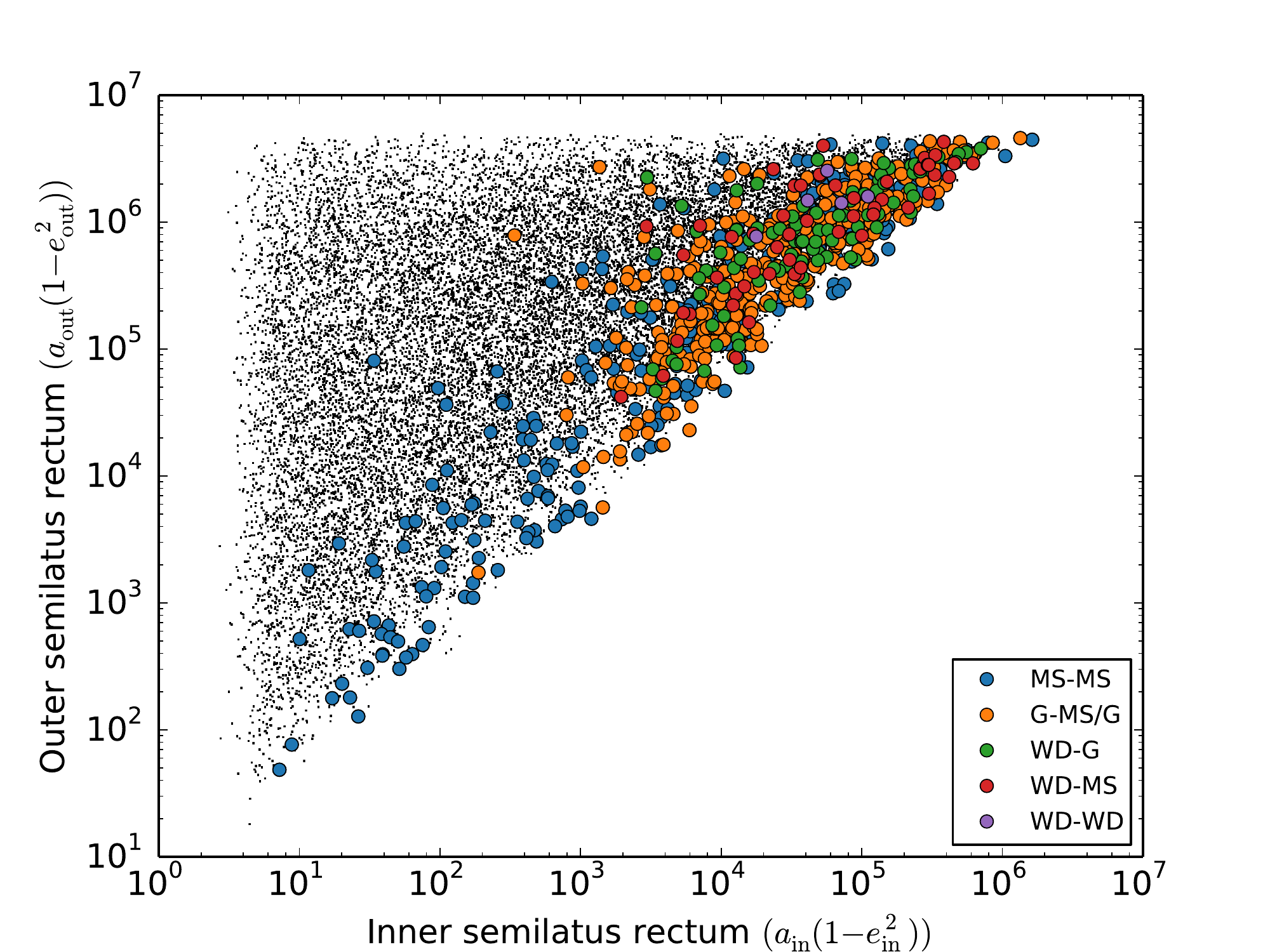}  \\ 
\\
  model E09 \\
\includegraphics[width=\columnwidth, clip=true, trim =0mm 0mm 0mm 11mm,]{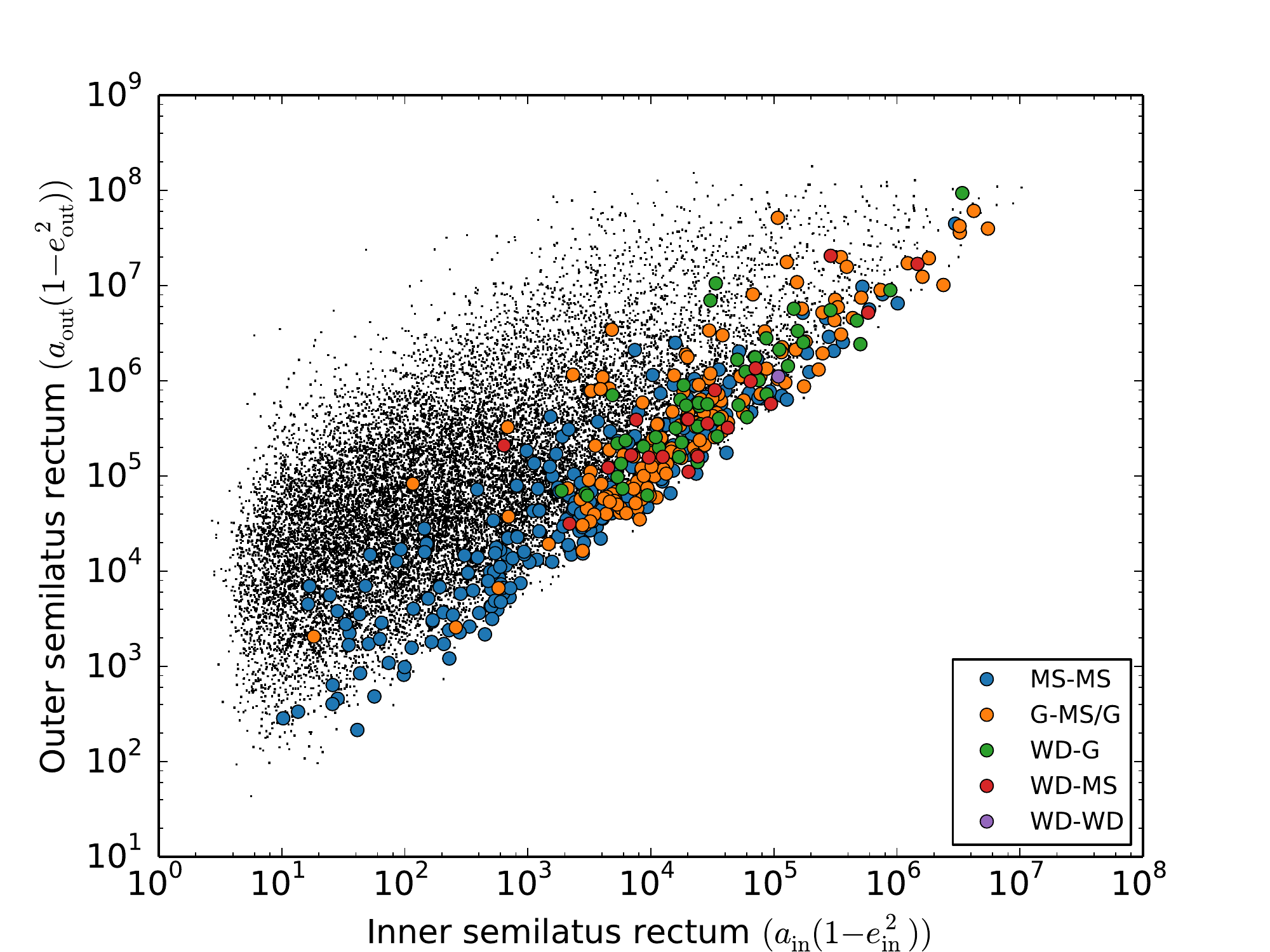}
	\end{tabular}
    \caption{Triples that become dynamical unstable during their evolution are born in a specific area of phase space close to the stability limit. The right lower part of the figure is empty as this is the 'forbidden' region of dynamically unstable systems. The different types of triples are colour-coded in the same way as in Fig.\,\ref{fig:pie_hier}. Black small dots represent triples that did not become dynamically unstable \citep[see][for their evolution]{Too20}.
    The x- and and y-axes show the initial distance the inner and outer orbit would circularise to if tides were efficient.  
    }
    \label{fig:pop_TRES}
    \end{figure}

From realistic simulations of triple evolution with  \texttt{TRES}, we find that a few percent of all triples become dynamically unstable due to their own internal evolution in a Hubble time. The percentage is highest for model OBin ($4.2\%$), and lowest for T14 ($2.3\%$) and E09 ($2.5\%$). A full exploration of the typical evolutionary pathways for stellar triples, and a comparison to binary evolution, is presented in \cite{Too20}. Common triple pathways include mass transfer between the stars of the inner binary, and mass transfer from the tertiary onto the inner binary (Fig.\,\ref{fig:overview}). 

The percentages mentioned above translate to a Galactic rate of 1-2 events every 1000 yrs. We expect the event rate to increase for triples with higher initial masses due to stronger stellar winds (especially early in the evolution on the MS) and due to the higher triple fraction. 
If we follow \cite{Moe17} and adopt a triple, binary and single star fraction of 75$\%$, 20$\%$ and 5$\%$ (instead of $15\%$, $50\%$, and $35\%$) as valid for O-type stars, the Galactic rate would already increase to 3-7 events per kyr.

{\ The systems that we focus on in this study are born in dynamically stable configurations, and cross the stability boundary simply as they evolve (hereafter destabilised systems). 
There are three main causes for this; 
three-body interactions, stellar winds, or a combination of the two. As stellar winds are most potent for evolved stars\footnote{In this paper we focus on stars in the low- and intermediate mass range, for which wind mass loss rates are small on the MS. However, for high-mass star evolution, stellar winds play a crucial role already on the MS.}, a significant fraction of triples crossing the stability boundary contain evolved stars (Fig.\,\ref{fig:overview}). While the giant phases only make up $\lesssim 10\%$ of a star's life, 30-40\% of destabilised triples have a giant\footnote{We consider a star a giant if the star has evolved off the MS and has not yet formed a compact object, i.e. ranging from the Hertzsprung gap to the asymptotic giant branch.} primary star at the onset of the instability. }

Fig.\,\ref{fig:overview} also shows that a large percentage of triples become dynamically unstable with all stars are on the MS. This constitutes $37\%$ for model OBin, and $60\%$ and $53\%$ for model T14 and E09 (Fig.\,\ref{fig:pie_hier}). The difference in the percentages is due to the characteristic orbital separations of each model \citep[Fig.\,\ref{fig:pop_TRES}, see also Fig.\,2 in][]{Too20}. 
 As outer orbits tend to be more compact in model T14 and E09, these models contain relatively more destabilised triples with MS primaries (MS,MS) instead of giant primaries (G,MS/G) or secondaries (WD,G). The latter cases occur in wide orbits ($a_{\rm in} \gtrsim 10^3\Ro$), such that Roche-lobe overflow before the instability can be avoided. Hereafter when we refer to the stellar types of the triple with '(X,Y)' or '((X,Y),Z)', X refers to the stellar type of the primary star that is the initially most massive star of the inner binary, Y refers to the secondary, and Z to the tertiary or outer star.

Lastly $2-6\%$ and $\lesssim 4\%$ of destabilised triples have a WD primary star, and a MS or WD secondary star respectively (i.e. 'WD-MS' and 'WD-WD'). At the onset of the instability, the stellar winds are negligible in these systems. The previous wind mass loss phase(s) has widened the inner orbit, driving the triple closer to the stability limit. However, contrary to the previously discussed evolution for 'G-MS/G' and 'WD-G' systems, the orbital widening itself is not enough to trigger an instability. In stead, the new configuration leads to stronger three-body effects and higher outer eccentricities that make the system enter the instability region.

We note that for all types [(MS,MS), (G,MS/G), (WD,G), (WD,MS) and (WD,WD)] three-body dynamics plays a role in the orbital evolution of the systems, and therefore the diminished dynamical stability. The far majority of systems are born in the octupole regime. The octupole parameter at birth $|\epsilon_{\rm oct,init}|>0.001$ for $97-99\%$ and  $|\epsilon_{\rm oct,init}|>0.01$ for $76-81\%$ of all triples that become dynamically unstable. This means that it is important to take three-body dynamics into account in the hierarchical evolution of the triples, even for (G,MS) and (WD,G).

The delay times between the formation of the triple and the onset of the instability range from a hundred Myr to several Gyr typically. Similarly to \cite{Too20}, we also find some tertiary-driven interactions that occur on shorter timescales. The issue is that if the delay time is a fraction of the star-forming timescale, one may wonder if the interaction would not have taken place before the zero-age main-sequence which we take as the start of our simulations. If we assume the primary's pre-MS timescale is 1$\%$ (0.1$\%$) of its MS timescale \citep{Bar02}, than it concerns $15-24\%$ ($10-17\%$) of the triples in the (MS,MS) channel. We do not find a significant difference in the outcome of the dynamically unstable phase (Sect.\,\ref{sec:outcome}) for these systems compared to the full population of triples with purely MS components.

\subsection{Evolution during the dynamically unstable phase}
\label{sec:outcome}

\subsubsection{Duration}
\begin{figure}
   \centering
   \includegraphics[width=\columnwidth]{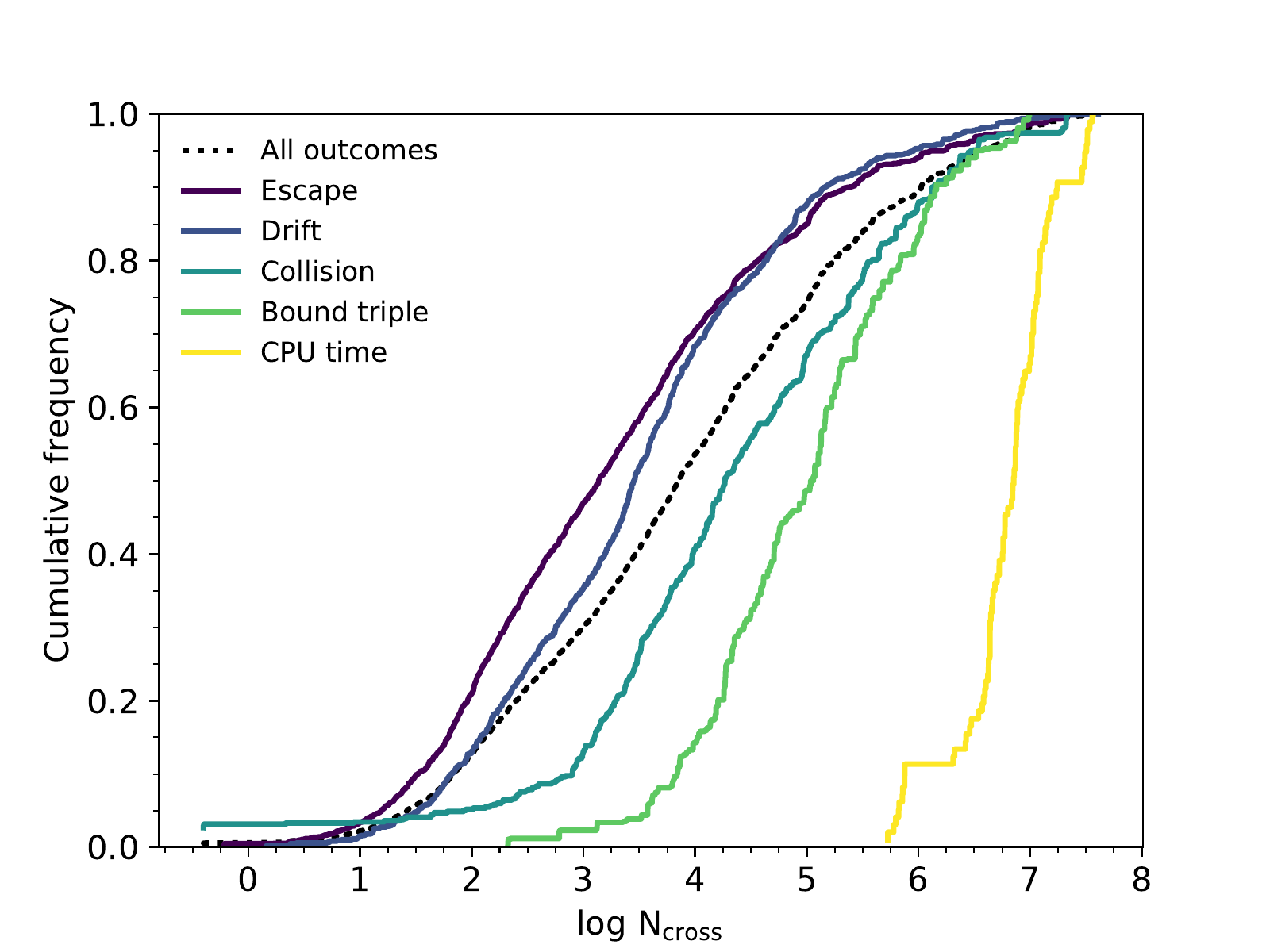} 
      \caption{Cumulative histogram of the duration of the dynamically unstable phase in crossing times ($t_{\rm dyn}/t_{\rm cross}$). The figures show that initially stable triples that have moved into the dynamically unstable regime can remain there for long times. The figure represents model OBin where the dynamics are modelled including stellar evolution. Other models show qualitatively similar behaviour. }
         \label{fig:t_cross}
   \end{figure}

One may naively expect the dynamical phase of an unstable triple to be short, that is of the order of several crossing times. However, the dynamical phase of the triples considered here, which are on the boundary of dynamical stability, typically lasts thousands or ten thousands of crossing times (Fig.\,\ref{fig:t_cross}). In other words, the dynamically unstable phase is not a short phase. As a result other physical processes can play a role during the 'dynamical' phase as well, as we show in the following sections.

Previous studies of unstable triples mainly focused on encounters between a binary and a single star in dense environments. 
Scatter experiments show that unstable triples have a `decay' time of order a hundred crossing times \citep[e.g.][]{Mikkola07, TB15}, and an algebraic tail towards very long lifetimes corresponding to long excursions of the single star with respect to the binary \citep{Orlov10}. The difference with our destabilised triples can be understood due to their higher angular momenta. Amongst others \cite{TB15} show that the duration of the dynamical phase increases with initial angular momentum, and also with mass ratio (more equal stellar masses).  Moderate mass ratios are expected for stable stellar triples \citep{Tok14b, Moe17}. On top of that, our hierarchical triples start off on the edge of instability, contrary to randomly sampled triples, which can start off much deeper inside the unstable regime.

\subsubsection{Outcomes}

\begin{figure*}
    \centering
    \begin{tabular}{|c|c|}

    \hline
    &\\
    \large{((MS, MS), MS)}& \large{((G, MS), MS) } \\
   & \\
    \includegraphics[width=\columnwidth]{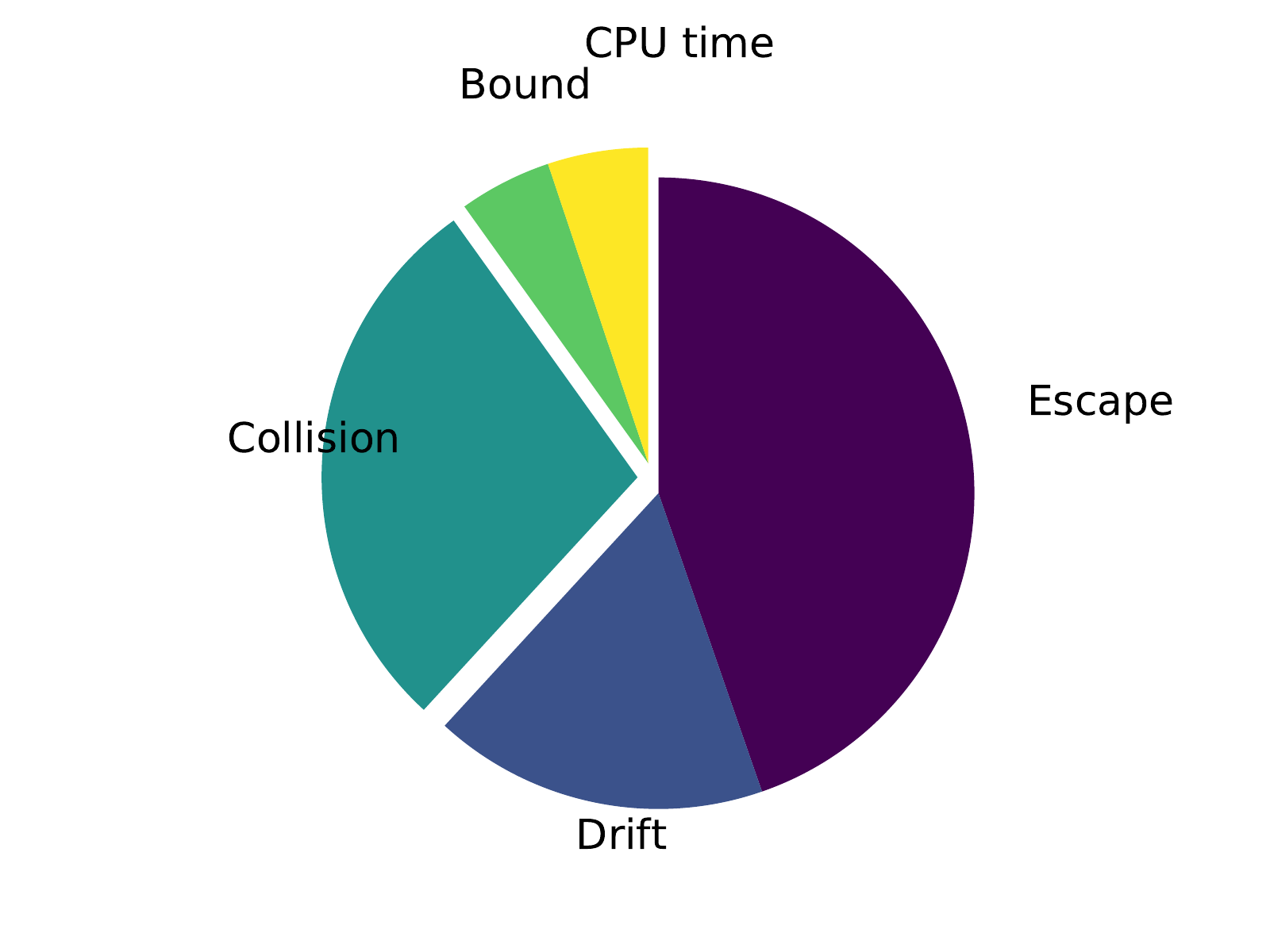} & 
    \includegraphics[width=\columnwidth]{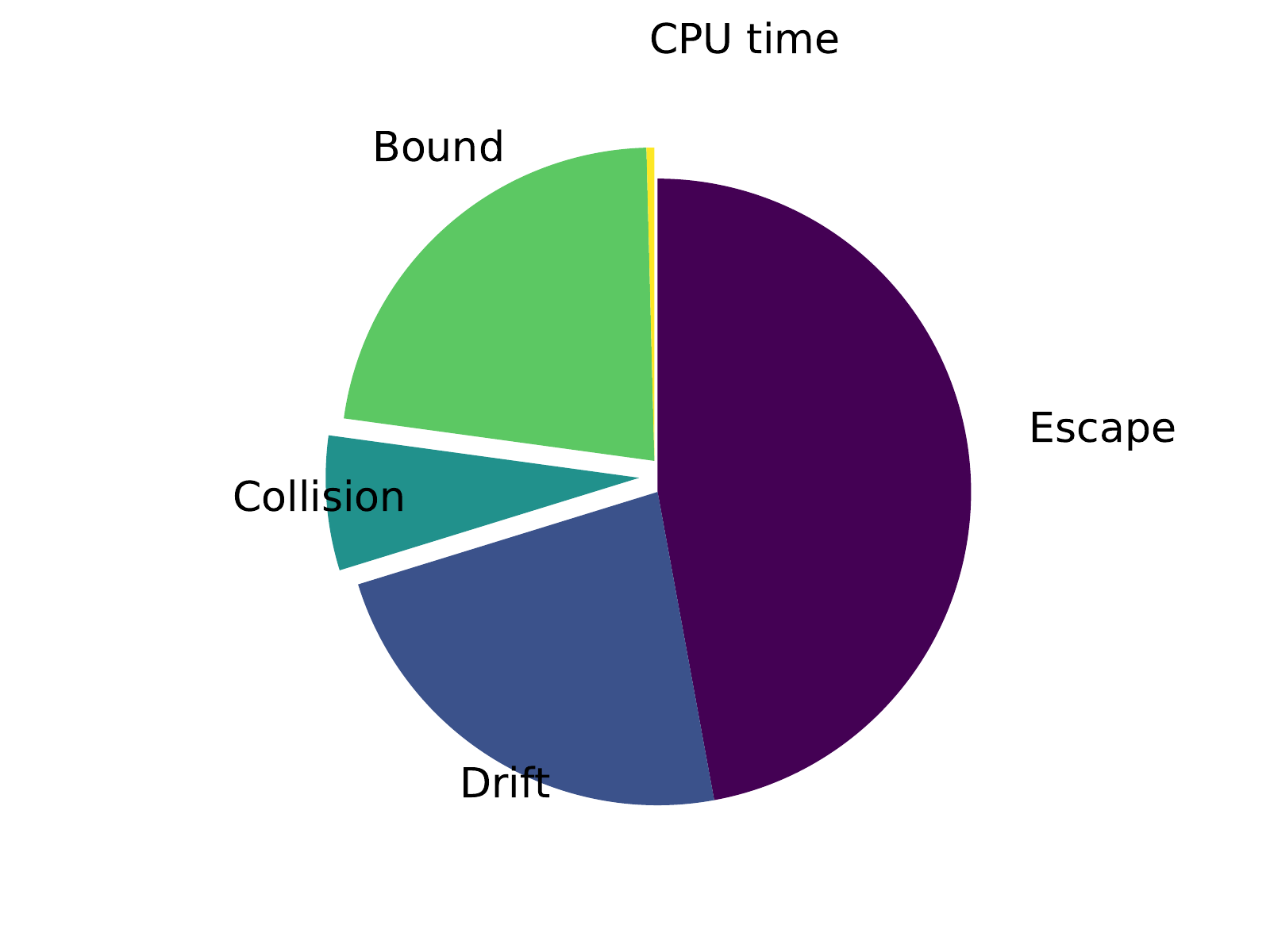} \\

\includegraphics[width=\columnwidth, clip=true, trim =20mm 0mm 20mm 35mm,]{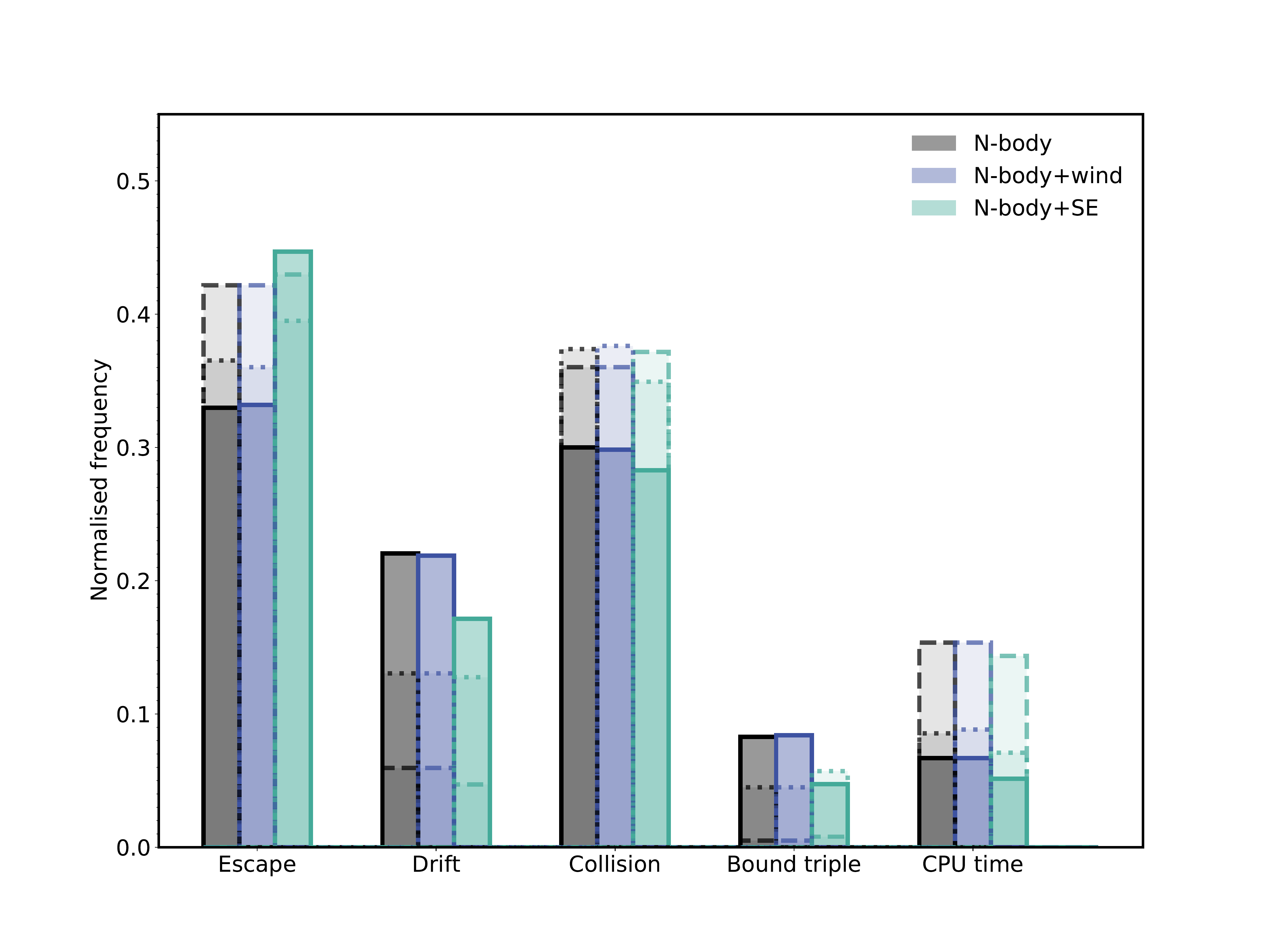}  &
\includegraphics[width=\columnwidth, clip=true, trim =20mm 0mm 20mm 35mm,]{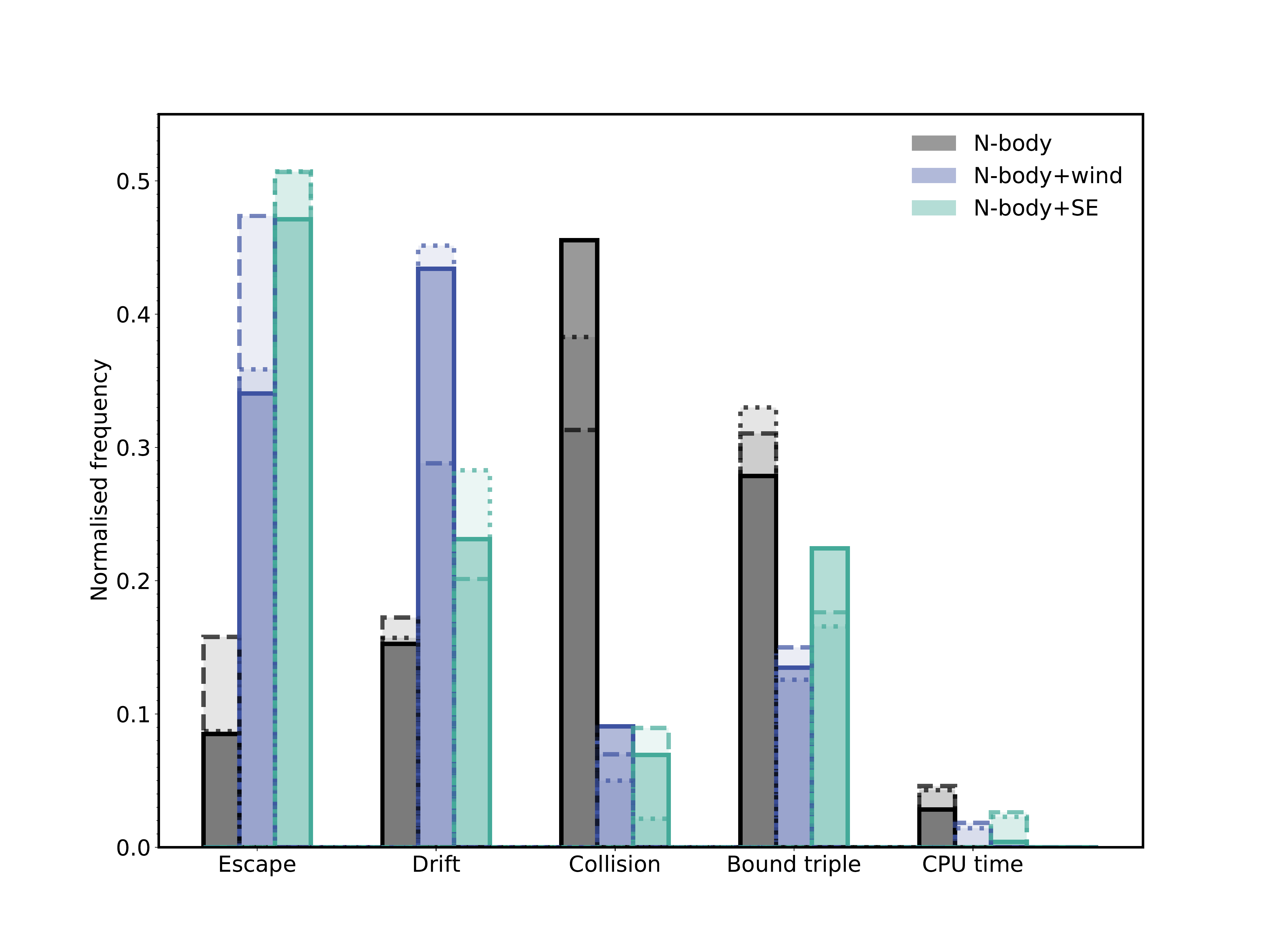}  \\
\hline
    \end{tabular}

    \caption{Outcomes of the dynamically unstable phase for the full population of triples that become dynamically unstable during their evolution. Pie-charts represent the N-body simulations with stellar evolution for model OBin. The histograms show the outcomes for all models where 
    different colours represent the different dynamical methods (Sect.\,\ref{sec:dyn}). Solid, dashed, and dotted line styles represent the models for the initial population of triples OBin, T14 and E09, respectively. The results for each model are normalised to unity.
    }
    \label{fig:outcomes_all}
    \end{figure*}

We find five possible outcomes of the dynamically unstable phase; 1) one of the stars escapes the system, 2) one of the stars drifts away from the system, 3) there is a collision between two of the stars, 4+5) at the end of the simulation the system is still bound and dynamically unstable. For 4) the system has evolved for a Hubble time for 5) the maximum CPU time has been reached (Sect.\,\ref{sec:dyn}). Both escaping and drifter stars are ejected from the system, but the former with high velocities such that the triple has become unbound, and the latter with low velocities such that the triple is technically still bound. 
The reason to consider drifters as an outcome is that a separation was reached of one parsec at which point perturbations from the Galactic environment cannot be neglected \citep{Jiang10}. In the following sections we discuss the characteristics of the various outcomes in detail.

All outcomes are prevalent (Fig.\,\ref{fig:overview}). However their relative frequency varies with stellar types and initial conditions (Fig.\,\ref{fig:outcomes_all}). 
 The left column of Fig.\,\ref{fig:outcomes_all} shows that unstable ((MS,MS),MS) triples frequently eject a stellar component and dissolve into a separate binary and single star. Collisions between MS stars are also common. Few unstable ((MS,MS),MS) survive for a Hubble time. There is little difference between the simulations with the pure N-body method and those with a constant wind mass loss rate for destabilised ((MS,MS),MS) triples as the typical wind mass loss rate for MS stars are small in the adopted mass range. The N-body simulations with stellar evolution show an enhancement in the fraction of escapers due to the evolution of the stars off the MS during the dynamical interaction.

The typical outcomes for unstable ((G,MS),MS) triples are very different from that of ((MS,MS),MS) triples. Furthermore, there are large differences depending on which method we use to model the dynamically unstable phase. This demonstrates that it is important to take stellar evolution into account during the dynamically unstable phase itself. 
The stellar winds from the giant primary not only cause the system's hierarchy to weaken during the hierarchical phase, but continue to effect the dynamics of the system after entering the dynamically unstable regime. The evolution during this phase is therefore not purely dynamical.
The effect can be seen in the relative large fraction of ejections and small fraction of long-lived bound triples in the N-body simulations that include stellar winds and evolution compared to that of the pure N-body method (bottom right panel of Fig.\,\ref{fig:outcomes_all}).

Another striking difference in the outcomes of unstable ((G,MS),MS) triples is the number of collisions (bottom right panel of Fig.\,\ref{fig:outcomes_all}). In the simulations with the pure N-body method, where the stellar parameters are assumed to be frozen in time throughout the dynamically unstable phase, collisions are remarkable frequent compared to the other methods. The typical timescale of the dynamically unstable phase is, however, longer than the lifetime of the giant star. In conclusion, for these systems it is therefore vital to take the change in stellar mass and radius into account.

\subsubsection{Ejections: escape and drift}  
\label{sec:ejection}

The destabilisation of a triple typically leads to  the ejection of one of its stars; either because a star escapes or drifts away. In our most sophisticated dynamical models that include stellar evolution, an escape occurs for 42-45\% of destabilised triples and a drift for 12-22\% (Fig.\,\ref{fig:overview}, see Tbl.\,\ref{tbl:rates} for other models).

\begin{figure}
   \centering
   \includegraphics[width=\columnwidth]{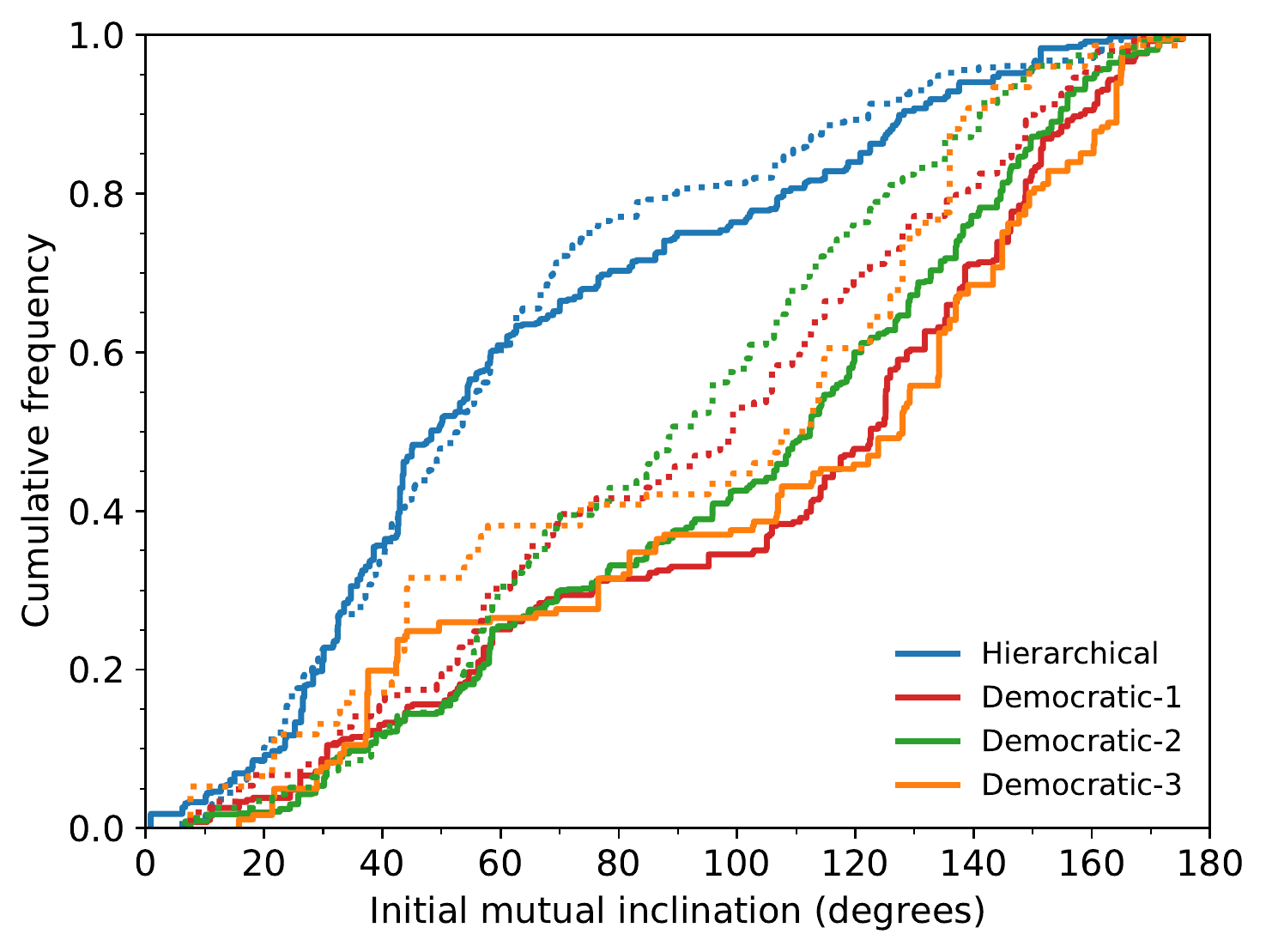}
      \caption{Cumulative histogram of the inclination at destabilisation for triples that will eventually eject one of the stellar components. Triples with prograde orbits tend to experience hierarchical encounters, whereas retrograde orbits tend to lead to democratic encounters.}
         \label{fig:eject_incl}
   \end{figure}

To investigate the dynamical interactions, we differentiate between triples for which the hierarchy is maintained, and those that undergo democratic resonances. We find a link between this and the orbital orientation at the onset of the instability. 
Triples with prograde\footnote{We refer to triples as being prograde if they have a relative inclination between the inner and outer orbit smaller than 90$^{\circ}$, and as retrograde otherwise. \label{footnote:prograde}}  orbits typically experience hierarchical encounters and eject their tertiary star (Fig.\,\ref{fig:eject_incl}). Those with retrograde orbits typically experience democratic encounters, during which most often the secondary star is ejected (Democratic-2), see also Tbl.\,\ref{tbl:dem}).
We expect that this is the case since retrograde orbits are more stable compared to prograde orbits when all other parameters are kept the same (Eq.\,\ref{eq:stab_crit}). Therefore, in order for a retrograde triple to cross the stability limit, the ratio of $a_{\rm out}/a_{\rm in}$ is therefore smaller compared to the prograde case, which makes is easier for a democratic encounter to ensue. 
Additionally, democratic encounters are more common when winds and stellar evolution are taken into account in the dynamical modelling compared to the pure N-body simulations. Stellar winds push the system deeper in the instability region.

 \begin{figure}
   \centering
   \includegraphics[width=\columnwidth]{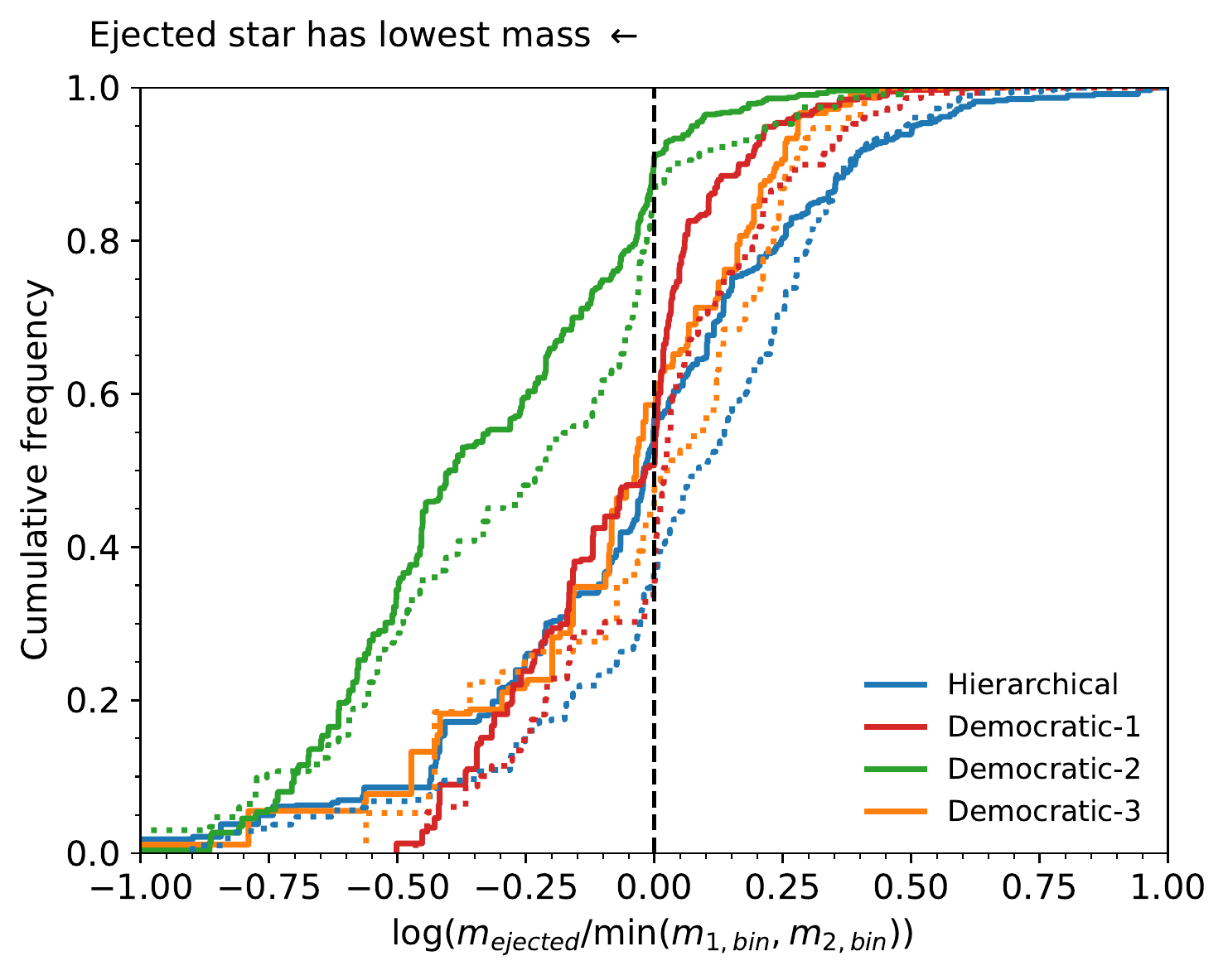}
      \caption{Which star is ejected from the destabilised triple? Is it the lowest mass star of the system as is often assumed? The figure shows a cumulative histogram of the mass of the ejected star over that of the lightest component of the inner binary. If this ratio is smaller than one (left of black dashed line), then the ejected star is the lightest component. Only for Democratic-2 encounters is it commonly the case that the lightest component is ejected. }
         \label{fig:eject_m}
   \end{figure}

Next we investigate which star is ejected from the triple. For their velocities, and the formation of runaway stars, see Sect.\,\ref{sec:runaway}. Naively, one may expect that this would be the lowest mass star predominantly. 
Fig.\,\ref{fig:eject_m} shows this is not the case for destabilised triples. Hierarchical encounters do not typically lead to the ejection of the lowest-mass star, neither do Democratic-1 and Democratic-3 encounters. The naive picture only holds for the Democratic-2 encounters; the lowest mass star is ejected in more than 85\% of the Democratic-2 encounters. 
The expectation of ejecting the lowest mass star can be traced back to the seminal work of \cite{Hills75}, who performed numerical scattering experiments between binaries and field stars with inner components of 3 solar mass each, and a tertiary of one solar mass. On the other hand, for truly ergodic, equal mass triples, the probability to eject any of the components is a third \citep{Spitzer87}. The destabilised triples considered here do not have extreme mass ratios $m_{\rm ejected}/$min$(m_{1,bin},m_{2,bin})$; 50\% of triples have mass ratios between $\sim$0.6-1.7.

The ejected star is typically on the MS. At the onset of the instability this comprises 79-87\% of the systems. At the moment of dissolution the fraction has slightly decreased to 72-82\% as stars evolve of the MS. For the majority of other systems, the ejected star was initially on a giant branch during the destabilisation (i.e. for 12-16\% of systems). However, at the moment of ejection it is very rare that the ejected star is still a giant star. At the moment of ejection,  17-28\% of ejected stars has already reached the WD stage. The formation rate of single WDs through this channel in the Milky Way is $(1-4)\times 10^{-4}$ per year, where as the birth rate of WDs from single stellar evolution is approximately a few 0.1 per year \citep{Ver13,Too17}. As the ejected stars are typically the original secondaries and tertiaries of the triples, the ejected WDs do not follow the same mass distribution as WDs from single stellar evolution.

\begin{figure}
   \centering
    \begin{tabular}{c}
     N-body simulations \\
     \includegraphics[width=\columnwidth]{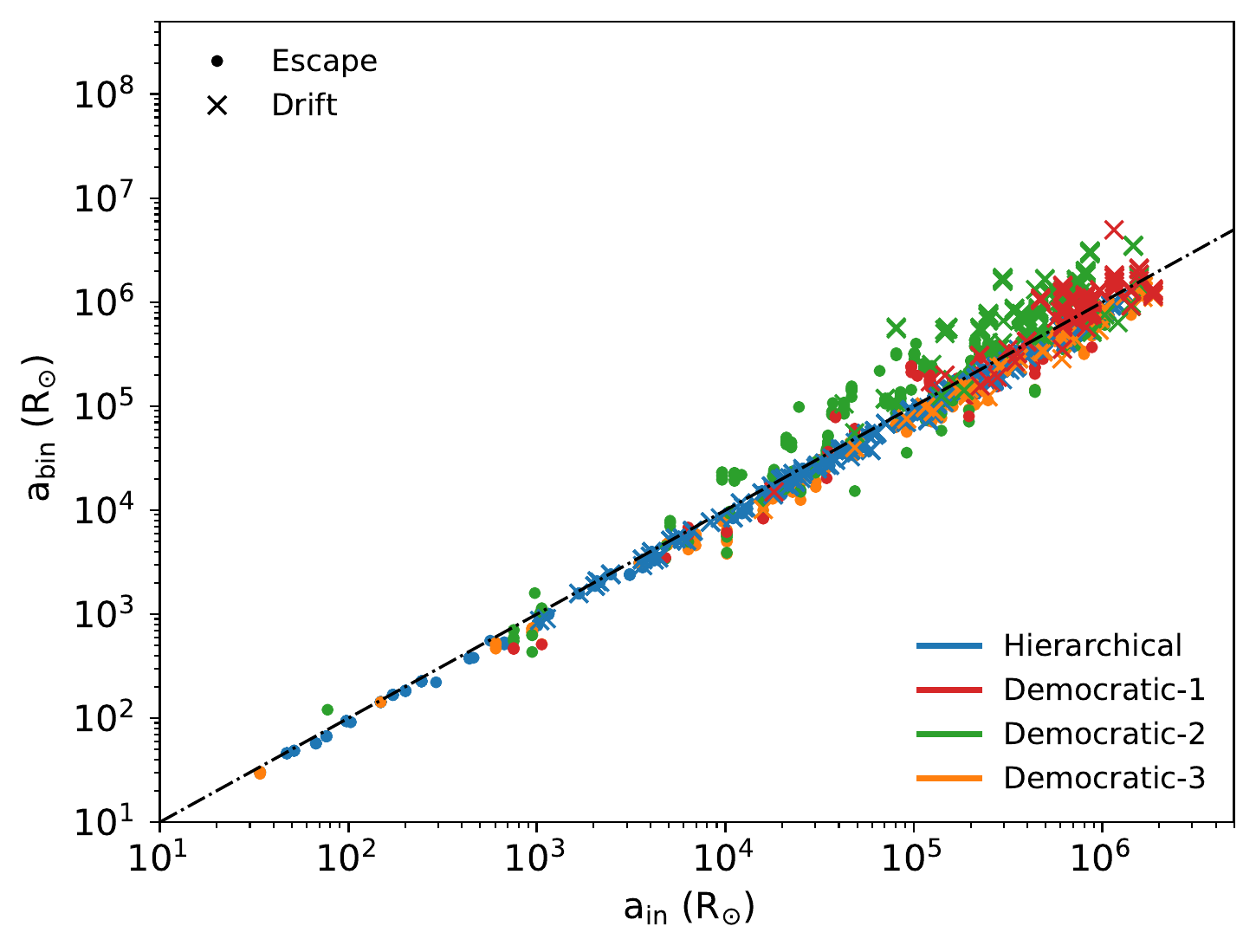}\\
     N-body + stellar evolution\\
   \includegraphics[width=\columnwidth]{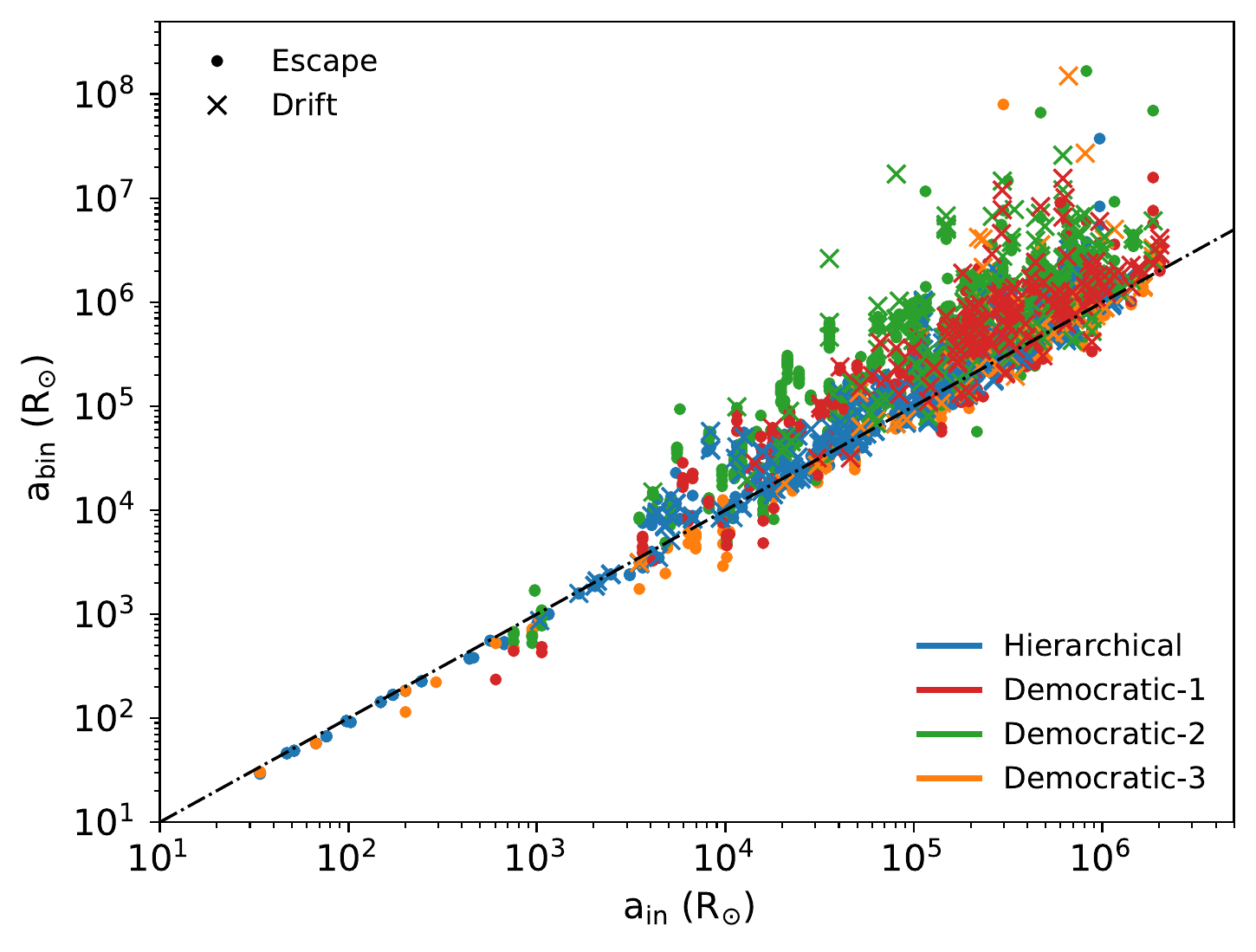}
\end{tabular}   
      \caption{Change in orbital separation for destabilised triples that unfold into a binary and single star. The x-axis shows the orbital separation of the inner orbit at the onset of the instability, and the y-axis shows the post-interaction orbital separation of the binary. The results of the pure N-body simulations are shown in the top panel, those of simulations that include stellar evolution are shown on the bottom. The black dash-dotted line represents no change in the orbital separation. The figure shows that besides adiabatic wind mass losses leading to orbital widening, ejecting a star from the inner orbit also tends to lead to orbital widening (see text). The two panels represent model OBin. Model T14 and E09 show qualitatively similar trends. 
      }
         \label{fig:eject_a}
   \end{figure}

\begin{figure}
   \centering
   \includegraphics[width=\columnwidth]{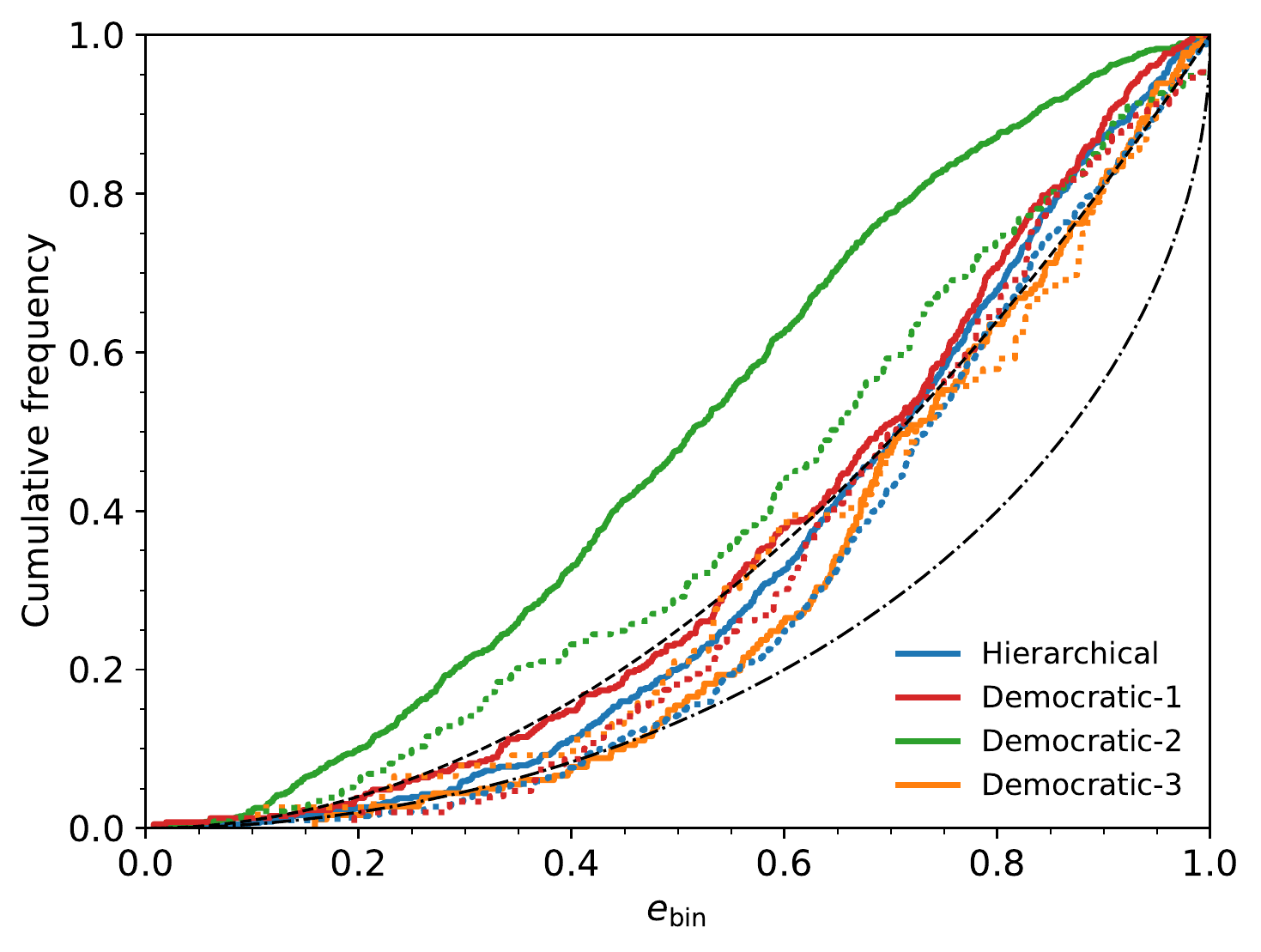}
      \caption{Cumulative histogram of the eccentricity of the remaining binary after the ejection. Democratic-2 encounters lead to lower eccentricities then hierarchical, Democratic-1, and Democratic-3 encounters. The black dashed and dash-dotted line represents a thermal and superthermal distribution, respectively. 
    }
         \label{fig:eject_e}
   \end{figure}

Next, we focus on the remaining binary and how the orbit changes due to the dynamical interaction. 
In the simplest scenario of purely dynamical ejections without stellar exchanges (i.e. hierarchical and Democratic-3 encounters),  we expect the orbital separation to shrink due to energy conservation. There are two exception two this scenario. 
Firstly, if stellar winds play an important role, than the orbital separation also experience widening. This is clearly visible as the difference between the top and bottom panel of Fig.\,\ref{fig:eject_a}. Secondly, if stellar exchanges take place than the orbit can widen as well. This can be seen in the top panel of Fig.\,\ref{fig:eject_a} in the red and green markers that lie above the black line. This can be understood in the following way: consider a triple that dissolves into a single star and a binary with a specific post-dissolution binding energy $E_{bin}$ and $m_2<m_3$. In a hierarchical interaction  $E_{bin}\equiv \frac{m_1m_2}{2a_{bin,hier}}$, where as for a Democratic-2 encounter $E_{bin}\equiv \frac{m_1m_3}{2a_{bin,dem-2}}$. From $m_2<m_3$, it follows that if the low-mass secondary is ejected the orbital separation $a_{bin,dem-2}>a_{bin,hier}$. Additionally, the post-interaction eccentricities of the remaining binary are shown in Fig.\,\ref{fig:eject_e}. Where as hierarchical interactions, as well as Democratic-3 encounters, lead to eccentricity distributions that are more or less thermal (black line), Democratic-2 encounters give rise to a distinct eccentricity distribution that is significantly sub-thermal, that is larger fraction at smaller eccentricities. 
 We observe that when the lowest-mass body escapes, that the remaining binary is not only more massive, but also statistically wider. Combined with the smaller eccentricities, these trends all tend to maximise the angular momentum of the binary. This implies that when the lightest body is ejected after a democratic resonance, it does so on a low angular momentum orbit.

Finally, destabilised triples with compact inner orbits ($a_{\rm in}\lesssim 10^4\Ro$), tend to have prograde orbits, and therefore keep their hierarchy and eject the tertiary star (Fig.\,\ref{fig:eject_a}). 
Given that the orbits are compact, the ejected star escapes from the triple; very few triples experience a drift, nor do they remain bound for a Hubble time. 
The retrogradely orbiting triples do not typically lead to democratic interactions and ejections, but to collisions instead (see also App.\ref{app:inc}).

In a small number of simulations, there was no remaining binary. Both orbits are dissolved during the interaction. This is not possible in purely dynamical situations due to energy conservation. However, it is known that impulsive winds can unbind a binary \citep{Ver11, Too17, TB22}, or in this case a triple. Not surprisingly, the particular triples have wide orbits ($a_{\rm in }\sim 10^5-10^6\Ro $), and the stars are close to their apocentre. As a result their orbital velocities are relatively low (and the impact of the winds relatively large). In the dynamical simulations with wind or stellar evolution, it comprises 0.04\% - 0.027\% of all destabilised triples. The rate is the highest in model OBin, as in this model we start the simulations with relatively many wide triples.

\subsubsection{Collisions}
\label{sec:collisions}

One of the exciting outcomes of a triple instability, is the collision of two stars. Typically the collision involves the two stars that initially were in the inner orbit, i.e the primary and secondary star. During the dynamically unstable phase leading up to the collision, the original hierarchy of the system is routinely preserved. 87-95\% of destabilised colliding systems never experience a democratic resonance. And even for those triples in our ensemble which are flagged as having lost their hierarchy,  stellar exchanges are rare (i.e. 1.6-2.4\% of all collisions). The inner orbits exhibits extreme behaviour with median eccentricities at the time of collision of more than 0.996 in all models. This is consistent with results from \cite{Bhaskar21}, who find that in the regime of similar mass stars ($m_2/m \geq 0.1$) and a semi-major axis ratio $a_{\rm{in}}/a_{\rm{out}} \geq 0.3$, the maximum eccentricity exceeds those estimated from secular theory, reaching up to 0.99 and beyond. 

Even though the hierarchy is usually maintained in the dynamically unstable phase, we see fluctuations in the configuration of the outer orbit. Not only the eccentricity of the outer orbit varies (also in cases without stellar winds), as expected in the eccentric Lidov-Kozai regime, but also the orbital separation varies. This illustrates that the double-averaging approximation, as used in  \texttt{TRES} during the hierarchical phase, breaks down in the dynamically unstable phase, justifying the transition to N-body simulations. Accordingly, there are variations in the mutual inclination of the orbits, commonly including orbital flips. It is the breakdown of the orbits that drives the collisions in an extraordinary efficient way. 

\begin{figure}
    \centering
\includegraphics[width=\columnwidth, clip=true, trim =0mm 0mm 0mm 0mm,]{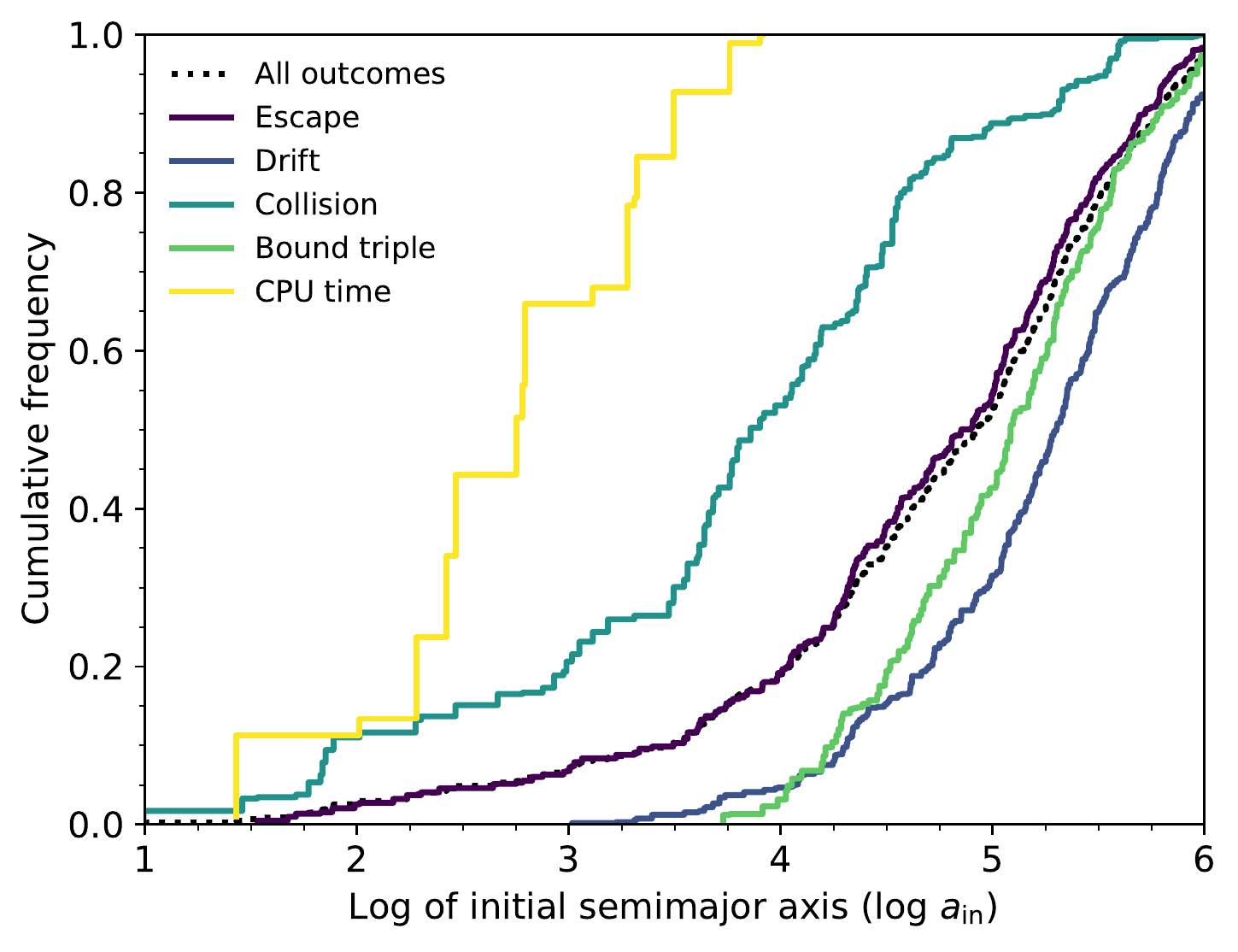}
    \caption{The origin of destabilised triples and their outcomes. Where as drifters and long-lived triples originate from triples with large initial orbital separations, colliding stars and CPU-limited systems  come from compact initial triples. 
    }
    \label{fig:outcome_origin}
    \end{figure}

The fraction of destabilised triples that experience a collision ranges between 13 and 35\% (Tbl.\,\ref{tbl:rates}). The pure N-body models give the highest fractions (32-35\%). The models that take stellar evolution into account during the dynamically unstable phase provide a much lower fraction, that is 13-23\% depending on the initial distributions of triples (Fig.\,\ref{fig:outcomes_all}). 
This can be easily understood as the radius of the evolving star decreases as it becomes a WD, which decreases the cross section for the collision. In fact, the fraction of triples with collisions in ((MS,MS),MS) systems does not change with the various dynamical modelling methods, whereas the fraction for ((G,MS),MS) systems decreases with a factor 3.5-18 when stellar evolution is taken into account (Fig.\,\ref{fig:outcomes_all}). The systems that avoid the collision typically have wide orbits ($a_{\rm in}\gtrsim 5\times10^4 \Ro$, Fig.\,\ref{fig:outcome_origin}) and dissolve into a binary and a single star through a stellar escape.

With our Galactic model as described in Sect.\,\ref{sec:hier}, we estimate a  collision rate of 2-3 per $10^4$ yr based on the N-body simulations with stellar evolution. The majority of these involve collisions between MS stars (Fig.\,\ref{fig:overview}); 76-94\% depending on the initial conditions. Collisions between a giant and a MS star occur in 2-6\%. More common are merger between WD and MS stars (4-15\%) that mostly come from ((G,MS),MS) systems whose giant primary has evolved into a WD by the time of the collision. We also find low rates of giant-WD and WD-WD collisions ($\sim$1\% and $\sim$0.5-2\% respectively).
Assuming mass is completely conserved during the merger, the merger remnant usually remains bound to the tertiary star. Less than 0.5\% of collisions in all models 
lead to an unbound pair. For a discussion on the remaining binary, see Sect.\,\ref{sec:obs}.

\subsubsection{Long-lived bound unstable triples}
\label{sec:bound}

At the end of our simulations, a small fraction of our destabilised triples is still bound, that is 7-23\%. The vast majority of these systems has preserved their initial hierarchy; only in 0.2-0.5\% of cases was  the system flagged as undergoing a democratic encounter. However, this did not lead to the disruption of the triple. 

\begin{figure}
    \centering

\begin{tabular}{c}
\includegraphics[clip=true, width=\columnwidth]{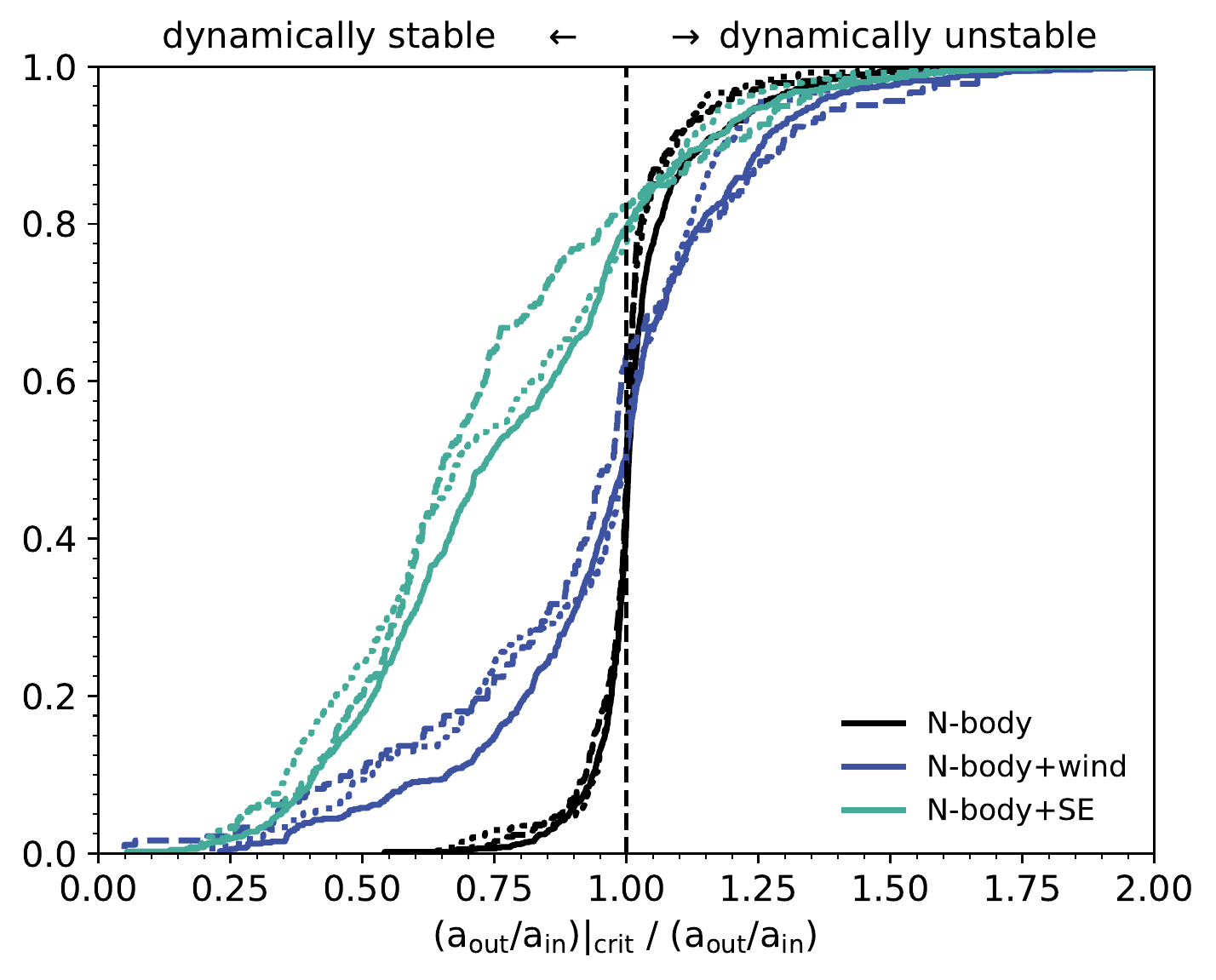} \\
\includegraphics[clip=true, width=\columnwidth]{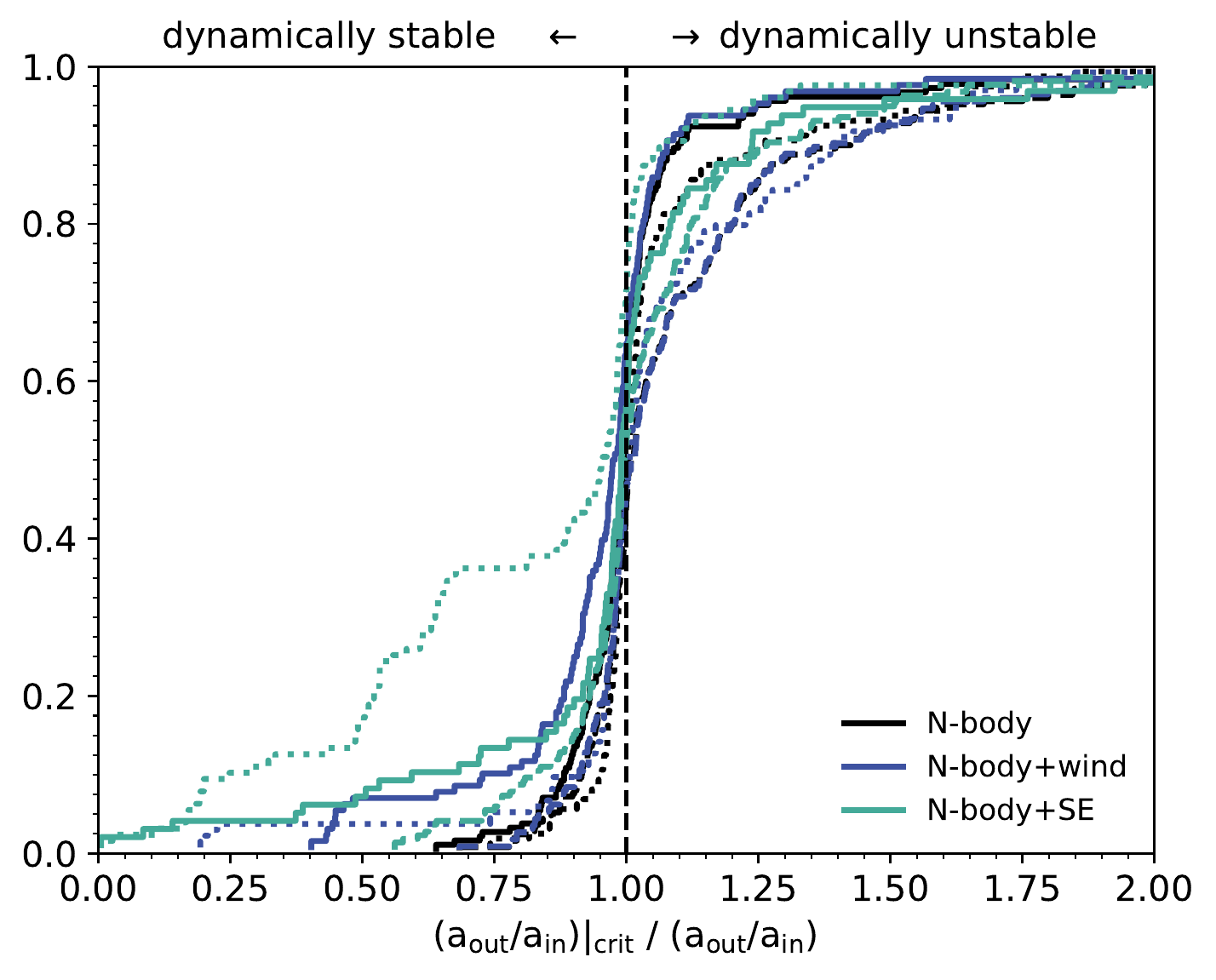} 
 	\end{tabular}
 	
 	\caption{For those destabilised triples  that remain bound, where are those systems compared to the stability limit? This is quantified by the ratio of $a_{\rm out}/a_{\rm in}|_{\rm crit}$ (that is the stability criterion of Eq.\,\ref{eq:stab_crit} over the current configuration (that is $a_{\rm out}/a_{\rm in}$). When the ratio is larger than 1, the system is considered to be in the dynamically unstable regime. If the ratio is smaller than 1, the system is dynamically stable. On the top we show the systems that remain bound after a Hubble time, on the bottom those systems that remain bound after simulating the dynamically unstable phase for 24 hours. 
 	Solid, dashed, and dotted line styles represent the models for the initial population of triples OBin, T14 and E09, respectively.
    }
    \label{fig:mardling}
    \end{figure}

Similar to the previous section on collisions, we find that the double averaged method breaks down for these systems. Besides strong variations in the inner and outer eccentricity, we also notice variations in the outer semi-major axis (also in cases without stellar winds). 
As a result of these variations, the triple can move back and forwards over the stability limit spending some time just inside the instability region, and some time just outside (top panel of Fig.\,\ref{fig:mardling}). In the pure dynamical simulations, 90\% of systems are within a factor 1.13-1.19 of $(a_{\rm out}/a_{\rm in})|_{\rm crit}$ (Eq.\,\ref{eq:stab_crit}). In the dynamical simulations with stellar winds and evolution, this increases to 1.7-2.1 and 2.4-2.8 for simulations including stellar winds and evolution, respectively. When including stellar winds and evolution, the additional wind mass loss can push the system deeper in the instability region (when mass is lost in the inner 'orbit') or out of the instability region (when the 'tertiary' star loses mass). The latter is reminiscent of the secular evolution freeze out studied by \cite{Mic14}.

The necessity of taking stellar evolution into account in the modelling of these systems is also apparent in the bottom panels of Fig.\,\ref{fig:outcomes_all}. The fraction of bound triples after a Hubble time decreases correspondingly as the stellar winds push the system further into the instability region (Sect.\,\ref{sec:ev}).

The long-lived triples  have wide orbits initially, that is initial inner orbital separations $\gtrsim 10^4\Ro$ (Fig.\,\ref{fig:outcome_origin}). Even though a wide orbit implies a relatively long crossing time, the median of the evolved time was still $10^5$ crossing times  (Fig.\,\ref{fig:t_cross}). This is significantly more than in the case of ejections or collisions, which indicates that the long-lived triples have remained near the instability border with only mild variations in its orbital elements.

At the onset of the dynamically unstable phase, the long-lived triples  typically have a stellar component on the giant branch that is (G,MS/G) and (WD,G). For example, this is the case for 76\%, 81\%, and 64\% of the triples in model OBin, T14, and E09 respectively for dynamical simulations including stellar evolution. At the end of the simulation, when a Hubble time has past, these giant components have become white dwarfs. If the other stellar components had masses above 1$\Mo$ initially, they too move off the MS and become white dwarfs by the end of the simulation. This is the case for 60-80\% of the ((G,MS,MS)) that remain bound in a Hubble time. Furthermore, all destabilised triples with (WD,WD) inner binaries remain bound within a Hubble time.

\subsubsection{CPU-time limited triples}  
\label{sec:cpu}

Lastly, we discuss the evolution of those destabilised triples for which the simulation of the dynamically unstable phase was terminated after simulating for 24 hours. 
The fraction of systems that is CPU-limited depends on the initial population of triples, and to a lesser degree on the modelling of the dynamically unstable phase. For model OBin it comprises 2.1-3.9\% of destabilised triples, 8.8-10.1\% for model T14, and 4.9-6.1\% for model E09. We have limited the simulation of each system in such a way that the total fraction less than 10\% for all models. In App.\,\ref{app:tcpu}, we demonstrate that in order to significantly reduce the fraction of CPU-time limited triples it would require over $10^3$ CPU-hours, which is beyond the scope of this project. Our rate estimates of the most common outcomes of destabilised triple evolution are therefore accurate to within an order of magnitude, while that of less frequent channels should be considered a lower limit. 

Many crossing times are simulated for the CPU-time limited triples, that is typically over $10^6$ (Fig.\,\ref{fig:t_cross}), however the systems do not change in a catastrophic way. The systems predominantly keep their hierarchy throughout the simulation (Tbl.\,\ref{tbl:rates}) and stay close to the stability limit (bottom panel of Fig.\,\ref{fig:mardling}). 75\% of systems are within a factor 1.10-1.24 of the stability criterion (Eq.\,\ref{eq:stab_crit}) for all models except one. The factor increases to 1.88 for the model that combines N-body dynamics with stellar evolution for the initial triple population of T14. In this model we find more systems that have moved into the dynamically stable regime (bottom panel of Fig.\,\ref{fig:mardling}). In addition, similarly to colliding systems and long-lived systems, the double averaged method breaks down for the cpu-limited systems, which demonstrates the necessity and justifies the transition from the secular simulations of \texttt{TRES} to the N-body calculations.

CPU-limited triples have initial inner orbital separations between $10^2-10^4\Ro$ and outer orbital separations $10^3-10^5\Ro$ (Fig.\,\ref{fig:outcome_origin}). As more such triples are created in model E09 compared to model OBin, the fraction of CPU-limited systems is higher in the former. In addition, the initial eccentricities of the outer orbits are remarkably high with a median value of 0.89-0.94 across all models. 
Given these relatively tight inner orbits, it is not surprising that the majority of CPU-limited triples are ((MS,MS),MS) systems (Tbl.\,\ref{tbl:rates}). The small fraction of ((G,MS),MS) triples decreases further when stellar winds and evolution are taken into account in the dynamical simulations.

\section{Observational exotica}
\label{sec:obs}

\subsection{Runaway stars}
\label{sec:runaway}
\begin{figure}
   \centering
   \includegraphics[width=\columnwidth]{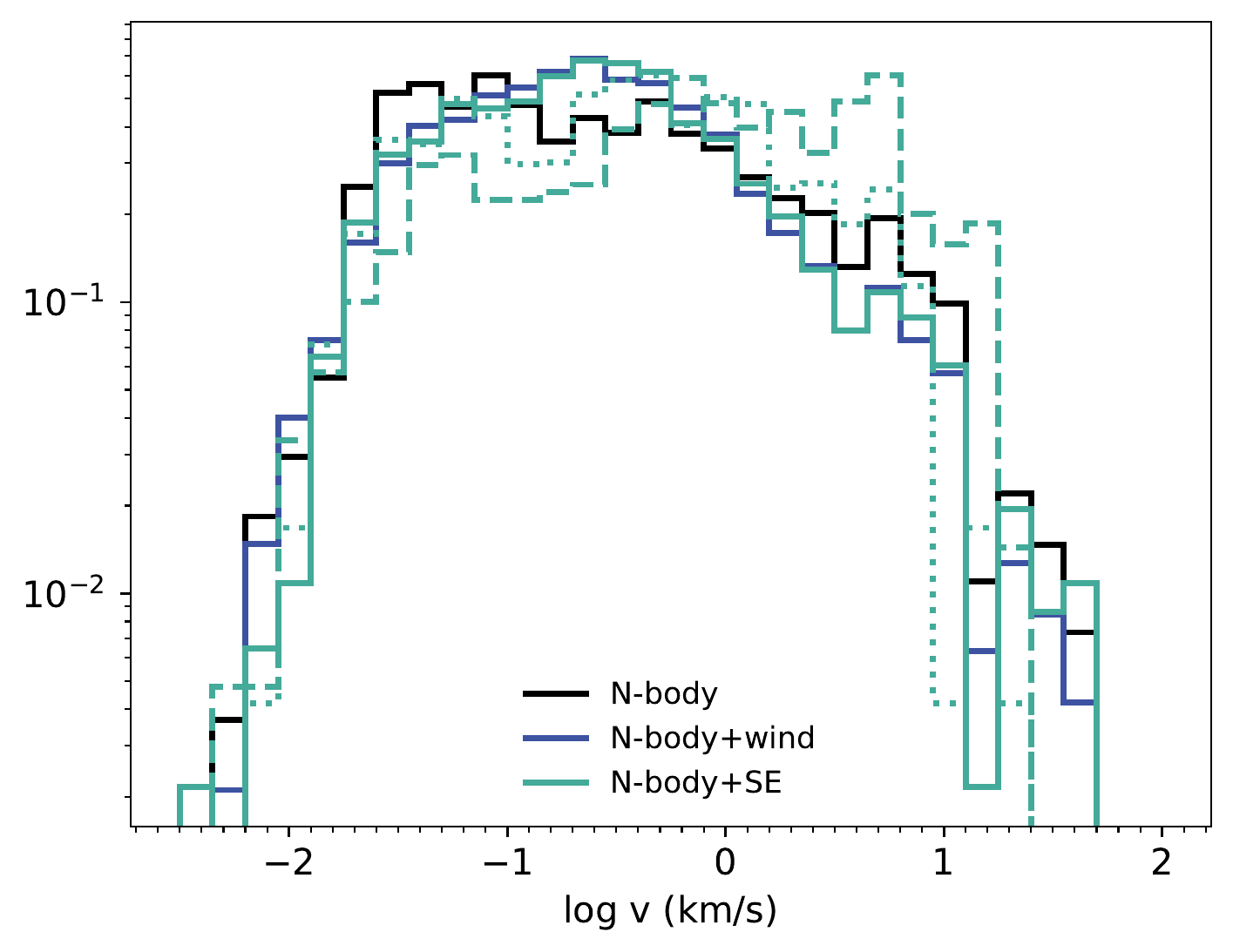} 
      \caption{ Histogram of the escape velocities of the escaping stars relative to the binary for different models. Different colours represent the different dynamical modelling methods, different line styles represent the different models for the initial population of triples; OBin (solid), T14 (dashed), E09 (dotted). }
         \label{fig:eject_vrel}
   \end{figure}

The destabilisation of a triple often leads to the ejection of one of the stars, and the formation of runaway and walkaway stars. These are stars that are moving through space with abnormally high velocities compared to their surroundings. Runaway stars are stars with velocities above 30(-40) km/s \citep{Bla61,DeD97, 2000ApJ...544..437P, Dra05,Eld11, Bou18,Ren19, Bis20}, whereas their slower cousins were nicknamed 'walkaway stars' by \citep{Ren19}. Runaway stars are thought to form as companions to massive stars collapsing in a supernova explosion \citep{Bla61}, or dynamically ejected from clusters in multiple body encounters \citep{Pov67}. In this paper we consider a third scenario, namely the destabilisation of an isolated triple. Fig.\,\ref{fig:eject_vrel} shows the velocity distribution of the ejected stars in our simulation relative to the remaining binary system. The velocity distribution peaks around a few km/s with a tail towards large velocities.  Drifters mainly contribute to velocities below 0.1km$/$s, and escaping stars above it.  Velocities up to several tens km/s are reached, placing these systems firmly in the regime of runaway stars. It is analogous to the escape velocities derived from scattering experiments of  three-body interactions with zero total angular momentum \citep[see e.g.][]{Ste95, Ste98}, or from N-body simulations for entire star clusters \citep{2011Sci...334.1380F}. As these interactions are typically hierarchical events in our simulations, the ejected star is typically the original tertiary of the system.

 The magnitude of the escape velocity is related to the orbital velocity before the destabilisation. For this reason, the fastest escaping stars originate from the most compact initial systems, and are on the MS during the ejection. We estimate the 
 terminal velocity $v_{\rm terminal}$ by 
 considering the average amount of energy that the tertiary extracts from the inner binary  (Appendix\,\ref{app:v_max}).  Assuming Solar masses and a minimum distance between the stars of 10\Rsolar\,(5\Rsolar), the maximum terminal velocity would be 67-157\,km/s (95-222\,km/s). 
 For more massive stars, the maximum velocity increases further. For 10\Msolar\, stars and a minimum distance of 10\Rsolar\,(50\Rsolar), the maximum terminal velocity reaches 95-222\,km/s (213-231\,km/s). In comparison, the expected velocities of runaway stars from the binary (supernova) channel are of the order of 20\,km/s \citep{Ren19}, as the orbital velocities decreases substantially during the pre-supernova mass transfer phase (i.e. orbital widening). 
As the observed runaway stars are typically O- and B-type stars\footnote{There is a strong observational selection effect to luminous and therefore massive stars. }, and massive stars are typically in compact orbits with inner periods of a few days \citep[$a_{\rm in}\sim 10-50\Ro$][]{San12,Moe17}, we suggest that destabilised massive triples have the potential to solve the mystery behind the origin of massive runaway stars.

Lastly, the ejection of stars  from destabilised triples (at all velocities), releases stars into the field that masquerade as 'normal' single stars. However, as the ejected stars are formed as part of a multiple, their mass distribution is likely different from that of a single star formed from a single cloud. In this way, the destabilisation of triples, is an interesting formation mechanism for single brown dwarfs , analogous to the hydrodynamical star formation calculations of unstable multiples \citep{Bat02, Del04}    

\subsection{MS-MS collisions}
\label{sec:colmsms}
Collisions between MS stars frequently occur in dense Galactic clusters \citep{Hills76, 1997A&A...328..143P, Hur01, Hur05, Umb08}. The rate of such collisions was estimated by \cite{Per12} to be of the order of $4\times 10^{-6} (N_{\rm GC}/20)$ per year, where $N_{\rm GC}\approx 20$ is the number of massive and dense clusters in the Milky Way. The Galactic rate of MS-MS collisions from destabilised triples is  $(2-2.2)\times 10^{-4}$ per year based on our simulations including SE during the N-body calculations. If you include collisions in stable and destabilised triples, the full rate of MS-MS collisions in triples would be even higher.
Stable triples can be driven to contact through  
 Lidov-Kozai cycles and the related high eccentricities without destabilising the system.
 In our simulations with \texttt{TRES} we label these systems highly-eccentric mass-transferring systems (Fig.\,\ref{fig:overview}). Their eccentricities at the onset of the mass transfer phase ranges from 0.4-1.0 with a median around 0.85-0.9 \citep[see Fig. 7 in][]{Too20}, whereas the eccentricities of the colliding destabilised triples are more extreme with a median value over 0.996. Based on the results of \cite{Too20}, the Galactic rate would be several $10^{-3}$ events per year. This is in good agreement with the estimate of \cite{He18} for the full MS-MS collision rate in triples, that is about $10^{-2}$ per year.

\cite{Kai14} proposed a third channel for producing stellar collisions. Their channel involves wide binaries that are perturbed by passing stars and/or the Galactic tide. They estimate a Galactic rate of $(1.3-10)\times 10^{-4}$ per year. This is similar to the collision rate we estimate for destabilised triples. However, if we include collisions in stable triples, then triple evolution is the dominant mechanism to lead to collisions in the Milky Way.

\begin{figure}
   \centering
   \includegraphics[width=\columnwidth]{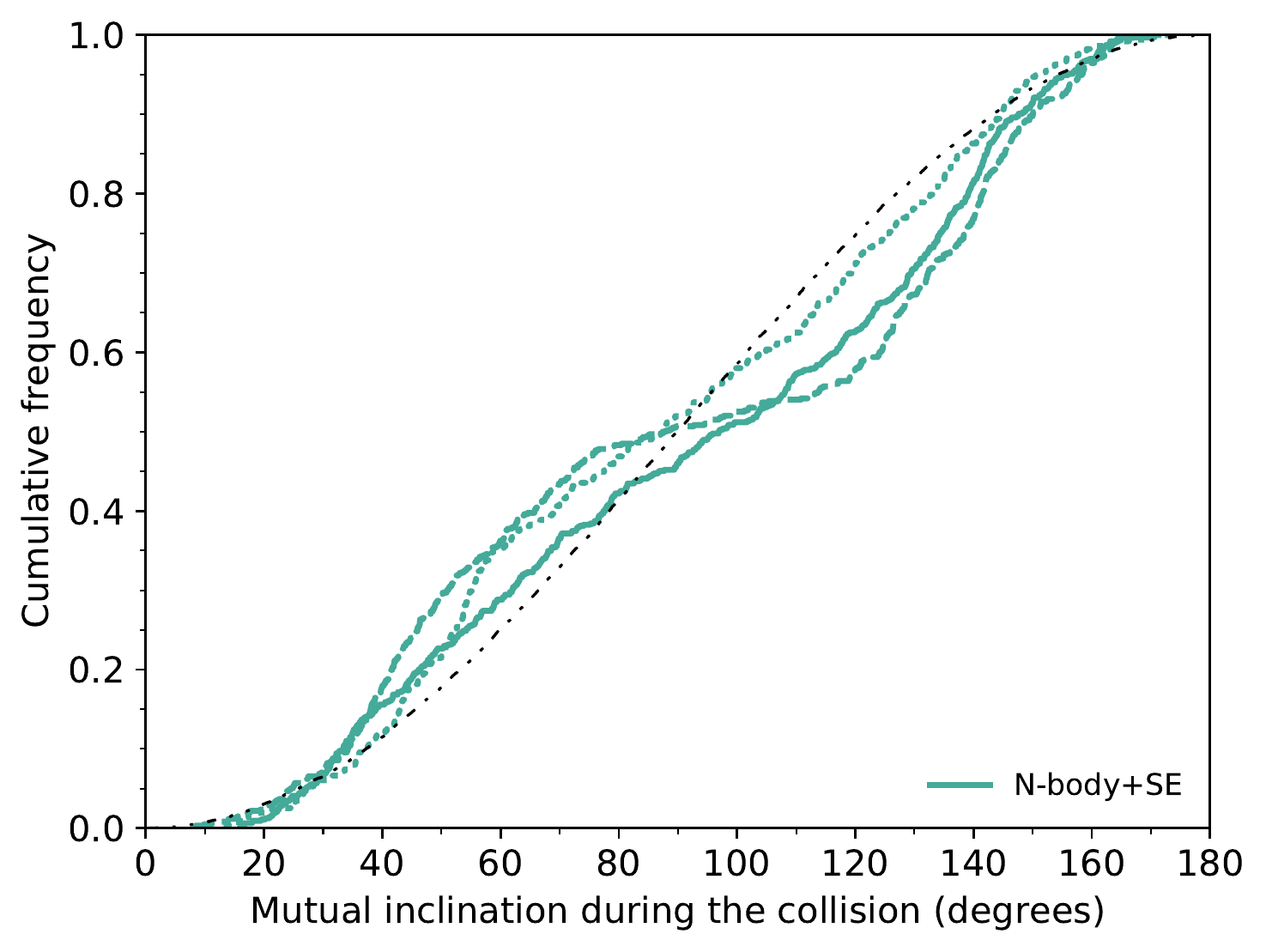}
      \caption{Distribution of inclinations at the time of the collisions. Green lines refer to the model where SE is included in the N-body simulations.  Solid, dashed, and dotted line styles represent the initial population of triples of model OBin, T14 and E09, respectively. The black dashdotdotted line represents a random distribution in the cosine of the inclination. }
         \label{fig:col_i}
   \end{figure}

In a cluster, MS-MS collisions can be recognised as (massive) blue stragglers \citep{Fre05,Lom95,Lom96, Sil97,Sil01,Sil02,Sil05}. 
Hydrodynamical simulations show that during the merger little mass is lost \citep[][]{Lom02, Dal06}, typically of the order of a few percent \citep{Gle08}. Some matter may be lost after the merger in order to spin down the star if the blue straggler is rapidly rotating, as expected for an off-axis collision \citep{Sil01,Sil05}.
If the remnant maintains its high rotation throughout the main-sequence phase, rotational mixing extends the lifetime of the blue straggler, as well as making it bluer and brighter \citep{Sil02, Gle08}. In case of head-on collisions and non- or slowly rotating remnants, substantial convective regions do not develop during the thermal relaxation phase, and no significant mixing occurs after the collision \citep{Sil01, Gle08}. Nevertheless, an extension of the lifetime of the remnant is expected due to hydrodynamical mixing \citep[e.g.][]{Dal06}. The amount of rejuvenation depends on how much hydrogen is carried into the core during the merger, which in turn depends on the entropy profiles of the colliding stars \citep[see e.g.][]{Lom96,Lom02, Gle08}. Unfortunately, collisional remnants from field triple evolution can not be  detected easily in a Hertzsprung-Russell (HR) diagram as is the case of cluster blue stragglers, which show two quite distinct tracks \citep{2009Natur.462.1028F} that indicate a different origin \citep{2019A&A...621L..10P}. 
Triple star evolution can, in such environments, also lead to a curious twin blue stragglers, such as WOCS ID 7782
\citep{2019ApJ...876L..33P}. In destabilised triples, the apparent age difference between the collision product and its outer tertiary star means the system will appear as a non-coeval binary. 
Moreover, the rotation of the remnant, in line with the orbital motion of the former inner binary, is typically inclined with respect to its orbital motion (Fig.\,\ref{fig:col_i}). 

The merger event can also be detected directly \citep{Sok03,Sok06, Smi11}, as in the case of V1309 Sco \citep{Tyl11}, V838 Mon \citep{Tyl06}, and even 
extragalactic events such as M85 OT2006-1 \citep{Kul07}. 
An observational link with triples has already been made, as \cite{Kam21} recently  showed that V838 Mon was a binary merger in a triple system. These intermediate luminosity optical transients (ILOTs) have a wide range in energies that span the range from classical novae to supernovae. This is likely related to the large range of possible accretion masses \citep[ e.g.][]{Sok11}. In particular, the subset of slowly-evolving red transients, also known as red luminous novae (RLN), are thought to originate from stellar mergers in common-envelope events \citep{Iva13} with giant donors \citep[see next section][]{Bla17,Bla21}.

\subsection{Collisions involving giants}
The first  estimate of the collision rate in destabilised triples was done by \cite{Per12}. They recognised the importance of stellar winds to drive the instability, and therefore the involvement of evolved stars in the TEDI. They estimate a Galactic rate of  collisions involving stellar giants (mostly AGB stars) of $10^{-4}$ per year. Our estimate of giant-MS collisions in destabilised triples is about an order of magnitude lower, that is $(0.4-1.8)\times 10^{-5}$ per year. The difference can likely be attributed to the evolution of the triples up to the destabilisation, in other words the initial conditions at the moment of the destabilisation. Whereas they assume a circular random distribution of inclinations, our systems avoid inclinations around 90$^\circ$. 
Systems with inclinations close to 90$^{\circ}$ experience strong gravitational perturbations and  eccentricity variations due to three-body dynamics, and would likely lead to collisions during the dynamically unstable phase. However, if three-body dynamics is taken into account in the hierarchical phase, as is the case with \texttt{TRES}, the systems interact prior to the destabilisation leading to a mass transfer phase or a collision earlier in the evolution (e.g. on the MS).  
As noted in Sect.\,\ref{sec:res_hier} for the hierarchical phase, the far majority of our destabilised triples are in and even born in the octupole regime, and three-body dynamics is an important contributor to the formation of the destabilisation. Despite the lower rate compared to \cite{Per12}, the events are still at least as common as collisions in Galactic clusters. 

Depending on the impact parameter of the collision, we envision three outcomes \citep{Bai99}: Either 1) the MS star merges with the giant, or 2) a compact binary is formed removing the envelope in a common-envelope like event \citep[for a review on common-envelope evolution see ][]{Iva13}, or 3) the MS star passes through the envelope removing (part of) the envelope on its way out. The former would lead to a giant star with an anomalously large envelope mass for its core and an anomalous internal rotation pattern, which could be detectable with asteroseismology.  
The latter scenario, may give rise to a sub-subgiant or a red straggler \citep{Gel17, Lei17}. Like blue stragglers, these stars are found in atypical areas of the HR-diagram of clusters. But whereas blue stragglers lie bluewards of the turn-off, sub-subgiants and red stragglers are observed to be redder than normal main-sequence stars. Sub-subgiants are fainter than normal (sub)giants, and red stragglers lie redward of the red giant branch (although often considered to be part of the sub-subgiant family). 
\cite{Lei17} showed that a rapid partial removal of the envelope mass leads to a rapid decrease in luminosity of the giant and the formation of a sub-subgiant. However, close encounters between a binary and a passing star are not frequent enough to explain the observed systems \citep{Lei17, Gel17b}, and so it would be interesting to see if destabilised triples could solve this mystery. Interestingly, \cite{Gel17} found that the number of sub-subgiants per unit mass increases towards lower-mass clusters, which is consistent with hierarchical triples being more prevalent in open clusters than in the dense globular clusters \citep{Lei13}.

In the second and third case, when most or all of the envelope mass is stripped from the giant, a planetary nebulae (PNe) may form, assuming that the matter is ionised sufficiently by the central object. Given the highly eccentric nature of the collisions and the close proximity of the tertiary star, the resulting PN can deviate from spherical symmetry, and even axial- and point-symmetry. The origin of the PN morphologies are sought in magnetic fields, binary motion and stellar rotation amongst others \citep[e.g.][]{Sok02,DeM09, DeM11,Zij15, Jon17}, however \cite{Bea17} estimate that one in six PN are too messy to be accounted for by the processes mentioned. \cite{Sok16} estimates that one in eight PN by triple evolutionary pathways, like the one studied in this paper. 
A direct connection between PN and triples was even made for PN Sh 2-71 \citep{Jon19}.

\subsection{Collisions involving white dwarfs}
After MS-MS collisions, the most typical collision involves a WD.
Our Galactic rate estimates are $(0.72-3.9)\times 10^{-5}$ per year for WD-MS collisions, $\lesssim3.4\times 10^{-6}$ per year for collisions between a WD and an evolved star, and $\lesssim1.7\times 10^{-6}$ per year for WD-WD collisions. As the WDs in this channel are post-AGB objects, they are composed of carbon-oxygen (CO) and/or neon, and have masses above 0.5M$_{\odot}$.

During a collision between a WD and a MS, the MS experiences strong tidal torques that can lead to the tidal disruption of the star (for 0.1-1\Msolar\, MS stars), and a significant disruption for more massive stars \citep{Reg87}. Hydrodynamical simulations show that a massive disk forms around the WD, liberating the binding energy of the MS \citep{Sha86, Reg87, Sok87, Roz89}. \cite{Mic21} estimate a bolometric luminosity of $10^7-10^9L_{\odot}$, which makes these events similar to kilonova in brightness. After the merger a giant-like star forms \citep{Roz89, Hur02}. Similar to what is discussed in Sect.\,\ref{sec:colmsms}, the newly formed giant likely has a anomalous core-to-envelope mass ratio, internal rotation pattern, and position in the HR-diagram.

In the case of a collision between two WDs, hydrodynamical simulations have shown that the shock can trigger nuclear reactions and the formation of a short-lived R Coronae Borealis star \citep{Web84,Lor10,Sch19,Sch21}, or lead to the total disruption of the star in a supernova-like explosion \citep{Ros09,Ras09,Kus13}.
Expected luminosities range from supernova Type Ia-like \citep[][and even matching spectra and lightcurves]{Ros09, Kus13} to underluminous supernovae \citep{Ras09}. \cite{Lor10} also showed that even in the case of a post-collision remnant, the fallback luminosities are close to $10^{48}$erg/s.

\subsection{Eta Carinae}
An interesting system to mention in the context of collisions and triples is Eta Carinae \citep{Dam97}. Eta Carinae is currently a binary system with two massive stars in a highly eccentric orbit, and is well known for its Great Eruption in the middle of the nineteenth century \citep{DeV52}. A link with triples was made by \cite{Por16} who proposed that the Great Eruption was caused by the merger of the stars in the inner binary due to the Lidov-Kozai cycles imposed by the tertiary companion \citep[see also][]{Too16}. Later on \cite{Smi18} suggested a formation channel involving a destabilised triple. In their scenario the instability is driven by the expansion of the inner orbit due to mass transfer. During the dynamically unstable phase, the original secondary and tertiary collide, causing the Great Eruption. 
 Detailed simulations of the scenario have been performed by \cite{Hir21}. These consist of hydrodynamical simulations of the merger, as well as pure N-body simulations of the dynamically unstable phase for one triple with 1000 initialisations of the initial orbital phase of the inner binary. Out of their runs resulting in a merger, only 8-11\% experienced a swap of the companions. 
While our simulations do not consider mass-transfer-driven instabilities, they show that collisions occur between the original primary and secondary (and not the tertiary) with few exceptions. Despite the similar mass ratio between our simulations and the proposed scenario for Eta Carinae, we only find collisions involving the tertiary in 1.6-2.4\% of all collisions. This does not change significantly when focusing only on ((MS,MS),MS) systems (i.e. 1.2-1.5\%) or compact orbits (i.e. 1.8-2.0\% for $a_{\rm in}<10^4\Ro$). This would imply that for the scenario to work for every Eta Carinae there should be 50 similar systems that have experienced a collision in the inner binary.

\subsection{Wide binary formation}

The majority of destabilised triples come out of the dynamically unstable phase as a wide binary system, either through ejecting a star or through a collision of two of the stars. That the latter is possible, in particular that the tertiary can remain bound, is demonstrated by OW Geminorum \citep{Egg02}, and see \cite{Egg13} for three other cases. 
This binary with a 3.45 yr period and eccentricity of 0.53 consists of two sub-giants with surprisingly different masses of 5.5\Msolar\, and 3.8\Msolar\, \citep{Gri93,Ter03, Gal08}. As the MS-lifetime is a strong function of stellar mass, two single stars with such different masses should not be on the giant branch simultaneously. \cite{Gal08} estimates the lower mass star to be at least twice as old (200 million years) as its companion. OW Geminorum posed a problem for standard stellar evolution theory if stellar interactions are not provoked. Indeed, \cite{Egg02} proposed a triple-star solution; OW Geminorum formed through a merger in a triple with a compact (2d) inner binary with a $\sim$4\Msolar\, and $\sim$2\Msolar\, star that was orbited by a $\sim$4\Msolar\, tertiary. If OW Geminorum formed through the destabilisation of a triple instead, we envision the following set-up: \\
- masses the same as suggested by \cite{Egg02}\\
- assuming the outer orbit has not changed much during the unstable phase,  the inner period at destabilisation would have been 50-100 days (Eq.\,\ref{eq:stab_crit}). This is wide enough for tides not to quench the three-body dynamics and (to a secondary) degree for the stars to evolve and develop winds. Both processes contribute to the breakdown of the stability.

  \begin{figure}
   \centering
\begin{tabular}{c}   
\includegraphics[width=\columnwidth]{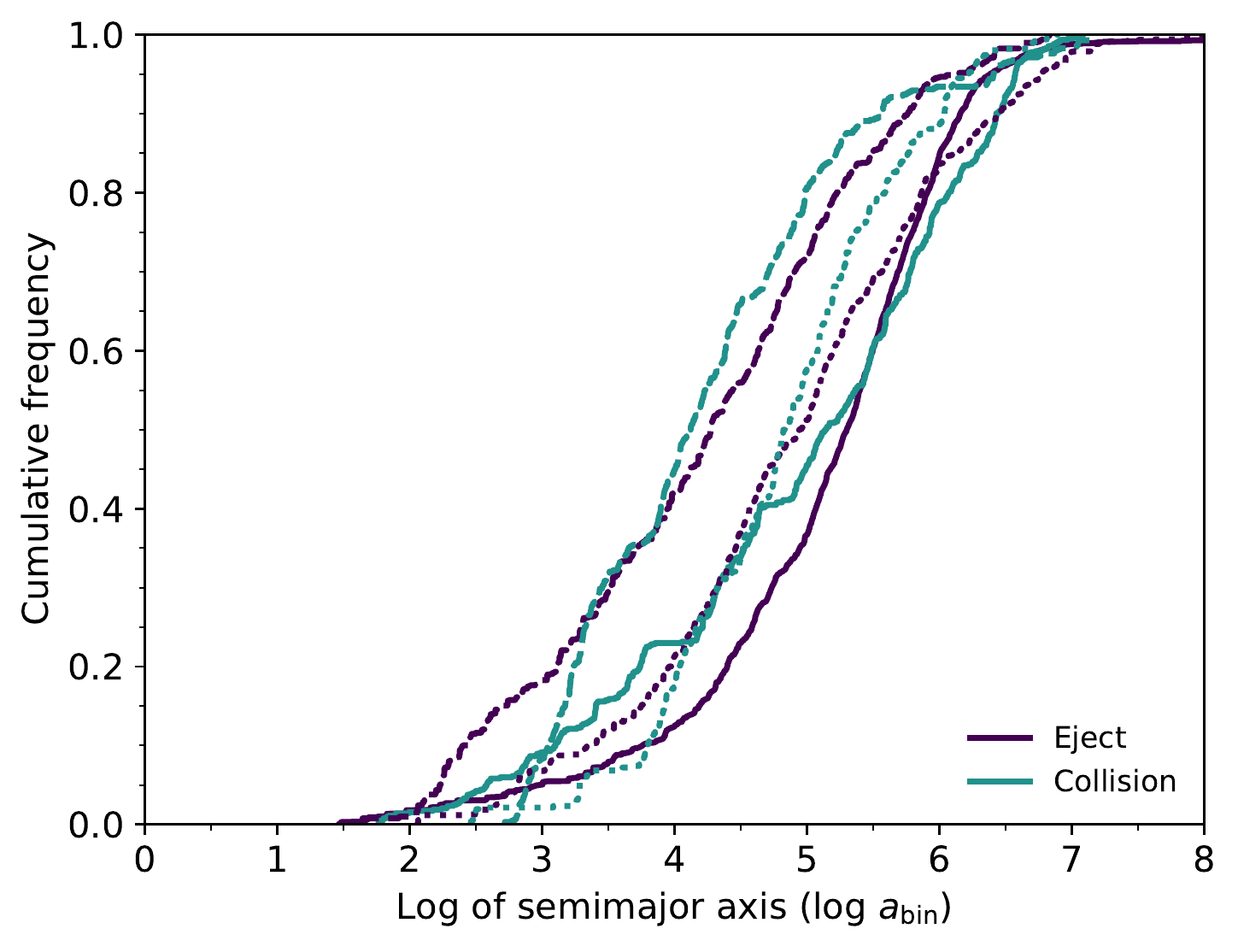} \\
\includegraphics[width=\columnwidth]{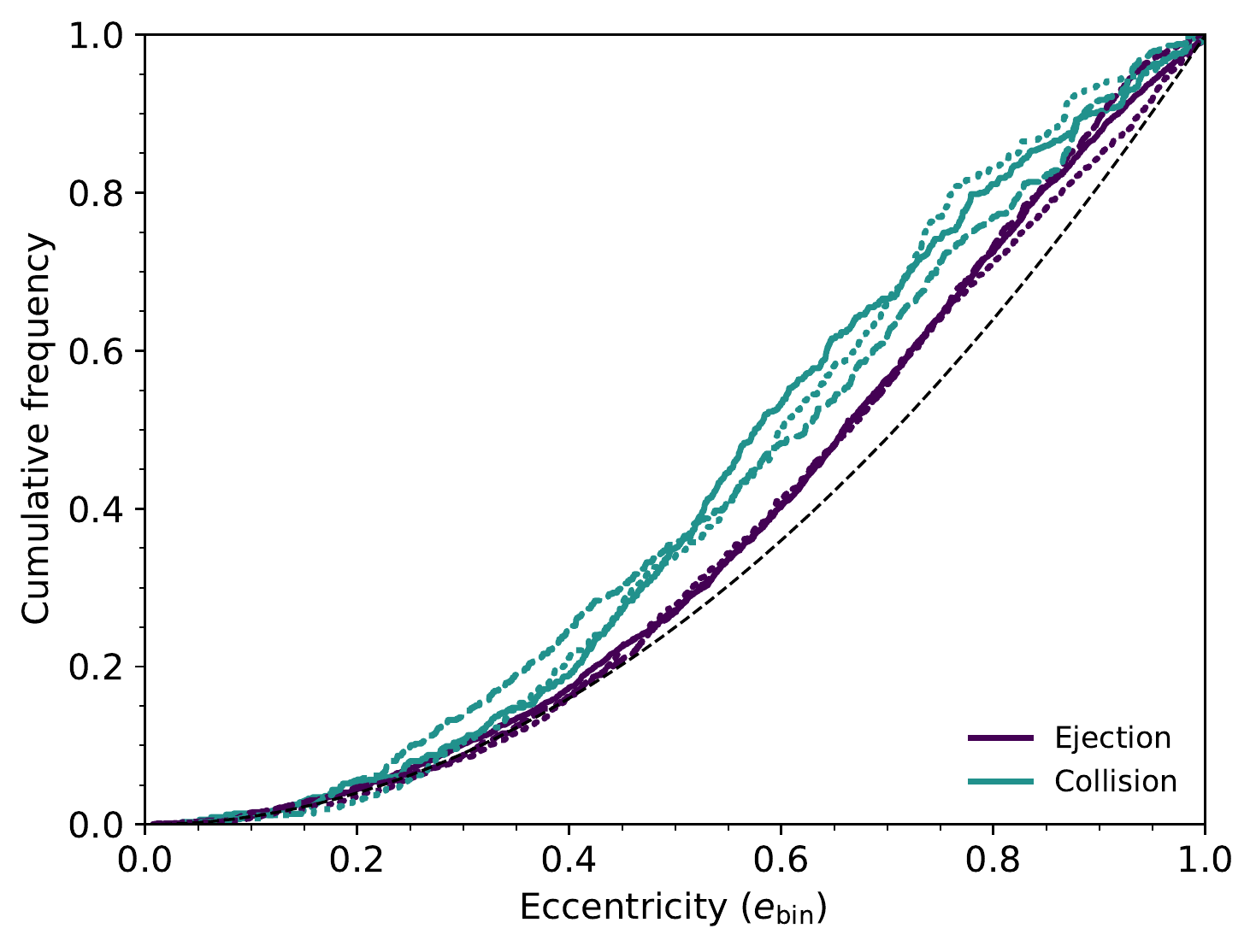} \\
\end{tabular}   
      \caption{Properties of the (isolated) binaries that are formed from destabilised triples. Solid, dashed, and dotted line styles represent the N-body + SE models for the initial population of triples OBin, T14 and E09, respectively.  The black dashed line represents a thermal distribution of eccentricities. 
      }
         \label{fig:bin}
   \end{figure}

The typical orbital separations and eccentricities of the remaining binaries are shown in Fig.\,\ref{fig:bin}. The eccentricity distributions are sub-thermal, and most strongly so for the post-collision binaries. The orbital separations of the newly found binaries are large, and span a broad range from $10^2-10^3\Ro$ upwards. The top panel of Fig.\,\ref{fig:bin} shows that the adopted initial orbital separation distribution has a strong imprint on the demographics of the binaries.  

As discussed above, post-collision binaries can be recognised as they harbor peculiar stars (e.g. blue, yellow and red stragglers) or by the apparent age difference (as for OW Geminorum). Blue stragglers are indeed typically found in wide binaries. \citet{Mat09} found that 76\% of blue stragglers in NGC 188 are part of a binary with typical orbits around a $10^3$d (or at least $10^3$d as longer periods are  hard to detect observationally). 
Moreover, the orbits have eccentricities between 0 and 0.8. This is surprising as the classical mass-transfer scenario in isolated binaries would have circularised the orbits \citep[but see][]{Bon08,Gel11,Der13}. The eccentricities indicate an origin in multiple stars. One scenario involves Lidov-Kozai cycles in combination with tidal friction \citep{Per09, Nao14}. Here we show that also collisions in destabilised triples contribute in their formation, albeit in the limit where Roche lobe overflow is neglected.

\subsection{Compact binary formation}
In this paper we focus on isolated, hierarchical stellar triples that become dynamically unstable as they evolve. We have shown that the duration of the dynamically unstable phase is much longer than what is generally expected from binary-single resonant interactions or triples drawn from Plummer spheres. Although triples on the edge of stability are considered unstable according to various stability criteria, they tend to be long lived reaching up to thousands of crossing times. And so, we argue that it is important to take stellar evolution into account during the unstable phase. Besides stellar evolution, dissipative processes  may also play a role. Examples are stellar tides, gravitational wave emission or general relativistic precession. These processes act most efficiently during strong encounters or as a cumulative effect during repeated close passages \citep{Kai14,Mic16,Mic19,Ham19,Mic20, Mic21}. As a result, we expect dissipative processes to be most important during democratic encounters \citep[see e.g.][]{Ginat2020}. In general, for ordinary triples in which dissipative processes play an important role in the dynamics, a compact inner binary forms that is kinematically decoupled from the tertiary \citep{Maz79, Kis98,Fab07,Liu15,Ant17,Rod18, Bat18}. Such a compact orbit may eventually lead to a merger of the two stars in the inner binary. 
How effective dissipative processes are for destabilised triples, which is outside of the scope of this paper, can be verified by using a direct N-body code, such as \texttt{TIDYMESS} \citep{TBTIDY}, which combines orbital evolution with stellar spins and gravitational tides between each pair of bodies in a self-consistent manner. 
It would likely lead to the formation of compact inner binaries at the expense of collisions and stellar escapes. The further evolution of the system may once again lead to a dynamical instability. As the stars in the newly-formed compact inner binary evolve, mass transfer may develop in the inner binary. If the mass transfer is stable and the orbit widens consequently, another dynamical instability may commence. On the other hand, if the stars in the compact inner binary are compact objects, a gravitational wave source can form that is detectable by LISA or even LIGO/Virgo  \citep[e.g.][]{Antognini14, Silsbee17, Antonini17, Hoang18, Rodri18, Bonetti19}. If any eccentricity remains in the orbit at the time of detection, the system would stand out compared to those formed through isolated binary evolution. The compact binary with compact objects  will eventually merge and give rise to electromagnetic transients such as supernova type Ia, gamma-ray bursts or kilonova \citep{Web84, Met10, Pak12, She18}.

\section{Discussion}
 \label{sec:discussion}

In this study we have constructed three different models for the initial population of triples that differ from one another based on their mass, mass ratio and period distribution. For all models we have assumed that the mutual inclinations of the orbits follow a circular uniform distribution. The observations of \cite{Tok17} support this assumption for outer projected separations $\gtrsim 10^5R_{\odot}$. For low-mass triples with outer projected separations $\lesssim 10^4R_{\odot}$ there is more alignment. Which destabilised triples would be affected by this and how? The mentioned range of affected orbital separations corresponds to the regime where the destabilised triples mainly consist of three MS stars (Fig.\,\ref{fig:pop_TRES}), which become unstable due to three body dynamics (i.e. the octupole term). Whereas the amplitude of Lidov-Kozai cycles are strongly dependent on the initial inclination (limiting the cycles to $39.2-140.8^{\circ}$ in the test-particle approximation), the eccentric Lidov-Kozai mechanism is accessible for a larger range of inclinations \citep{Nao14,Nao16}. Moreover, the inclination distribution of the destabilised triples in our simulations avoid inclinations around 90$^{\circ}$ in the first place, as these typically lead to mass transfer \citep[][see e.g. their Fig.15]{Too20}. And so, if in real life compact triples have more alignment, it would mean our synthetic compact destabilised triples would have retrograde orbits less frequently than currently modelled. These are the systems that predominantly experience collisions, where as their prograde counterparts predominantly dissolve into a binary and an escaping star.   
\subsection{Limitations of the model}

The simulations performed in this paper consist of two parts. The hierarchical phase is simulated with \texttt{TRES}, which is based on the double averaged method, up to the stability limit. The subsequent evolution is simulated with the N-body approach. 
In general the stability limit that we adopted \citep[Eq.\,\ref{eq:stab_crit},][]{Mar99, Aar01} is considered
to be conservative\footnote{That is, it may take longer than a Hubble time for the instability to manifest itself} \citep{Myl18, He18}. 
This is further corroborated by the fractions of bound systems left in our various N-body simulations and the large number of crossing times simulated (Fig.\,\ref{fig:t_cross}). However, it is good that we transition from \texttt{TRES} to the N-body simulations, as we notice the break down of the double-averaging method in sections\,\ref{sec:collisions}-\ref{sec:cpu}. This behaviour leading to variations in the outer orbital separations, as well as higher maximum eccentricities we would not have caught with the double-averaging approach with quadropole and octupole terms \citep[see e.g.][]{Ant12,Kat12,Luo16,Gri18,Bha21}.

This leads one to wonder: what if we would have stopped the \texttt{TRES} simulations earlier? The experiment in  appendix\,\ref{app:inc} shines light on this matter, in particular model\,B~\&~D. These models show the difference in outcome for pure N-body simulations of retrograde triples on the stability limit (model\,B) or slightly above it (model\,D). Whereas over 60\% of triples in model\,B experience a catastrophic event (i.e. collision or ejection), this is only the case for 2\% in model\,D, for which the far majority of triples is long-lived and remain bound (Tbl.\,\ref{tbl:inc}). This behaviour confirms that the triples are indeed not dynamically unstable in model\,D. If the double averaged method including the octupole term underestimates the maximum eccentricity for a given triple as it approaches the stability limit \citep{Ant12,Kat12, Luo16, Gri18, Bha21}, then the triple would have destabilised earlier in the evolution compared to what is assumed in this work. This is mostly important for the ((MS,MS),MS) systems, as the instability for systems with giants are largely driven by their stellar winds. However, if the destabilisation of a ((MS,MS),MS) system occurred earlier in its evolution, the outcome of the dynamically unstable phase would  not be very different.

\subsection{Limitations of the direct model}\label{sec:limit}

Our direct integrations include both the orbital and the mass-radius evolution of the three stars. 
These two physical ingredients cause the destabilisation of hierarchical triple star systems.
When two stars overlap, we consider them to have collided, that is the sticky sphere approximation. 
Our direct model has some limitations, as we do not include relativistic effects, tides, or mass transfer during close passages within the Roche limit. 
These effects could potentially play a key role in the formation of close binaries and other phenomena.

The interplay between von Zeipel-Lidov-Kozai cycles and relativistic precession can lead to various eccentricity behaviours. If the time scale for relativistic precession becomes significantly shorter than the time scale for von Zeipel-Lidov-Kozai cycles, then it is possible for the maximum attained eccentricity to be quenched. However, in the case when the two time scales resonate, eccentricity excitation can be re-triggered \citep{Ford00, Nao13, Will17, Lim20, Hansen20}. 
Depending on the radii of the objects in the inner binary, 
a change in the maximum attained eccentricity could directly influence our measured collision rate. 
However, for triples on the edge of stability, for which $a_{\rm{out}} \sim 3 a_{\rm{in}}$, the inner orbit is strongly perturbed by the tertiary star. 
On the other hand, the relativistic precession time scale of the inner orbit does not depend on the outer orbit.
Hence, for destabilised triples in which all orbital elements of the inner and outer binary vary in time, it is unclear whether relativistic precession acts
to quench eccentricity oscillations in the inner binary, or rather acts as a perturbation to a chaotic system, leading to loss of memory of its initial condition. 
The latter scenario on the effect of relativity in a non-hierarchical, chaotic triple system was recently studied by \cite{TB2021}. They find that relativistic perturbations to the Newtonian force,
grow exponentially due to the exponential sensitivity in the chaotic three-body problem. This results in a sensitive dependence in the lifetime of the triple, but the
statistical outcome remains unchanged, that is a dissolution or head-on collision. Only when the velocities start to approach the speed of light, do gravitational wave captures
become prominent outcomes, which mostly applies to compact stellar remnants.   

Tidal dissipation tends to circularise eccentric binaries into an orbit whose semi-major axis is about twice the initial pericentre separation. 
The damping of the eccentricity prevents direct collisions at pericentre in the sticky sphere approximation, and instead leads to the production of close binaries, or compact triples if the tertiary remains. On the other hand, for destabilised triple systems which lose their hierarchy and become democratic, it is still uncertain how tides affect the outcome. 
The equilibrium tide model used in our secular calculations, assumes that the tidal lag is always small compared to the orbital time scale \citep[e.g.][]{Mignard79, Hut81}. For high eccentricity, this will not always hold during close pericentre passages, where
the orbital time scale can be very short. In this case, non-linear tides is a better description (such as the Maxwell model, e.g. \cite{acor14}). Furthermore, if the orbital time scale resonates with internal oscillation modes of the star as in dynamical tide models \citep[e.g.][]{Fuller12}, this triggers a strong enhancement of the tidal dissipation. This would lead us to expect that the equilibrium tide model underestimates the amount of tidal dissipation during very close pericentre passages. As a consequence, high eccentricities in the inner orbit of a triple are less effectively damped, thereby prolonging the possibility of a collision or dynamical dissolution. Our estimated fractions for these two outcomes can thus be considered as upper limits, and a benchmark test for future studies with more sophisticated tidal models.

Another limitation of our numerical setup is the simplified handling of close encounters and collisions. 
We consider two stars to have collided if their mutual distance is less than the sum of their radii, that is they overlap. 
If their separation is only slightly larger however, tidal effects and Roche lobe overflow events occur instead. 
For two solar mass stars in a circular orbit, this happens within a separation of about 1.25 times the sum of their radii, 
while for a giant and a white dwarf a separation of up to 2.5 times the sum of their radii suffices.
In our simulations these effects are not taken into account, which could potentially introduce a bias. 
This can be tested by including physical models for tidal effects and eccentric mass transfer (episodes),
whose influence on the eccentricity fluctuations of the inner binary is to be compared to the chaotic fluctuations caused by the tertiary. The inner binaries' semi-major axis could also be affected if orbital energy is dissipated by tides or if there is mass transfer and/or mass loss.
However, (episodic) mass transfer events are obviously challenging to model efficiently in a population synthesis study,
but apart from that, the workings of mass transfer in very eccentric inner binaries is still poorly understood. 
The inclusion of relativity, tides, and other effects in the direct simulation of destabilised triples is left for a future study, whose benchmark will be the current study. 
Here, our aim is to determine the influence of different stellar evolution models on the outcome of destabilised triple systems.

\section{Conclusions}
\label{sec:concl}

We have presented a numerical study on the evolution of destabilised, hierarchical triple-star systems, that is `triples on the edge'. 
The current study follows-up on \cite{Too20} who studied the typical evolution of hierarchical triples, and focuses on those triples that start out in a stable configuration \citep{Aar01}, but become dynamically unstable due to stellar wind mass loss and/or strong dynamical perturbations between the inner and outer orbit. 
The initial hierarchical phase is modelled in the secular regime with the multi-physics code \texttt{TRES}, while the unstable phase is treated by direct N-body integration. To determine the influence of stellar evolutionary effects during the unstable phase, we consider 1) a model with non-evolving stars, 2) a model with constant mass loss rates, and 3) a realistic model of stellar evolution following the code \texttt{SeBa} \citep{Por96, Too12}. A comprehensive overview of our main results regarding the evolution of destabilised triples is given below.

\begin{itemize}
    \item {\it Hierarchy:} 2-4\% of all low and intermediate mass triples become destabilised. 
    Of these destabilised triples, 80-83\% preserve their initial hierarchy throughout the evolution in the case when stellar winds are neglected, while 54-69\% do so when stellar evolution is included. This is in contradiction with the commonly adopted picture that unstable triples always experience a chaotic, democratic resonant interaction. 
    \item {\it Lifetime:} The duration of the unstable phase has a median of $10^{3-4}$ crossing times, reaching up to millions. This is in contradiction with the assumption that the lifetime of unstable triples is relatively short, so that long-term effects can be neglected. The extended duration allows for other physical effects, such as stellar evolution, to play a crucial role in the evolution.
    \item {\it Lifetime:} In fact for a number of systems, the lifetime is sufficiently long, in excess of $10^6-10^7$ crossing times, such that it exceeds our maximum CPU time of 24 hours. It comprises 
     2-4\% of destabilised triples for model OBin, 9-10\% for model T14, and 5-6\% for model E09. Our rate estimates can therefore be regarded as lower limits, while for the most common outcomes of destabilised triple evolution they accurate to within the order of magnitude. 
\item {\it Ejections:} The most probable outcome is a dissolution of the triple into a single star and binary:  42-45\% leads to unbound escapers, and 12-22\% to slow drifters.
    \item {\it Ejections:} Two ejection modes are: 1) the commonly known democratic ejection during which the initial hierarchy is lost and the lightest body tends to escape (57-70\% for escapers, 24-44\% for drifters), and 2) hierarchical slingshot of the tertiary irrespective of its mass (30-43\% for escapers, 56-76\% for drifters).
\item {\it Runaway stars:}     
   Our simulations produce runaway and walkaway stars with speeds up to several tens km/s. In appendix \ref{app:v_max}, we calculate a maximum velocity of one to a few hundred km/s. The main factor determining the ejection speed is the compactness (or total energy) of the initial triple. We suggest that destabilised massive triples have the potential to solve the mystery behind the origin of the observed (massive) runaway stars.
    \item {\it Collisions:} Collisions are common and occur in 13-24\% of destabilised triples. Almost all of them occur between the original components of the inner binary, only in 1.6-2.4\% is the tertiary star involved. This is consistent with the fact that the initial hierarchy is preserved until the moment of the collision (88-95\%), and contradicts the idea that collisions during democratic encounters dominate (only 5-12\%).
    \item {\it Collisions:} Collisions typically involve main-sequence stars (77-94\%). This is in contradiction with the expectation that giant stars are usually involved, since due to their enhanced radii they form large targets during democratic resonant encounters. The relatively short time-frame of the giant phase, and the random walk of the inner eccentricity, result in collisions being effectively avoided. Once the giant has evolved to a much smaller white dwarf, collisions can still take place due to the much longer time frame afterwards. Indeed, we find a higher fraction of collisions between a white dwarf and main-sequence star (4-14\%) than between a giant and main-sequence star (2-6\%).
    \item {\it Collision rates:} We estimate the following Galactic event rates for collisions, which originated from destabilised, hierarchical triple systems: $(2-2.2) \times 10^{-4}\,\rm{yr^{-1}}$ for MS-MS stars, $(0.72-3.9) \times 10^{-5}\,\rm{yr^{-1}}$ for WD-MS stars, $(0.4-1.8) \times 10^{-5}\,\rm{yr^{-1}}$ for G-MS stars, $3.4 \times 10^{-6}\,\rm{yr^{-1}}$ for G-WD stars, and $1.7 \times 10^{-6}\,\rm{yr^{-1}}$ for WD-WD pairs. When including collisions in stable triples, we find that triple evolution is the dominant mechanism for stellar collisions in the Milky Way.
\end{itemize}

The results listed above complement previous results on the dynamical evolution of unstable triples from binary-single scattering experiments. Destabilised, hierarchical triple systems can evolve not only along the democratic track, but also along the hierarchical track. If the evolutionary path takes the triple through a democratic resonance, then the initial hierarchy is lost and the main trends found in extensive binary-single scattering experiments in the literature apply. However, a large fraction of destabilised hierarchical triples preserve their hierarchy, resulting in different ejection and collision scenarios. When studying triples on the edge of stability, both these pathways are prevalent and affect the outcome statistics.

Based on our statistical analysis, we present two rules of thumb. These can provide a general prediction for the fate of a specific triple, based on the initial inclination of the system.  Specifically, they concern whether a triple will destabilise along the hierarchical or the democratic track. Obviously detailed and accurate predictions are better made with the N-body approach described in Sect.\,\ref{sec:method}, but -taken with care- the rules of thumb provide a new and fast insight in the problem of destabilised triples.

\begin{itemize}
    \item First rule of thumb: prograde triples\footnote{See footnote \ref{footnote:prograde}}
    tend to evolve towards hierarchical collisions and ejections, that is the tertiary star is ejected irrespective of its mass. 
    \item Second rule of thumb: triples with retrograde orbits tend to evolve towards democratic encounters and loss of initial hierarchy. This commonly leads to ejection of the lowest-mass body. However, compact retrograde systems experience hierarchical collisions.
\end{itemize}

These statistical rules indicate that the inclination plays a key role in the energy exchange between the inner and outer orbit of the triple. Since these are triples on the edge of stability, the pericentre distance of the outer orbit is comparable to the inner binary separation. As a consequence, the inner binaries' orbital speed and the tertiary's pericentre speed can be comparable in magnitude. Since the two velocity vectors point in approximately the same direction for prograde triples, the relative velocity will be small, which in turn maximises the interaction time in which energy can be exchanged. In the retrograde case, the velocity vectors point in opposite directions, and the interaction time is minimzed. Based on this simplified analysis, we thus expect tertiaries to be able to escape through a `hierarchical slingshot' mostly in prograde triples. Further studies are required to reveal the role of inclination, and other potential indicators, in shaping the dynamical evolution of destabilised, hierarchical triple star systems and corresponding observables.

\begin{acknowledgements}
We thank Nathan Leigh 
for the discussions on stellar collisions.
      This project was supported by funds from the European Research Council (ERC) under the European Union’s Horizon 2020 research and innovation program under grant agreement No 638435 (GalNUC). ST acknowledge support from the Netherlands Research Council NWO (VENI 639.041.645 grants and VIDI 203.061 grants). 
TB acknowledges support from ENGAGE SKA RI, grant POCI-01-0145-FEDER-022217, funded by COMPETE 2020 and FCT, Portugal. This work is funded by FCT/MEC through national funds and when applicable co-funded by FEDER – PT2020 partnership agreement under the project UID/EEA/50008/2019. The calculations were performed using the LGMII (NWO grant \#621.016.701). \newline

\textit{Data availability.} A data reproduction package for this paper can be found at the following DOI: 10.5281/zenodo.6319686
\end{acknowledgements}

\bibliographystyle{aa} 
\bibliography{teske_breaking_triples}

\begin{appendix} 
\section{Overview of formation rates}
\label{app:rates}
In this appendix, we give a detailed overview of the formation rates for each combination of outcomes, type of interaction, and model. The possible outcomes are discussed in detail in Sect.\,\ref{sec:ejection}-\ref{sec:cpu}. 
In Tbl.\,\ref{tbl:rates}, we give the formation fractions for the models of the initial populations of triples (OBin, T14, and E09) as well as for the dynamical models (pure N-body, N-body with constant winds, and N-body with stellar evolution). Tbl.\,\ref{tbl:dem} focuses on ejections and whether the democratic encounters lead to ejections of the primary, secondary, or tertiary star.

\begin{table*}
\caption{Fraction of systems formed for each outcomes as a function of all destabilised triples for the given model. There are three models of the initial populations of triples and three models for the dynamical simulations. 'Hier.' and 'Dem.' refer to hierarchical and democratic encounters respectively. Fractions are given for the full population, and for the two largest groups of destabilised triples, that is consisting of three MS stars and one giant star and two MS stars.} 

\label{tbl:rates}      
\centering        
\begin{tabular}{|ll|ccc|ccc|ccc|}
\hline
&& \multicolumn{3}{c|}{OBin} & \multicolumn{3}{c|}{T14} & \multicolumn{3}{c|}{E09} \\
Outcomes &  &All& Hier. & Dem. & All & Hier. & Dem. & All & Hier. & Dem. \\
\hline
\multicolumn{2}{|c|}{All stellar types} &&& &&& &&& \\
\multirow{3}{*}{All}
& N-body &1.000 &0.80 &0.20 &1.000 &0.82 &0.18 &1.000 &0.83 &0.17 \\
& + wind &1.000 &0.54 &0.46 &1.000 &0.67 &0.33 &1.000 &0.65 &0.35 \\
& + SE &1.000 &0.58 &0.42 &1.000 &0.69 &0.31 &1.000 &0.69 &0.31 \\
\hline
\multirow{3}{*}{Escape}
& N-body &0.174 &0.44 &0.56 &0.307 &0.61 &0.39 &0.233 &0.50 &0.50 \\
& + wind &0.307 &0.29 &0.71 &0.432 &0.45 &0.55 &0.338 &0.36 &0.64 \\
& + SE &0.433 &0.30 &0.70 &0.451 &0.43 &0.57 &0.415 &0.41 &0.59 \\
\hline
 \multirow{3}{*}{Drift}
& N-body &0.209 &0.65 &0.35 &0.111 &0.85 &0.15 &0.152 &0.80 &0.20 \\
& + wind &0.361 &0.38 &0.62 &0.171 &0.63 &0.37 &0.260 &0.51 &0.49 \\
& + SE &0.221 &0.56 &0.44 &0.115 &0.76 &0.24 &0.194 &0.70 &0.30 \\
\hline
\multirow{3}{*}{Collision}
& N-body &0.341 &0.93 &0.07 &0.326 &0.88 &0.12 &0.347 &0.93 &0.07 \\
& + wind &0.151 &0.90 &0.10 &0.231 &0.87 &0.13 &0.224 &0.95 &0.05 \\
& + SE &0.134 &0.90 &0.10 &0.241 &0.88 &0.12 &0.196 &0.95 &0.05 \\
\hline
\multirow{3}{*}{Bound triple}
& N-body &0.237 &0.99 &0.01 &0.155 &1.00 &0.00 &0.207 &1.00 &0.00 \\
& + wind &0.154 &0.99 &0.01 &0.074 &0.99 &0.01 &0.127 &1.00 &0.00 \\
& + SE &0.192 &0.99 &0.01 &0.105 &1.00 &0.00 &0.146 &0.99 &0.01 \\
\hline
\multirow{3}{*}{CPU time}
& N-body &0.039 &1.00 &0.00 &0.101 &1.00 &0.00 &0.061 &1.00 &0.00 \\
& + wind &0.027 &1.00 &0.00 &0.091 &1.00 &0.00 &0.051 &1.00 &0.00 \\
& + SE &0.020 &1.00 &0.00 &0.088 &1.00 &0.00 &0.049 &1.00 &0.00 \\
\hline
\multicolumn{2}{|c|}{Only ((MS,MS),MS)} &&& &&& &&& \\
\multirow{3}{*}{All}
& N-body &0.369 &0.73 &0.27 &0.559 &0.81 &0.19 &0.529 &0.79 &0.21 \\
& + wind &0.369 &0.73 &0.27 &0.559 &0.81 &0.19 &0.529 &0.79 &0.21 \\
& + SE &0.369 &0.68 &0.32 &0.559 &0.78 &0.22 &0.529 &0.79 &0.21 \\
\hline
\multirow{3}{*}{Escape}
& N-body &0.122 &0.49 &0.51 &0.236 &0.64 &0.36 &0.193 &0.55 &0.45 \\
& + wind &0.123 &0.48 &0.52 &0.236 &0.64 &0.36 &0.190 &0.55 &0.45 \\
& + SE &0.165 &0.45 &0.55 &0.240 &0.62 &0.38 &0.209 &0.58 &0.42 \\
\hline
 \multirow{3}{*}{Drift}
& N-body &0.081 &0.62 &0.38 &0.033 &0.99 &0.01 &0.069 &0.79 &0.21 \\
& + wind &0.081 &0.63 &0.37 &0.033 &0.99 &0.01 &0.069 &0.79 &0.21 \\
& + SE &0.063 &0.65 &0.35 &0.026 &0.83 &0.17 &0.067 &0.81 &0.19 \\
\hline
\multirow{3}{*}{Collision}
& N-body &0.111 &0.94 &0.06 &0.201 &0.88 &0.12 &0.198 &0.95 &0.05 \\
& + wind &0.110 &0.94 &0.06 &0.201 &0.88 &0.12 &0.199 &0.95 &0.05 \\
& + SE &0.104 &0.94 &0.06 &0.208 &0.89 &0.11 &0.185 &0.95 &0.05 \\
\hline
\multirow{3}{*}{Bound triple}
& N-body &0.031 &0.97 &0.03 &0.003 &1.00 &0.00 &0.024 &1.00 &0.00 \\
& + wind &0.031 &0.97 &0.03 &0.003 &1.00 &0.00 &0.024 &1.00 &0.00 \\
& + SE &0.018 &0.96 &0.04 &0.004 &1.00 &0.00 &0.030 &1.00 &0.00 \\
\hline
\multirow{3}{*}{CPU time}
& N-body &0.025 &1.00 &0.00 &0.086 &1.00 &0.00 &0.045 &1.00 &0.00 \\
& + wind &0.025 &1.00 &0.00 &0.086 &1.00 &0.00 &0.047 &1.00 &0.00 \\
& + SE &0.019 &1.00 &0.00 &0.080 &1.00 &0.00 &0.038 &1.00 &0.00 \\
\hline
\multicolumn{2}{|c|}{Only ((G,MS),MS)} &&& &&& &&& \\
\multirow{3}{*}{All}
& N-body &0.365 &0.86 &0.14 &0.308 &0.84 &0.16 &0.268 &0.85 &0.15 \\
& + wind &0.365 &0.38 &0.62 &0.308 &0.48 &0.52 &0.268 &0.37 &0.63 \\
& + SE &0.365 &0.48 &0.52 &0.308 &0.53 &0.47 &0.268 &0.47 &0.53 \\
\hline
\multirow{3}{*}{Escape}
& N-body &0.031 &0.44 &0.56 &0.049 &0.57 &0.43 &0.023 &0.38 &0.62 \\
& + wind &0.124 &0.16 &0.84 &0.146 &0.24 &0.76 &0.096 &0.12 &0.88 \\
& + SE &0.172 &0.18 &0.82 &0.156 &0.21 &0.79 &0.136 &0.20 &0.80 \\
\hline
 \multirow{3}{*}{Drift}
& N-body &0.056 &0.70 &0.30 &0.053 &0.70 &0.30 &0.042 &0.70 &0.30 \\
& + wind &0.158 &0.29 &0.71 &0.089 &0.48 &0.52 &0.121 &0.30 &0.70 \\
& + SE &0.084 &0.52 &0.48 &0.062 &0.70 &0.30 &0.076 &0.55 &0.45 \\
\hline
\multirow{3}{*}{Collision}
& N-body &0.166 &0.90 &0.10 &0.096 &0.87 &0.13 &0.103 &0.87 &0.13 \\
& + wind &0.033 &0.76 &0.24 &0.021 &0.89 &0.11 &0.013 &0.97 &0.03 \\
& + SE &0.025 &0.74 &0.26 &0.028 &0.90 &0.10 &0.006 &0.93 &0.07 \\
\hline
\multirow{3}{*}{Bound triple}
& N-body &0.102 &1.00 &0.00 &0.096 &1.00 &0.00 &0.089 &1.00 &0.00 \\
& + wind &0.049 &0.99 &0.01 &0.046 &0.99 &0.01 &0.034 &1.00 &0.00 \\
& + SE &0.082 &0.99 &0.01 &0.054 &0.99 &0.01 &0.044 &0.99 &0.01 \\
\hline
\multirow{3}{*}{CPU time}
& N-body &0.010 &1.00 &0.00 &0.014 &1.00 &0.00 &0.011 &1.00 &0.00 \\
& + wind &0.000 &- &- &0.006 &1.00 &0.00 &0.004 &1.00 &0.00 \\
& + SE &0.001 &1.00 &0.00 &0.008 &1.00 &0.00 &0.006 &1.00 &0.00 \\
\hline

\end{tabular}
\end{table*}

\begin{table*}
\caption{Fraction of democratic encounters in destabilised triples leading to ejections
compared to the total number of democratic encounters for a given model. Either the primary is ejected (Democratic-1), the secondary (Democratic-2), or tertiary (Democratic-3). Other outcomes do not typically come from a democratic encounter.
Table headers are similar to Tbl.\,\ref{tbl:rates}
}
\label{tbl:dem}      
\centering       
\begin{tabular}{|ll|ccc|ccc|ccc|}
\hline
&& \multicolumn{3}{c|}{OBin} & \multicolumn{3}{c|}{T14} & \multicolumn{3}{c|}{E09} \\
Outcomes &  &Dem.-1 & Dem.-2 & Dem.-3 & Dem.-1 & Dem.-2 & Dem.-3 & Dem.-1 & Dem.-2 & Dem.-3 \\
\hline
\multicolumn{2}{|c|}{All stellar types} &&& &&& &&& \\
\hline
\multirow{3}{*}{Escape}
& N-body &0.17 &0.55 &0.28 &0.11 &0.61 &0.29 &0.22 &0.65 &0.13 \\
& + wind &0.34 &0.51 &0.14 &0.32 &0.52 &0.16 &0.36 &0.55 &0.09 \\
& + SE &0.27 &0.60 &0.13 &0.23 &0.59 &0.18 &0.33 &0.59 &0.08 \\
\hline
 \multirow{3}{*}{Drift}
& N-body &0.28 &0.55 &0.17 &0.00 &0.95 &0.05 &0.21 &0.71 &0.09 \\
& + wind &0.35 &0.57 &0.08 &0.31 &0.66 &0.03 &0.37 &0.59 &0.03 \\
& + SE &0.33 &0.51 &0.17 &0.41 &0.50 &0.09 &0.35 &0.50 &0.16 \\
\hline
\multicolumn{2}{|c|}{Only ((MS,MS),MS)} &&& &&& &&& \\
\hline
\multirow{3}{*}{Escape}
& N-body &0.06 &0.61 &0.33 &0.05 &0.60 &0.36 &0.17 &0.66 &0.17 \\
& + wind &0.06 &0.61 &0.33 &0.05 &0.60 &0.36 &0.18 &0.65 &0.17 \\
& + SE &0.13 &0.62 &0.25 &0.07 &0.57 &0.36 &0.25 &0.59 &0.16 \\
\hline
 \multirow{3}{*}{Drift}
& N-body &0.03 &0.70 &0.27 &0.00 &1.00 &0.00 &0.41 &0.49 &0.11 \\
& + wind &0.04 &0.69 &0.27 &0.00 &1.00 &0.00 &0.41 &0.49 &0.11 \\
& + SE &0.17 &0.46 &0.37 &0.09 &0.64 &0.27 &0.47 &0.26 &0.26 \\
\hline
\multicolumn{2}{|c|}{Only ((G,MS),MS)} &&& &&& &&& \\
\hline
\multirow{3}{*}{Escape}
& N-body &0.14 &0.63 &0.23 &0.25 &0.58 &0.17 &0.13 &0.84 &0.03 \\
& + wind &0.42 &0.51 &0.07 &0.52 &0.42 &0.06 &0.47 &0.51 &0.02 \\
& + SE &0.27 &0.67 &0.06 &0.31 &0.61 &0.08 &0.30 &0.66 &0.04 \\
\hline
 \multirow{3}{*}{Drift}
& N-body &0.27 &0.66 &0.08 &0.00 &0.97 &0.03 &0.00 &0.97 &0.03 \\
& + wind &0.42 &0.55 &0.03 &0.40 &0.60 &0.00 &0.40 &0.58 &0.01 \\
& + SE &0.31 &0.60 &0.09 &0.46 &0.50 &0.04 &0.31 &0.63 &0.06 \\
\hline

\end{tabular}
\end{table*}

\section{Inclination dependence}
\label{app:inc}

The hierarchical triple star systems in this study start out in a stable configuration according to the stability criterion of Eq.\,\ref{eq:stab_crit}  \citep{Mar99, Aar01}. Due to stellar winds and dynamical perturbations between the inner and outer orbits, the level of stability of the triple evolves in time. A fraction of triples destabilises in the sense that at some moment during the evolution, the system crossed the adopted stability criterion. This is the moment where we change our integration method from the secular model to a direct one. The adopted stability criterion has an inclination dependence, such that triples with a mutual inclination between the inner and outer orbits larger than 90 degrees, that is `retrograde triples', are allowed to be more compact than their prograde counterparts. Here, compact means that the inner and outer orbits can be closer together. Hence, if we regard the ensemble of destabilised triples, there is a potential bias in the separation ratio due to the inclination. 

Having explored the rich variation in outcomes of the unstable phase of triple star evolution, we find that inclination plays an important role. For the pure N-body simulations without winds, prograde triples tend to produce more hierarchical ejections, while retrograde triples typically lead to democratic resonances. 
The models with stellar winds included make this distinction less strong, but it is still there. The aim of this appendix is to test whether the dependence on inclination is due to the inclination dependence in the stability criterion itself, or more interestingly, due to a real physical distinction in the orbital dynamics. 

We setup an experiment in which we focus on a specific triple star system with the following properties. Firstly, for the stellar masses and radii, we assume $m_1 = 1\,M_\odot$, $m_2 = 0.5\,M_\odot$, $m_3 = 0.75\,M_\odot$, and  $r_1 = 1\,R_\odot$, $r_2 = 0.5\,R_\odot$, $r_3 = 0.75\,R_\odot$. Secondly, for the inner and outer orbit, we adopt $a_{\rm in} = 10^4\,R_\odot$, $e_{\rm in} = 0.7$, and $e_{\rm out} = 0.7$.

\noindent The lightest body in our triple is therefore the secondary star.
We define four variations of this triple system in the following way:

\begin{itemize}
    \item Model A: i = $45^\circ$, $a_{\rm out}$ such that the system is on the stability limit,
    \item Model B: i = $135^\circ$, $a_{\rm out}$ such that the system is on the stability limit,
    \item Model C: i = $45^\circ$, $a_{\rm out}$ such that the system is on the stability limit if the inclination would be $i=135^\circ$,
    \item Model D: i = $135^\circ$, $a_{\rm out}$ such that the system is on the stability limit if the inclination would be $i=0^\circ$,
\end{itemize}

\noindent where $i$ is the mutual inclination between the inner and outer orbit.
Models A and B correspond to prograde and retrograde triples respectively, which are put exactly on the stability limit with their respective inclinations. In model C, we put the prograde triple on the more compact separation of its retrograde counterpart, while model D consists of a retrograde triple with the wider separation of a perfectly prograde triple. 
By comparing the outcomes of these four models, we determine the difference between prograde and retrograde orbits, and the effect of the inclination dependence of the stability criterion. 
We generate 200 realisations for models A, B and C, and 100 realisations for model D for which the angular elements are randomly generated. We adopt the same stopping conditions as used in the main text, except that we now introduce a maximum integration time of $10^5$ initial outer periods. We summarise the results in Tbl.~\ref{tbl:inc}. 

We first compare models A and B, which are the prograde and retrograde triples respectively, and which start off on the conventional stability limit. We confirm the trend found in the main study, that prograde triples produce more hierarchical interactions. These interactions typically lead to the dissolution of the triple. All solutions show that the tertiary star is ejected from the system - even though the secondary star has a lower mass. 
Simultaneously, democratic encounters are only found for retrograde systems. We also confirm the trend that for small $a_{\rm in}$, retrograde orbits produce many collisions.

Next, we investigate the influence of the inclination dependence in the stability limit. We first compare models A and C, which are the prograde models, with model C having a more compact outer orbit. As expected, we find that in the more compact case, there is a significant increase in the fraction of escapers. For all of these escaping solutions, the initial hierarchy was preserved throughout.
Comparing models B and D, which are the retrograde models with model D having a wider outer orbit, we find that triples in model D are too wide, as almost all of them remain bound and preserve their hierarchy for at least $10^5$ outer orbital periods. 
These results confirm that the outcome of a triple's evolution depends sensitively on the inclination, which should therefore be taken into account in any stability limit for hierarchical triple systems.  

By comparing model B and C, we are comparing prograde and retrograde triples with the same compactness, determined by the retrograde inclination. In this case, we still observe a difference between prograde and retrograde triples, in the sense that prograde triples mainly produce hierarchical escapers, while retrograde triples preferentially result in collisions and long-lived systems. This result thus supports the findings of the main study, that is the difference between the prograde and retrograde triples is physical and not a result of the inclination dependence in the stability limit.

\begin{table}
\caption{Fraction of ensemble that results in a specific outcome for each triple model defined in App.~\ref{app:inc}. The super- and subscripts refer to the fraction which remained hierarchical or became democratic, respectively.  }
\begin{center}
\begin{tabular}{ |l|l|l|l|l| } 
\hline
Outcome & Model A & Model B & Model C & Model D \\
\hline
Escape & $0.335^{1.00}_{0.00}$ & $0.160^{0.44}_{0.56}$ & $0.780^{1.00}_{0.00}$ & $0.000^{1.00}_{0.00}$ \\ 
 & & & & \\
Drift & $0.390^{1.00}_{0.00}$ & $0.075^{1.00}_{0.00}$ & $0.205^{1.00}_{0.00}$ & $0.000^{1.00}_{0.00}$ \\
 & & & & \\
Collision & $0.055^{1.00}_{0.00}$ & $0.380^{1.00}_{0.00}$ & $0.015^{1.00}_{0.00}$ & $0.020^{1.00}_{0.00}$ \\
 & & & & \\
Bound triple & $0.220^{1.00}_{0.00}$ & $0.385^{1.00}_{0.00}$ & $0.000^{1.00}_{0.00}$ & $0.980^{1.00}_{0.00}$ \\ 
\hline
\end{tabular}
\label{tbl:inc}
\end{center}
\end{table}

\section{Maximum terminal velocity}\label{app:v_max}

Most of the triples simulated in this work unfold into an unbound binary-single pair. Furthermore, these break-ups occur in a hierarchical fashion, that is energy from the inner binary is transferred to the tertiary during one or multiple pericentre passages. This eventually leads to the ejection of the tertiary, irrespective of its mass. Here, we provide semi-analytical estimates of the expected terminal velocity of tertiary stars, with an extrapolation to more compact systems than covered in the main text.  

The initial specific energy of the inner binary is given by:

\begin{eqnarray}
    \epsilon_{in,0} = -\frac{G\left( m_1+m_2 \right) }{2a_{\rm in}},
\end{eqnarray}

\noindent with $m_1$ and $m_2$ the masses of the primary and secondary respectively, $a_{\rm in}$ the initial inner semi-major axis, and $G$ the gravitational constant. After the ejection of the tertiary, the orbit of the inner binary will have shrunken (increased its binding energy). The smallest possible orbit is when the separation equals the sum of the radii of the two bodies, the associated energy being:

\begin{eqnarray}
    \epsilon_{in,1} = -\frac{G\left( m_1+m_2 \right) }{2\left( R_1 + R_2\right)} < \epsilon_0,
\end{eqnarray}

\noindent with $R$ the stellar radius. For wider binaries however, this amount of shrinkage is generally not achieved. Instead, the inner orbit shrinks by an average factor, $f_{\rm{shrink}}$. We estimate this shrinkage factor from our simulations, by regarding the model without stellar evolution, and comparing the binary separation before and after the ejection. We find the shrinkage factor to vary between 0.6-1.0. Here we adopt an average value 0.8. As an alternative model we adopt a value of 0.6. In terms of $f_{\rm{shrink}}$, the final energy is

\begin{eqnarray}
    \epsilon_{in,2} = -\frac{G\left( m_1+m_2 \right) }{2 f_{\rm{shrink}} a_{\rm in}} < \epsilon_0.
\end{eqnarray}

We now assume that the tertiary star extracts energy from the inner binary during one  pericentre passage. Since the total energy of the triple remains conserved, the energy lost by the inner orbit is gained by the outer orbit:

\begin{eqnarray}
    \Delta \epsilon_{out} = -\Delta \epsilon_{in} = \epsilon_{in,0} - \max{\left( \epsilon_{in,1}, \epsilon_{in,2}\right)}.
\end{eqnarray}

The specific energy of the initial outer orbit is:

\begin{eqnarray}
    \epsilon_{out} = \frac{1}{2} v_p^2 - \frac{G\left( m_1 + m_2 + m_3\right)}{r_p},
\end{eqnarray}

\noindent with $m_3$ the mass of the tertiary, and $r_p$ and $v_p$ the initial pericentre distance and speed of the outer orbit respectively. Taking the derivative with respect to $v_p$, we can approximate the gain in speed by:

\begin{eqnarray}
    \Delta v_p = \frac{\Delta \epsilon_{out}}{v_p}.
\end{eqnarray}

The total new speed is then calculated by adding the initial pericentre speed of the bound, outer orbit, $v_p$, and the gained speed, $\Delta v_p$. The terminal speed is obtained from 

\begin{equation}
    v_{\infty}^2 = \left(v_p+\Delta v_p \right)^2 - v_{esc}^2,
\end{equation}

\noindent with $v_{\infty}$ the terminal speed, and $v_{esc}$ the escape speed. Given three Sun-like stars, an initial inner orbital separation $a_{in}=10(5)\,\rm{R_\odot}$, an initial pericentre distance of the outer orbit $r_p=3a_{in}$ (roughly on the edge of stability), and initial outer eccentricity of 0.7, we estimate a terminal speed of $v_\infty = 67(95)\,\rm{km/s}$. In case of a stronger orbital shrinkage (i.e. $f_{\rm{shrink}}=0.6$), we find 157(222)\,km/s. 
For more massive stars of $10\,\rm{M_\odot}$, and an inner separation of $a_{in}=50(10)\,\rm{R_\odot}$, we find a terminal speed of $v_\infty = 95(213)\,\rm{km/s}$. Assuming strong orbital shrinkage (again $f_{\rm{shrink}}=0.6$), the terminal speed is 222(231)\, km/s.  We conclude that massive and compact triples, which destabilise hierarchically, have a high likelihood of being the birth sites of runaway stars.  

\section{CPU time stopping condition}\label{app:tcpu}

 We evolve each destabilised triple system by direct integration, until a stopping condition is fulfilled. Physically motivated conditions are: 1) a dissolution into a binary and single, 2) an ionisation into three singles, 3) a collision between any two bodies, or 4) age reached a Hubble time. A practical condition is: 5) CPU time has reached a pre-defined maximum. We initially set the maximum CPU time to 10 minutes, and simulate the full ensemble of triples. We find that a rather large fraction of the ensemble reach the CPU time barrier, rather than fulfilling a physical stopping condition. In an iterative manner, we simulate the unfinished triples again, but with an increased maximum CPU time. Our final maximum is 24 hours, which results in a percentage of unfinished simulations of about 10\% or smaller (depending on the stellar evolution model and initial condition). 

\begin{figure}
   \centering
   \includegraphics[width=\columnwidth]{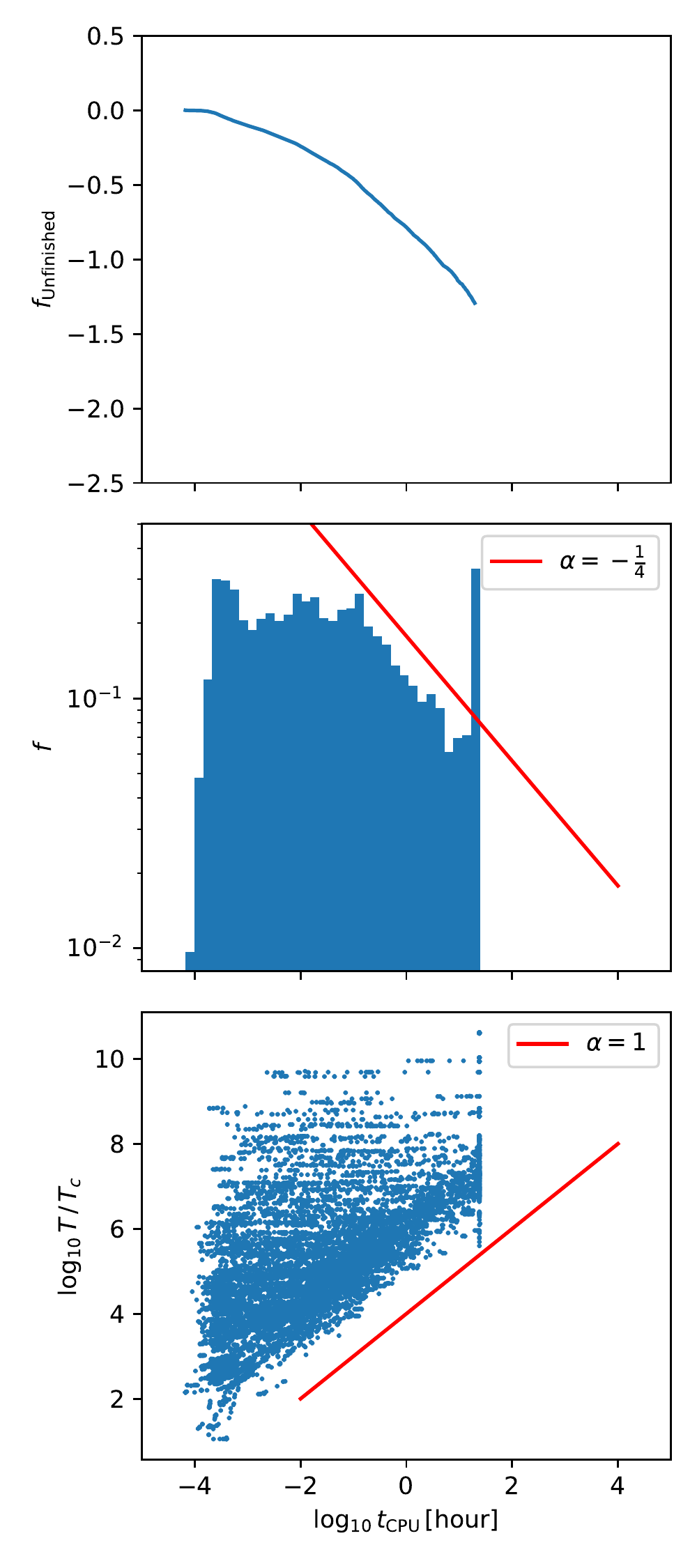}
      \caption{ For the ensemble of triples which included full stellar evolution, we plot 1) the fraction of unfinished simulations vs. CPU time (top), 2) histogram of CPU times (middle), and 3) simulated physical time normalised by the crossing time vs. CPU time (bottom). The red lines represent estimates of the slopes in the data, with coefficient $\alpha$.} 
         \label{fig:tcpu}
   \end{figure}

 Ideally, we would reduce the fraction of unfinished simulations to zero, but this is unfeasible. In Fig.~\ref{fig:tcpu}, we plot the fraction of unfinished simulations as a function of maximum CPU time. Extrapolating the curve by eye and assuming a similar final slope, we observe that in order to reach 1\% of unfinished simulations, some require $\sim 10^{3}$ hours to finish. 
The origin of these simulations that take a very long time to finish is the algebraic tail of long-lived, unstable, triple systems \citep{Orlov10}. In Fig.~\ref{fig:tcpu}, we also plot the correlation between simulated physical time and the maximum CPU time. We observe that the lower limit of the data grows approximately linearly. Extrapolating this trend, we observe that the lifetime reaches up to $10^{10}$ crossing times, with CPU times reaching up to $10^3$ hours and more. We emphasise that these CPU times are for a simulation of a single triple system, while the aim of our study was to perform a population synthesis. 
Our practical stopping condition of a maximum CPU time of 24 hours, is thus based on reducing the fraction of unfinished simulations to below 10\%. Considering that the aim of our study is to measure an order of magnitude estimate of outcome statistics, the completeness of $90\%$ or more is sufficient for our purposes.  

\end{appendix}

\end{document}